\documentclass[useAMS,usenatbib,letterpaper]{mnras}

\usepackage{epsfig}
\usepackage{amsmath}
\usepackage{natbib}
\usepackage{aas_macros,amssymb,graphicx,times,xcolor}
\title{Science from a glimpse: Hubble SNAPshot observations of massive galaxy clusters} \author{A. Repp and H. Ebeling}

\author[]{
Repp A.\ \& Ebeling H.\\
Institute for Astronomy, University of Hawaii, 2680 Woodlawn Drive, Honolulu, HI 96822, USA}

\date{MNRAS 479, 844--864 (2018)}

\begin{document}
\label{firstpage}
\pagerange{\pageref{firstpage}--\pageref{lastpage}}
\maketitle

\begin{abstract}
\textit{Hubble Space Telescope} SNAPshot surveys of 86 X-ray selected galaxy clusters at $0.3 < z < 0.5$ from the MACS sample have proven invaluable for the exploration of a wide range of astronomical research topics. We here present an overview of the four MACS SNAPshot surveys conducted from Cycle 14 to Cycle 20 as part of a long-term effort aimed at identifying exceptional cluster targets for in-depth follow up by the extragalactic community.  We also release redshifts and X-ray luminosities  of all clusters observed as part of this initiative. To illustrate the power of SNAPshot observations of MACS clusters, we explore several aspects of galaxy evolution illuminated by the  images obtained for these programmes. We confirm the high lensing efficiency of X-ray selected clusters at $z>0.3$. Examining the evolution of the slope of the cluster red sequence, we observe at best a slight decrease with redshift, indicating minimal age contribution since $z\sim 1$. Congruent to previous studies' findings, we note that the two BCGs which are significantly bluer ($\geq 5\sigma$) than their clusters' red sequences reside in relaxed clusters and exhibit pronounced internal structure. Thanks to our targets' high X-ray luminosity, the subset of our sample observed with \textit{Chandra} adds valuable leverage to the X-ray luminosity--optical richness relation, which, albeit with substantial scatter, is now clearly established from groups to extremely massive clusters of galaxies. We conclude that SNAPshot observations of MACS clusters stand to continue to play a vital pathfinder role for astrophysical investigations across the entire electromagnetic spectrum.
\end{abstract}

\section{Introduction}
Massive clusters ($M{\ga}10^{15}$ M$_\odot$, $L_{\rm X}{\ga}10^{45}$ erg s$^{-1}$, 0.1--2.4 keV, $\sigma{\ga}1000$ km s$^{-1}$) are extremely rare systems that constitute sensitive probes of cosmological parameters even at modest redshifts
(e.g., \citealp{Mantz2008, Mantz2010a, Mantz2014}).
\nocite{Mantz2010b}
As the largest and most massive gravitationally bound systems in the Universe, they also represent ideal laboratories for astrophysical studies of the interactions and properties of dark matter, gas, and galaxies (e.g., \citealp{Markevitch2004, Bradac2008, Merten2011, Ebeling2014, Linden2014, Harvey2015}) as well as for investigations into the evolution of brightest cluster galaxies (BCGs), the most luminous stellar aggregations in the Universe (e.g., \citealp{Quillen2008, HL2012, Stott2012, Werner2014,Green2016}). Clusters also play a central role in attempts to trace large-scale cosmic flows to distances well beyond those accessible to galaxy surveys (e.g., \citealp{LauerPostman1994, HudsonEbeling1997, Kocevski2007, Kashlinsky2010, PC2014}). Finally, and importantly, massive clusters act as extremely powerful gravitational telescopes that allow us to detect and characterize distant background galaxies out to redshifts that would otherwise be inaccessible to observation (see \citealp{KneibNatarajan2011} for a review). As a result, the most massive galaxy clusters (and among these the most distant ones) are prized targets for a wide range of extragalactic research.

\begin{figure}
\leavevmode
\epsfxsize=8.5cm
\epsfbox{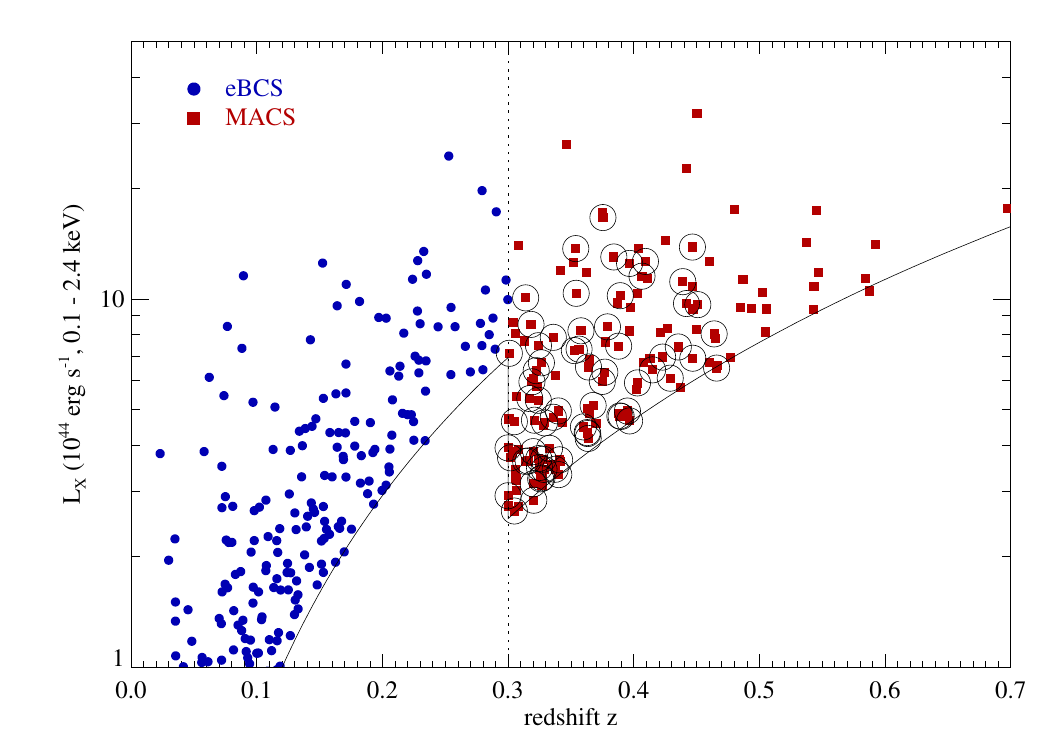}
\caption{X-ray flux limited cluster samples compiled from RASS data. Limited to systems at $z>0.3$, the MACS sample contains many of the most X-ray luminous (and hence most massive: \citealp{Stanek2006,Reichert2011,Zhang2017}) clusters in the universe. Clusters circled in black have been observed as part of our \textit{HST} SNAPshot programmes and are the subject of this paper. } \label{fig:lx-z} \end{figure}

X-ray luminous clusters have proven to be the most powerful gravitational `lenses', since they are -- by virtue of the X-ray selection process -- three-dimensionally bound, rarely affected by projection effects, and intrinsically massive (e.g., \citealp{Horesh2010}).  In addition, the frequency of strong-lensing features increases dramatically with cluster redshift \citep{Meneghetti2003}. Consequently, the cluster targets of choice have been provided by the Massive Cluster Survey (MACS; \citealp{Ebeling2001, Ebeling2007, Ebeling2010, Mann2012}). Like its counterparts at lower redshift, the BCS and REFLEX \citep{Ebeling1998, Ebeling2000, Bohringer2004}, MACS uses X-ray selection and optical follow-up observations to identify galaxy clusters among the tens of thousands of X-ray sources detected in the \textit{ROSAT} All-Sky Survey \citep[RASS;][]{Voges1999, Boller2016}. By focusing exclusively on systems at at $z>0.3$ and surveying a vast solid angle (in excess of 22,000 deg$^2$), MACS pursued a singular goal: a comprehensive search for the rarest, most massive clusters (Fig.~\ref{fig:lx-z}; see \citealp{Ebeling2001} for more details). Thanks to the very large solid angle covered, MACS \citep[like the \textit{Planck} cluster sample;][]{Planck2016} is complementary to cluster surveys covering tens, hundreds, or a few thousand square degrees; the latter, which include groundbased surveys exploiting the Sunyaev-Zel'dovich effect \citep{SZ1972} using, for instance, the South Pole Telescope or the Atacama Cosmology Telescope \citep{Williamson2011, Marriage2011}, primarily probe the population of average-mass clusters, due to the much smaller solid angles covered.

We here provide an overview of an observational programme that unites two valuable resources: the MACS legacy sample of the most massive clusters in the Universe at $z>0.3$, and the unparalleled resolution and sensitivity of the \textit{Hubble Space Telescope} (\textit{HST}). Complementing the pivotal research facilitated by \textit{HST}'s deep observations of galaxy clusters (exemplified most recently by the Hubble Frontier Fields initiative; Lotz et al.\ 2016, and references therein), the observatory's ability to acquire `snapshot' images of a large number of objects, randomly selected from a large target pool, has proven uniquely efficient in surveying entire sub-populations of objects. The most impressive testimony to date of the power of SNAPshot observations of very X-ray luminous clusters is provided by our \textit{HST} programmes GO-10491, -10875, -12166, and -12884, which imaged 86 clusters from the MACS sample, 28 of them in all of the four passbands chosen for this project (F606W, F814W, F110W, F140W). All proprietary rights were waived for these programmes, providing the scientific community with immediate access to all data.

In this paper we present and briefly explore several scientific applications of the MACS SNAPshot images, both to emphasize the wide range of astrophysical research topics one can address with these data and to stress their value for the selection of targets for detailed follow-up study. In keeping with these goals, our overview is not exhaustive, either in breadth or in depth: it does, however, aspire to convince the reader that SNAPshot surveys of massive galaxy clusters yield outstanding returns in terms of `science per exposure time.'

We structure this paper as follows: in Section~\ref{sec:data} we describe the data and their reduction, and remind the reader of the cluster-morphology classifications originally devised by \citet{Ebeling2007}.  In Section~\ref{sec:lens} we identify gravitational arcs produced by these powerful cluster lenses, before exploring, in Section~\ref{sec:BCG}, properties of the BCGs such as dust, signs of recent star formation, and their offset in colour from their host clusters' red sequence. In Section~\ref{sec:rs_slope} we then investigate the evolution of the cluster red sequence since $z \sim 0.5$, before, in Section~\ref{sec:Xcorrel}, investigating the correlation between cluster X-ray luminosity and optical richness. In Section~\ref{sec:legacy} we discuss the legacy value of this dataset and the release of the redshifts for all clusters in our sample. We present our conclusions in Section~\ref{sec:concl}.

All magnitudes are measured and reported in the AB system \citep{OkeGunn1983}; coordinates are quoted in the J2000.0 epoch. In our adopted  concordance $\Lambda$CDM cosmology ($\Omega_m=0.3$, $\Omega_\Lambda=0.7$, and $h=0.7$), one arcsecond on the sky corresponds to distances of 4.45 kpc and 5.76 kpc at $z=0.3$ and $z=0.45$, respectively.

\section{Observations and Data Analysis}
\label{sec:data}
\subsection{Groundbased Imaging}
\label{sec:uh22m}
As described in \citet{Ebeling2001}, the MACS project used images in three optical passbands obtained with the University of Hawaii 2.2-m telescope to confirm the presence of galaxy overdensities at the location of X-ray selected cluster candidates. Cluster candidates that meet the MACS X-ray selection criteria but which lie too far south ($\delta < -40^\circ$)  to be imaged in the optical from Maunakea were observed with facilities in the southern hemisphere. Due to the MACS team's limited access to the required resources, this southern extension of MACS (dubbed `SMACS') remains incomplete. An overview of the SMACS sample will be presented in a separate paper.

A brief overview of this imaging campaign also appears in \citet{Mann2012} who use data from X-ray observations in conjunction with the aforementioned UH 2.2-m images to sort MACS clusters into morphological classes reflecting their apparent dynamical state. This classification system, introduced by \citet{Ebeling2007}, assigns clusters to one of four classes ranging from the  most relaxed (1) to the most  disturbed (4), based on the system's appearance in the X-ray and optical wavebands. Class-1 clusters feature a compact core and excellent alignment between the peak of the X-ray emission and the location of the cluster's sole BCG; a Class-2 designation reflects reasonable X-ray/optical alignment and approximately concentric X-ray contours; Class 3 exhibits obvious small-scale structure and nonconcentric X-ray contours; and Class 4 systems show poor X-ray/optical alignment, multiple X-ray peaks, and no obvious BCG. Since not all of the clusters in this paper's MACS subsample have X-ray data, we expand this classification scheme to also assign morphology classes using optical data alone\footnote{For more sophisticated techniques and parameterisations see \citet{Wen2013} and references therein.}. Aiming again to coarsely subdivide the range from fully relaxed to extremely disturbed, we base our optical classification on the symmetry of the cluster galaxy distribution and the degree to which it is dominated by a single bright galaxy. Similar to the scheme used in the X-ray-based classification, the morphological extremes are thus characterized by a single agglomeration of galaxies dominated by a single and central BCG (Class 1) and an association of several distinct subclusters featuring multiple BCGs of comparable brightness (Class 4). A comparison of optical and X-ray classifications performed independently by both authors found them to be well correlated and to differ by at most one class, with one exception\footnote{The lone exception is SMACS0234.7$-$5831. Its X-ray morphology indicates that it is a fully relaxed system, but  a superimposed foreground group creates an optical appearance that led us (erroneously) to classify it as moderately disturbed.}. For seven SMACS clusters we forgo an optical morphology classification, since the availability of images in only one optical filter precludes a reliable discrimination between cluster galaxies and foreground objects.

\subsection{Hubble Space Telescope}

Our dataset consists of \textit{HST} images of the 86 MACS and SMACS clusters (Table~\ref{tab:clusters})
imaged by the aforementioned \textit{Hubble} SNAPshot programmes\footnote{GO-10491, GO-10875, GO-12166, GO-12884: PI Ebeling}. The integration times per cluster for each of the chosen four passbands appear in Table~\ref{tab:exposures}. 

\begin{table}
  \begin{minipage}{8.5cm}
  \centering
  \caption{Passbands and integration times}
  \label{tab:exposures}
  \begin{tabular}{ccc} \hline
Instrument	&  Filter	&  Integration\\\hline
ACS		&  F606W	& 1200 s\\
ACS		&  F814W	& 1440 s\\
WFC3	&  F110W	& 706 s\\
WFC3	&  F140W	& 706 s\\
\hline
 \end{tabular}
\end{minipage}
\end{table}

Note that, since the target list for these SNAPshot programmes excluded (by design) exceptional MACS clusters imaged previously with \textit{HST} -- such as the 12 MACS clusters at $z > 0.5$ \citep{Ebeling2007} -- the sample presented here is not representative of MACS as a whole.
In addition, due to the opportunistic nature of the SNAPshot programme, many of our targets do not have images in all four passbands (Table~\ref{tab:clusters} 
lists the passbands in which images were obtained for each cluster). Furthermore, images of a given target in different passbands are, in general, not aligned, since they were acquired at times and orientations dictated by \textit{HST}'s scheduling requirements.

The work described here focuses largely on images acquired with the Advanced Camera for Surveys (ACS, \citealp{ACS}) in the F606W and F814W bands. However, Table~\ref{tab:clusters}
also lists all near-infrared images obtained with the Wide-Field Camera 3 (WFC3, \citealp{WFC3}) in the F110W and F140W passbands, where the lack of atmospheric absorption and emission puts space-based observatories at a huge advantage over groundbased telescopes (we note specific applications of our WFC3 data in Section~\ref{sec:legacy}).

Although one can derive various colours from combinations of passbands used by our SNAPshot programmes, we here focus on observations in the F606W and F814W filters, both to identify and parametrize the cluster red sequence (Section~\ref{sec:rs_slope}) and to screen BCGs for internal features. To this end, we redrizzle\footnote{using \texttt{DrizzlePac}: \url{http://drizzlepac.stsci.edu}} all images (where possible) to the F814W reference frame and pixel scales, and then run \texttt{SExtractor}\footnote{Version 2.19.5} \citep{SExtractor} in dual-image mode, using F814W as the detection image. For clusters without an F814W image we use \texttt{SExtractor} in single-image mode on SNAPshots in other passbands -- on the F606W image if available, otherwise on the F110W image; for these clusters we confine ourselves to identifying the BCGs and second-brightest red sequence members (G2) (see Section~\ref{sec:BCG} and Table~\ref{tab:BCG_G2}) with the help of colour information provided by groundbased images obtained during the compilation of the MACS sample \citep{Ebeling2001}. We express all photometry in terms of \texttt{SExtractor} isophote magnitudes.
\begin{figure}
\leavevmode
\epsfxsize=8.5cm
\epsfbox{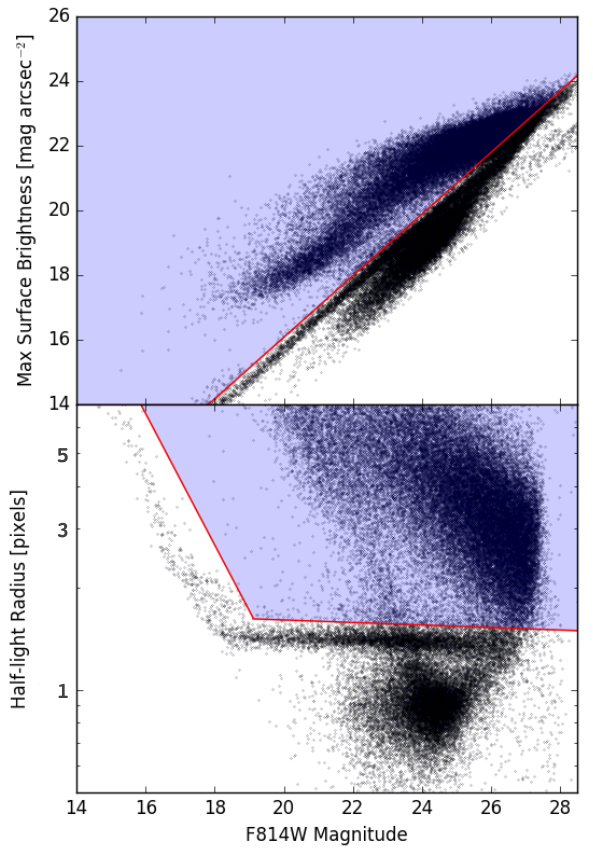}
\caption{Star-galaxy separation based on surface brightness and half-light radius. The diagonal line of sources in the top panel and the (approximately) horizontal line of in the bottom panel correspond to the locus occupied by point sources (stars).  We reject any source situated less than $2\sigma$ above this star line in either panel.} \label{fig:stargal} \end{figure}

In clusters with an F814W image, we then separate galaxies from stars and artefacts. 
As shown in Fig.~\ref{fig:stargal}, point sources (stars) occupy a well defined region (`star line') in the magnitude/surface brightness and magnitude/half-light radius planes. Any object on these lines is unresolved, and any detection below these lines is more compact than a point source and thus unphysical. We therefore select as galaxies only objects located at least $2\sigma$ above the star line in both panels.

The final two columns of Table~\ref{tab:clusters}
provide qualitative assessments of the cluster morphology in terms of relaxation state (see Section~\ref{sec:uh22m} for an overview).

\section{Gravitational Lensing}
\label{sec:lens}

One of the primary motivations of our SNAPshot programme was to survey MACS clusters for signatures of strong gravitational lensing in order to tentatively identify the most powerful cluster lenses. We here present a brief overview of the spectacular lensing features discovered in the course of this project; a discussion of some in-depth investigations enabled by our SNAP surveys appears in Section~\ref{sec:legacy}.

\begin{figure*}
\includegraphics[width=0.9\textwidth]{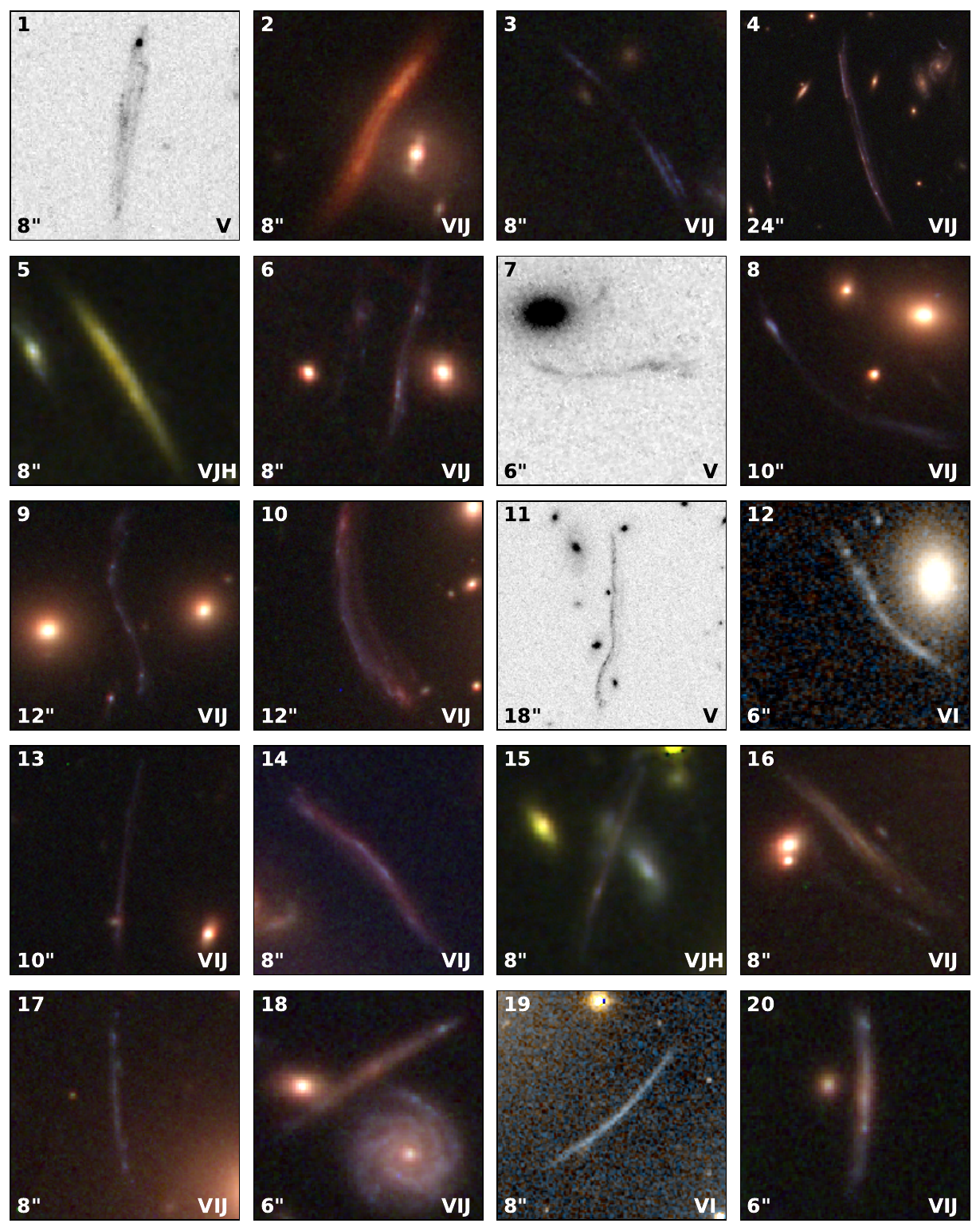}
\caption{A sample of the arcs identified by the algorithm of \citet{Horesh2005}. Each colour image is `true colour' in that the RGB brightnesses correspond to the specific intensities in the bandpasses used in constructing the image. All images utilize the same intensity stretch in order to communicate the range of brightnesses of these arcs. Note that all images do not represent the same solid angle on the sky; instead, the notation at the lower left of each image specifies the length (in arcseconds) of one side of the square. The coordinates (and lensing clusters) of each arc appear in Table~\ref{tab:arcs}, indexed by the numbers at the upper left of each image. The letters at the lower right of each image specify the photometric bands approximating the \textit{HST} filters used in constructing the image: $V=\mathrm{F606W}$; $I=\mathrm{F814W}$; $J=\mathrm{F110W}$; $H=\mathrm{F140W}$. Where three filters appear, they correspond to the blue, green, and red components, respectively, of the image. Where only two filters appear, they correspond to the blue and red components (respectively), the green component being the average of the intensity in the two filters. North is up, west is right.}
\label{fig:arcs}

\end{figure*}

\begin{figure*}
\includegraphics[width=0.9\textwidth]{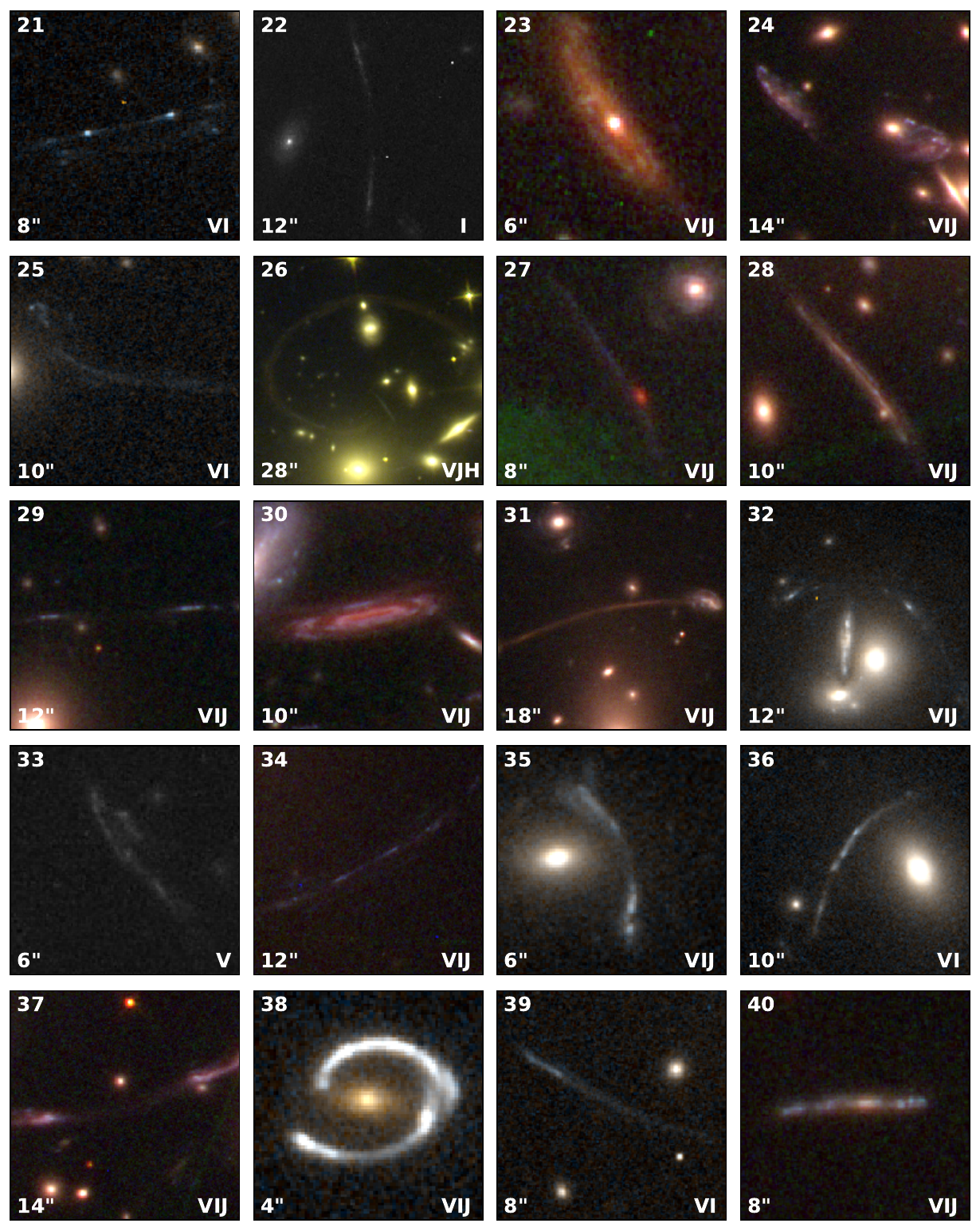}
\caption{A sample of the arcs \textit{not} detected by our implementation of the algorithm of \citet{Horesh2005}. See caption of Fig.~\ref{fig:arcs} for explanation of colour and stretch. The side length of each image appears at its lower left corner, and the letters at the lower right specify the photometric bands approximating the \textit{HST} filters used in constructing the image: $V=\mathrm{F606W}$; $I=\mathrm{F814W}$; $J=\mathrm{F110W}$; $H=\mathrm{F140W}$. The coordinates and host cluster of each arc appear in Table~\ref{tab:arcs}, indexed by the numbers at the upper left of the image. North is up, west is right.}
\label{fig:arcs2}
\end{figure*}

Figs.~\ref{fig:arcs} and \ref{fig:arcs2} show the most dramatic gravitational arcs produced by the gravity of our massive cluster targets. We separate this gallery into arcs detected by the automatic algorithm of \citet{Horesh2005} (Fig.~\ref{fig:arcs}), and other arcs (Fig.~\ref{fig:arcs2}) that, although often just as impressive to the eye, were missed by our implementation of the algorithm due to low surface brightness, over-zealous deblending, or a low length-to-width ratio. The colour images in these figures use `true colour' in the sense that the brightnesses of the RGB components are proportional to the specific intensities in the relevant filters. In addition, we display each image at the same dynamic stretch so that differences in image brightness consistently reflect actual differences in magnitude. Table~\ref{tab:arcs} lists the coordinates of each of the arcs and the corresponding cluster lens. In the following we discuss in more detail the implementation and findings of automated arc-finding algorithms.


\begin{table}
\begin{tabular}{llcc}
\hline\\[-6mm]
Arc            & Host Cluster & Right Ascension	&  Declination\rule{0cm}{15pt} \\ \hline\\[-4mm]
1  &  MACSJ0242.5$-$2132  &  02:42:37.32  &  $-$21:32:20.6 \rule[-3pt]{0cm}{18pt}\\
2  &  MACSJ0257.6$-$2209  &  02:57:39.01  &  $-$22:09:26.0 \rule[-3pt]{0cm}{0ex}\\
3  &  MACSJ0451.9+0006  &  04:51:53.44  &  $+$00:06:41.5 \rule[-3pt]{0cm}{0ex}\\
4  &  MACSJ0451.9+0006  &  04:51:57.13  &  $+$00:06:16.0 \rule[-3pt]{0cm}{0ex}\\
5  &  MACSJ0520.7$-$1328  &  05:20:40.33  &  $-$13:28:28.6 \rule[-3pt]{0cm}{0ex}\\
6  &  MACSJ0712.3+5931  &  07:12:17.60  &  $+$59:32:16.2 \rule[-3pt]{0cm}{0ex}\\
7  &  MACSJ1105.7$-$1015  &  11:05:47.54  &  $-$10:15:07.7 \rule[-3pt]{0cm}{0ex}\\
8  &  MACSJ1115.2+5320  &  11:15:17.55  &  $+$53:19:04.4 \rule[-3pt]{0cm}{0ex}\\
9  &  MACSJ1115.2+5320  &  11:15:18.26  &  $+$53:19:49.5 \rule[-3pt]{0cm}{0ex}\\
10  &  MACSJ1133.2+5008  &  11:33:14.20  &  $+$50:08:39.2 \rule[-3pt]{0cm}{0ex}\\
11  &  MACSJ1206.2$-$0847  &  12:06:10.74  &  $-$08:48:04.1 \rule[-3pt]{0cm}{0ex}\\
12  &  MACSJ1354.6+7715  &  13:54:09.67  &  $+$77:15:56.9 \rule[-3pt]{0cm}{0ex}\\
13  &  MACSJ1354.6+7715  &  13:54:24.47  &  $+$77:15:30.3 \rule[-3pt]{0cm}{0ex}\\
14  &  MACSJ1354.6+7715  &  13:54:41.98  &  $+$77:15:24.9 \rule[-3pt]{0cm}{0ex}\\
15  &  MACSJ1526.7+1647  &  15:26:45.11  &  $+$16:47:45.5 \rule[-3pt]{0cm}{0ex}\\
16  &  MACSJ1738.1+6006  &  17:38:08.04  &  $+$60:06:09.0 \rule[-3pt]{0cm}{0ex}\\
17  &  MACSJ2051.1+0215  &  20:51:09.91  &  $+$02:16:16.6 \rule[-3pt]{0cm}{0ex}\\
18  &  MACSJ2135.2$-$0102  &  21:35:10.57  &  $-$01:02:30.1 \rule[-3pt]{0cm}{0ex}\\
19  &  SMACSJ0549.3$-$6205  &  05:49:12.54  &  $-$62:06:18.4 \rule[-3pt]{0cm}{0ex}\\
20  &  SMACSJ2031.8$-$4036  &  20:31:46.22  &  $-$40:37:06.0 \\[-1mm]
\hline
21  &  MACSJ0032.1+1808  &  00:32:12.10  &  $+$18:07:53.4 \rule[-3pt]{0cm}{0ex}\\
22  &  MACSJ0034.4+0225  &  00:34:27.36  &  $+$02:25:18.2 \rule[-3pt]{0cm}{0ex}\\
23  &  MACSJ0140.0$-$0555  &  01:40:01.47  &  $-$05:55:09.7 \rule[-3pt]{0cm}{0ex}\\
24  &  MACSJ0152.5$-$2852  &  01:52:34.75  &  $-$28:53:51.8 \rule[-3pt]{0cm}{0ex}\\
25  &  MACSJ0308.9+2645  &  03:08:56.31  &  $+$26:45:06.3 \rule[-3pt]{0cm}{0ex}\\
26  &  MACSJ0520.7$-$1328  &  05:20:41.96  &  $-$13:28:34.7 \rule[-3pt]{0cm}{0ex}\\
27  &  MACSJ0947.2+7623  &  09:47:08.18  &  $+$76:23:24.1 \rule[-3pt]{0cm}{0ex}\\
28  &  MACSJ0947.2+7623  &  09:47:15.13  &  $+$76:23:03.0 \rule[-3pt]{0cm}{0ex}\\
29  &  MACSJ1142.4+5831  &  11:42:22.70  &  $+$58:31:31.7 \rule[-3pt]{0cm}{0ex}\\
30  &  MACSJ1142.4+5831  &  11:42:24.75  &  $+$58:31:16.4 \rule[-3pt]{0cm}{0ex}\\
31  &  MACSJ1142.4+5831  &  11:42:26.35  &  $+$58:32:53.4 \rule[-3pt]{0cm}{0ex}\\
32  &  MACSJ1142.4+5831  &  11:42:26.98  &  $+$58:30:47.3 \rule[-3pt]{0cm}{0ex}\\
33  &  MACSJ1206.2$-$0847  &  12:06:11.26  &  $-$08:47:43.7 \rule[-3pt]{0cm}{0ex}\\
34  &  MACSJ1319.9+7003  &  13:20:06.03  &  $+$70:04:26.7 \rule[-3pt]{0cm}{0ex}\\
35  &  MACSJ1354.6+7715  &  13:54:08.75  &  $+$77:15:50.5 \rule[-3pt]{0cm}{0ex}\\
36  &  MACSJ1452.9+5802  &  14:52:50.37  &  $+$58:01:35.7 \rule[-3pt]{0cm}{0ex}\\
37  &  MACSJ2135.2$-$0102  &  21:35:11.81  &  $-$01:03:35.3 \rule[-3pt]{0cm}{0ex}\\
38  &  MACSJ2135.2$-$0102  &  21:35:12.70  &  $-$01:01:44.0 \rule[-3pt]{0cm}{0ex}\\
39  &  MACSJ2149.3+0951  &  21:49:20.06  &  $+$09:51:26.1 \rule[-3pt]{0cm}{0ex}\\
40  &  SMACSJ0234.7$-$5831  &  02:34:39.09  &  $-$58:31:36.5 \rule[-8pt]{0cm}{0ex}\\[-2mm] \hline
\end{tabular}
\caption{Host clusters and coordinates for the gravitational arcs shown in Figs.~\ref{fig:arcs} and \ref{fig:arcs2}. Arc numbers correspond to the labels in the upper left corners of the images in the figures. Arcs 1--20 were detected by our implementation of \citet{Horesh2005}'s algorithm, whereas arcs 21--40 were not. \label{tab:arcs}}
\end{table}

The main incentive for the development of arc-finding algorithms is the desire to efficiently and objectively compare the observed incidence of gravitational arcs with that predicted by numerical simulations of large-scale structure in various cosmologies. The results of such comparisons differ greatly: while the study of \citet{Bartelmann1998} finds the observed incidence to be orders of magnitude lower than the one predicted by their simulations for a $\Lambda$CDM cosmology, \citet{Zaritsky2003} and \citet{Gladders2003} report observed arc incidences that significantly exceed the very same predictions. Attempts to better understand these discrepancies considered the effects of cluster triaxiality \citep{Oguri2003}, source redshift \citep{Dalal2004}, and cluster mergers \citep{Torri2004}. Simulations by \citet{Meneghetti2010} predicted that strong lensing ability would correlate with X-ray luminosity and depend on cluster orientation; comparison with 12 MACS clusters \citep{Meneghetti2011} still indicated a discrepancy between actual and observed lensing measures, but the discrepancy was much less than previous analytical models had predicted. (See also the review by \citealp{Meneghetti2013}.)

It was against this background that \citet{Horesh2005} developed an arc-finding algorithm in order to objectively compare
the observed frequency of giant arcs with that of arcs produced by simulated clusters lensing real backgrounds. The ensuing study led them to conclude that the observed and simulated arc incidences are in fact consistent. (See also \citealp{Horesh2011} for a similar study using the Millennium Simulation.) Using their algorithm to detect giant arcs in \textit{HST} imaging data, \citet{Horesh2010} find that X-ray selected clusters (their sample overlaps with ours) give rise to 6--8 times more giant arcs per cluster than those from the optically selected comparison sample, suggesting a difference in cluster mass of an order of magnitude.

While \citeauthor{Horesh2005}'s code is no longer publicly available, we attempt to reconstruct it from the published description in order to quantify the arc production of our cluster sample. We note that the algorithm is sensitive to the background-estimation procedure as well as to various parameter settings in the data-reduction pipelines. These differences can affect the inclusion of marginal pixels in an arc, resulting in slightly different length-width ratios; in some cases, the arc will be split into two or more pieces, neither of which passes the length-to-width threshold. While our results are thus not directly comparable with those of \citet{Horesh2010} (even where their cluster sample overlaps ours), the algorithm provides a simple means of investigating whether lensing efficiency evolves with redshift.

\begin{figure}
    \leavevmode\epsfxsize=9cm\epsfbox{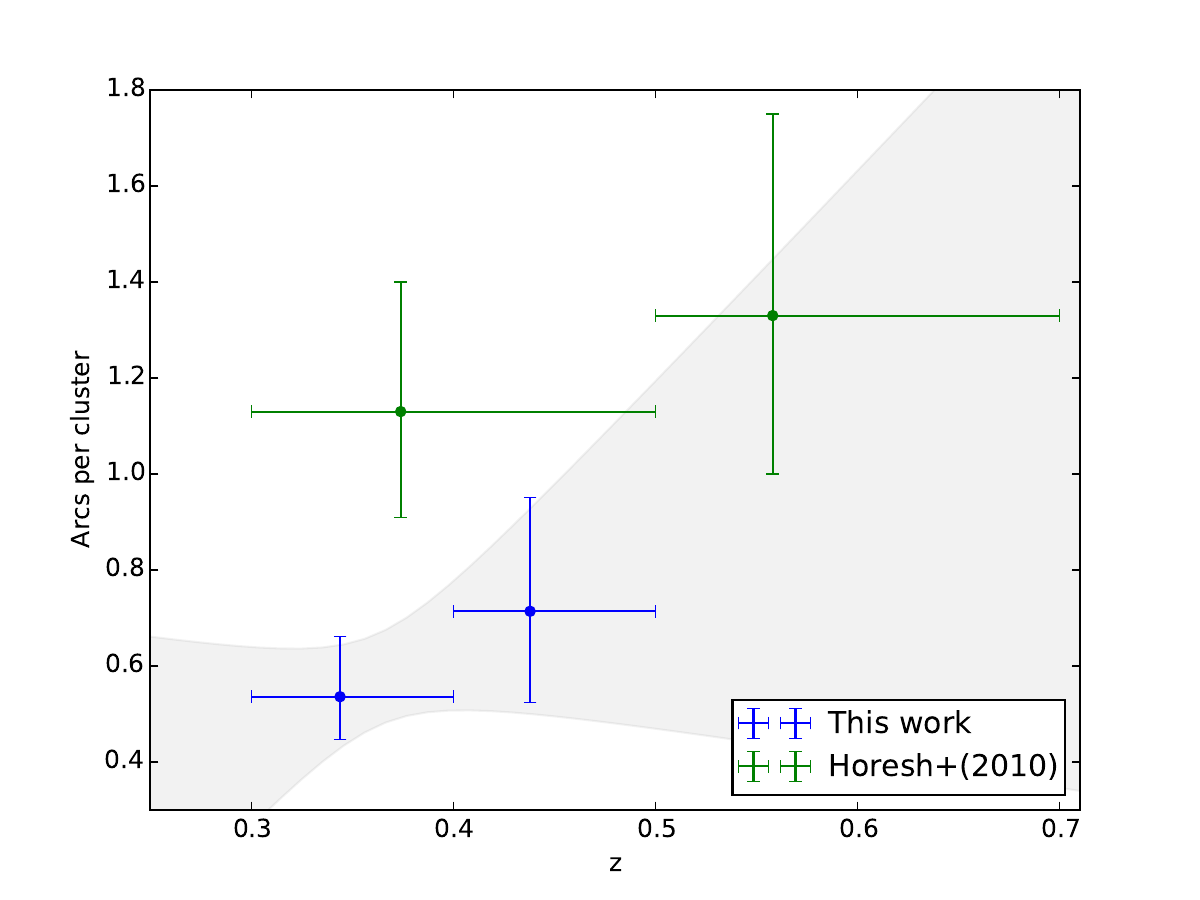}
    \caption{Arc-production efficiency of 77 MACS clusters, defined as the number of giant arcs ($\ell/w \ge 8$) detected by \citet{Horesh2005}'s algorithm divided by the number of clusters in the redshift bin. Vertical error bars indicate 68 per cent Poisson confidence intervals, and horizontal error bars indicate bin widths; data points appear at the mean redshift of the clusters in a given bin. We also show the results from \citet{Horesh2010}'s analysis. Note that one cannot directly compare the two sets of results due to the sensitivity of the algorithm to differences in the background estimation and the image-processing pipelines. The shaded area marks the $1\sigma$ error region for a linear fit to our two data points. Neither our data nor those of \citet{Horesh2010} rule out a redshift-independent arc-production efficiency.}
\label{fig:lens_eff}
\end{figure}

Sorting the clusters into redshift bins $[0.3, 0.4)$ and $[0.4, 0.5)$, we obtain the arc-production efficiencies (arcs per cluster) shown in Fig.~\ref{fig:lens_eff}. In agreement with the findings of \citet{Horesh2010}, more distant clusters tend to produce more arcs than nearer clusters, but the trend is not statistically significant within the redshift range probed by MACS. Note that our results are not in conflict with those of \citet{Xu2016} who, using a more robust (and not publicly available) arc-finding algorithm, examine the lensing efficiency of clusters observed by the CLASH programme and find a slight -- but again statistically insignificant -- decrease of lensing efficiency with redshift\footnote{Note that \citet{Xu2016} find lensing efficiences over four times greater than ours, an impact of using both a different algorithm and a less restrictive $\ell/w$ criterion.}.
Since lensing efficiency correlates with cluster mass, selection biases are bound to result in mild increases of the lensing efficiency with redshift for flux-limited cluster samples such as MACS (more distant clusters tend to be more massive) and may well create the opposite trend in samples without quantifiable selection criteria such as the one compiled for CLASH.

\section{BCG Properties} \label{sec:BCG} 

Although identifying powerful cluster lenses was the primary goal of our SNAPshot surveys, characterizing (at high angular resolution) the galaxy population of the cluster lenses themselves was an important secondary aim. Of particular interest in these contexts are two special classes of cluster members: galaxies observed in the process of being accreted by the cluster (either from the surrounding field or during a cluster merger), and (at the opposite end of the evolution spectrum) the giant ellipticals in the cluster centres, the BCGs. Insights into the properties of the former class of galaxies (often dubbed `jellyfish' galaxies) from our SNAPshot observations appear in \citet{Ebeling2014}, \citet{McPartland2016}, and Ebeling et al.\ 2018 (in preparation). We therefore focus here on the properties of MACS BCGs as viewed in \textit{HST} SNAPshots.

We use \textit{HST} photometry and colours, i.e., \texttt{SExtractor}-derived magnitudes, to identify the BCG as well as the second-brightest cluster galaxy (G2) for each cluster in our sample. We resort to groundbased imaging (see Section~\ref{sec:uh22m}) where colour information is not available from \textit{HST} and also routinely scrutinize our $7\arcmin\times 7\arcmin$ groundbased images to ensure that no brighter cluster member is present outside the field of view of our \textit{HST} images. We note that all BCGs thus identified have been spectroscopically confirmed as cluster members. For seven SMACS clusters for which neither ground- nor space-based colour information is readily available, our BCG identifications should be considered tentative; we also do not identify G2.
Table~\ref{tab:BCG_G2} lists our BCG and G2 magnitudes and coordinates as well as the passband in which the magnitude is calculated (F814W where available; F606W next in preference; and F110W if neither ACS band is available).  

\subsection{BCG morphology}

\begin{figure*}
\includegraphics[width=0.95\textwidth]{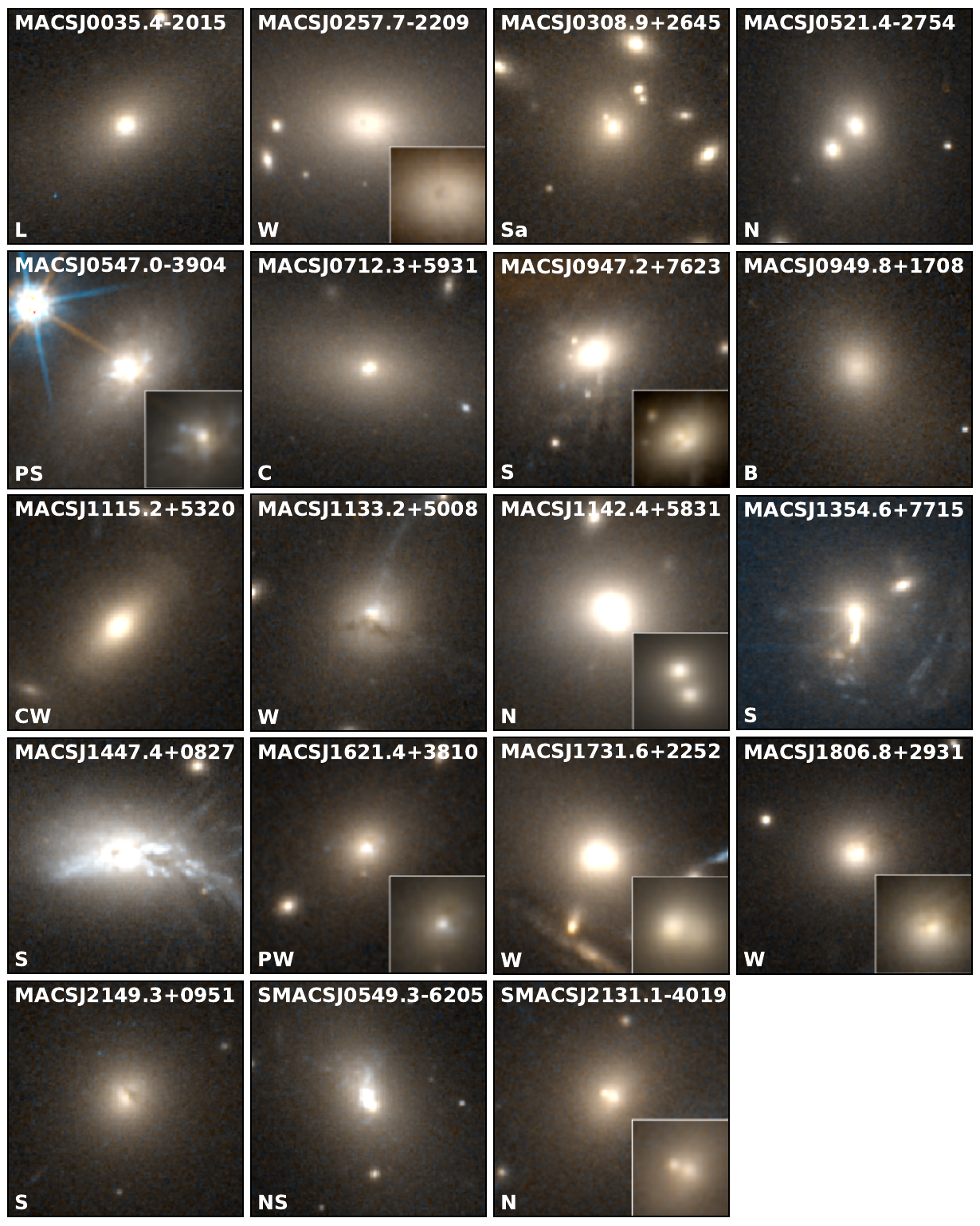}
\caption{Cut-outs (40 kpc each side) of BCGs showing various features of interest. All images use F110W, F814W, and F606W data for the r, g, and b channel, respectively. The code(s) in the lower left corner indicate(s) the presence of the following, as determined visually: B -- low surface brightness; C -- compact core; L -- layers/shells; N -- multiple nuclei; P -- point source; S/W -- strong/weak structure; Sa -- satellite galaxies. Main images displayed with equal brightness stretch. Insets display the central 10 kpc at a different brightness to show detail. North is up, and west is right.}
\label{fig:BCG_cutouts}
\end{figure*}

\begin{figure*}
\includegraphics[width=0.95\textwidth]{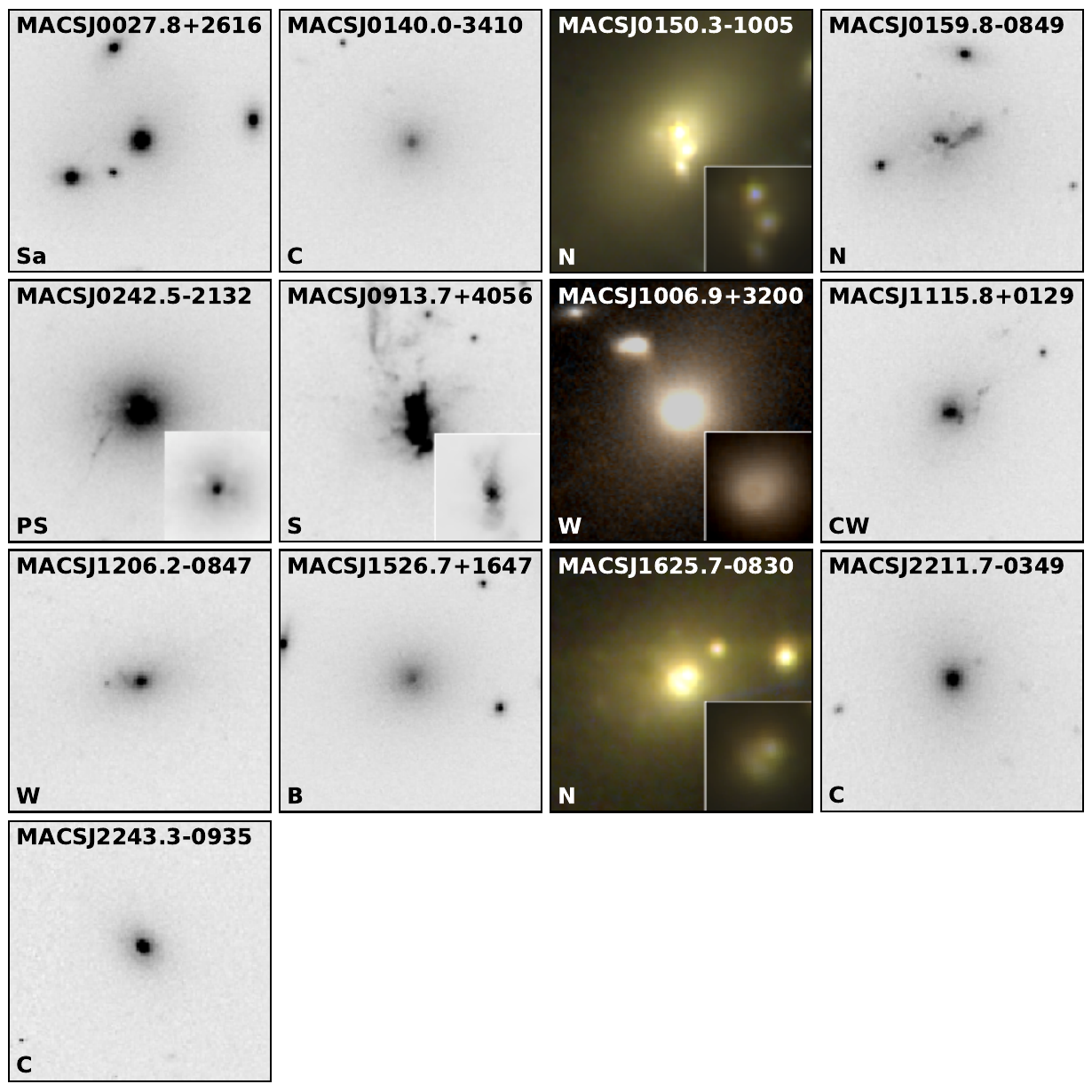}
\caption{Cut-outs (40 kpc each side) of BCGs showing various features of interest. The code(s) in the lower left corner indicate(s) the presence of the following, as determined visually: B -- low surface brightness; C -- compact core; N -- multiple nuclei; P -- point source; S/W -- strong/weak structure; Sa -- satellite galaxies. Main images displayed with equal brightness stretch. Insets display the central 10 kpc at a different brightness to show detail.  Monochrome images are from F606W; for MACSJ1006.9+3200, red is from F814W, blue is from F606W, and green is from the average of the two; for the remaining two images, red is from F140W, green from F110W, and blue from F814W. North is up, and west is right.}
\label{fig:BCG_cutouts2}
\end{figure*}

In an approach similar to that in Section~\ref{sec:lens}, we present a gallery of BCG images, as obtained by our SNAPshot programmes, in Fig.~\ref{fig:BCG_cutouts} (F606W/F814/F110W) and Fig.~\ref{fig:BCG_cutouts2} (various filters); all images span 40 kpc per side at the respective cluster redshift. As in Figs.~\ref{fig:arcs} and \ref{fig:arcs2}, the images in Fig.~\ref{fig:BCG_cutouts} and Fig.~\ref{fig:BCG_cutouts2} are shown at uniform stretch and contrast; the smaller inset images, however, use display parameters that have been adjusted to emphasize faint structural features in the very cores of these systems. Our gallery does not show all BCGs but only those that exhibit deviations from a regular, smooth surface brightness distribution. In both figures we highlight features of interest, such as central point sources, multiple nuclei, shells, unusually low surface brightness, compact (but not pointlike) cores, and irregular structure (most commonly caused by dust lanes or star bursts). We also mark a few BCGs as featuring satellite galaxies, but we note that the underlying apparent overdensity of galaxies near the BCG could be simply a projection effect.

We draw particular attention to extreme examples of pronounced and irregular structure indicative of active star formation; these include the BCGs of MACSJ0159.8$-$0849, MACSJ0242.5$-$2132, MACSJ0547.0$-$3904, MACSJ0913.7 +4056, MACSJ0947.2+7623 \citep[aka RBS797;][]{Schindler2001}, MACSJ1133.2+5008, MACSJ1354.6+7715, MACSJ1447.4+0827, and SMACSJ0549.3$-$6205. Also worth mentioning are the unusual, faint, central depressions in the surface brightness of the BCGs of MACSJ0257.7$-$2209 and MACSJ1006.9+3200, only visible at the custom stretch of the respective inset images in Figs.~\ref{fig:BCG_cutouts} and \ref{fig:BCG_cutouts2}; these are similar to features highlighted by \citet{Laine2003} in their study of \textit{HST} images of BCGs in nearby clusters ($z<0.06$). We consider the possibility that these features might be the result of a dynamical process like core scouring (see, e.g., \citealp{Begelman1980, Thomas2014}), in which mergers produce a (temporarily) binary black hole that ejects stars from the core. However, for both of these BCGs the luminosity decrease in the dark spot weakens in redder passbands, suggesting that the cause of the depression is extinction by dust clouds or rings. We thus speculate that MACSJ0257.7$-$2209 and MACSJ1006.9+3200 might be high-redshift equivalents of NGC3311, the BCG of A1060 \citep{Laine2003}. A more quantitative comparison with the data and classifications presented by \citet{Laine2003} is clearly warranted in view of the similarity of the features identified as dust rings, dust spirals, and circumnuclear disks in their work.

Overall, 14 out of 47 clusters with \textit{HST} SNAPshot images in both ACS filters (F606W and F814W) exhibit evidence of various types of activity, including cannibalism, star-formation, and dust, underlining yet again that, in the cores of massive clusters, giant ellipticals can be far from `red and dead.'

\subsection{Colour offsets and BCG dominance}

In spite of its dominant position among cluster members, the BCG does not always lie on the cluster red sequence. Both a steeper Kormendy relation \citep{Bildfell2008} and evidence of recent star formation in the BCGs of cool-core clusters (e.g., \citealp{Johnstone1987, Edge2001, Edwards2007}) suggest a different assembly history for these galaxies compared to that of other cluster ellipticals. \citet{Bildfell2008}'s analysis of the Canadian Cluster Comparison Project finds that about 25 per cent of the clusters host BCGs offset from their cluster's red sequence by 0.5 to 1.0 magnitudes in $(g' - r')$. In the following we explore whether similar offsets are present in the BCGs of our target clusters and, if so, how such colour offsets correlate with other BCG or host cluster properties.

\begin{figure}
    \leavevmode
    \epsfxsize=9cm\epsfbox{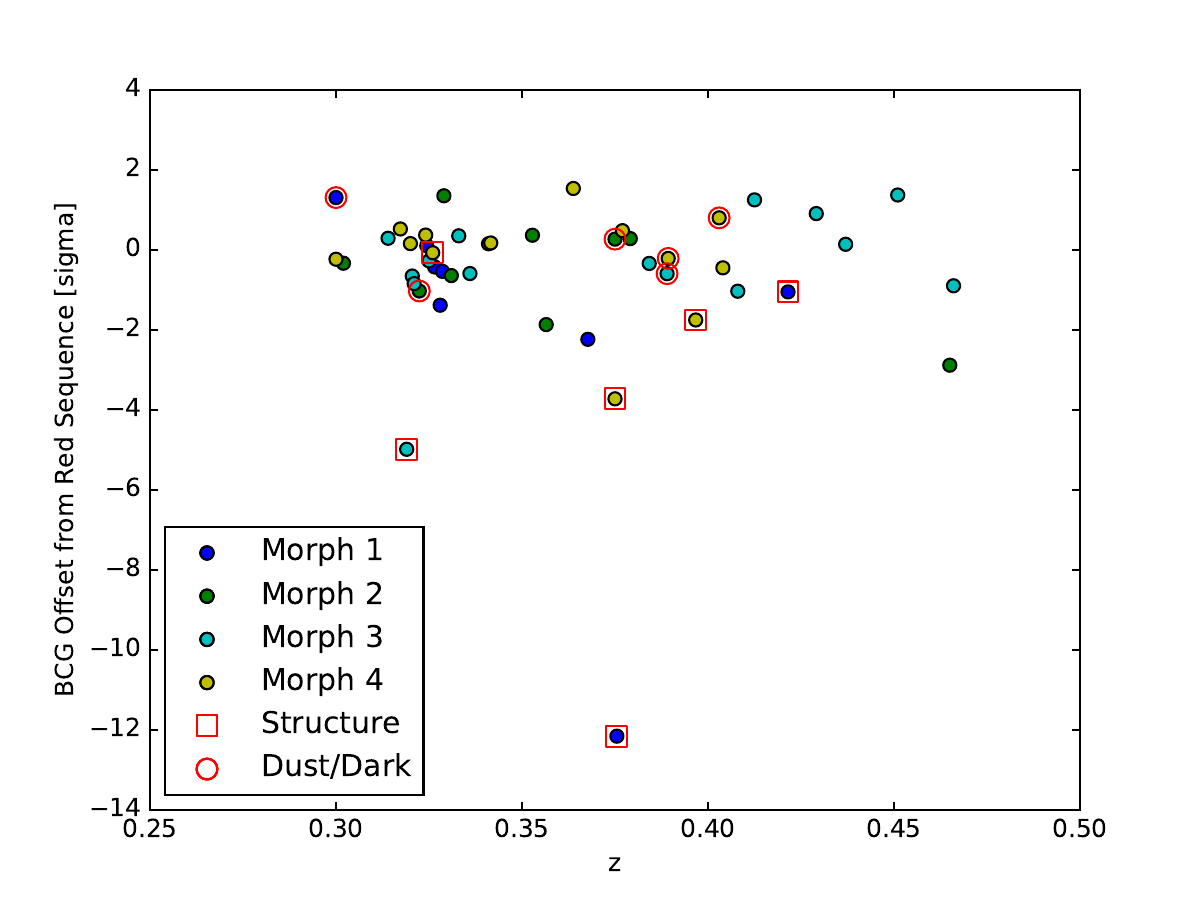}
    \caption{Colour offset (from the red sequence) of BCGs for 46 MACS clusters, in units of the
    $1\sigma$ red-sequence width, with morphology class as noted in the legend. BCGs with noted structural features (see Figs.~\ref{fig:BCG_cutouts}, \ref{fig:BCG_cutouts2}) are marked.  The largest offsets belong to the BCGs of MACSJ1447.4$+$0827 ($-12\sigma$), MACSJ0547.0$-$3904 ($-5.0\sigma$), and SMACSJ0549.3$-$6205 ($-3.7\sigma$). Although the BCG of MACSJ0947.2$+$7623 does not appear in this figure due to contamination of its F814W image, it would most likely be a fourth extreme outlier.}
    \label{fig:BCG_offset_sig}
\end{figure}

As noted earlier (see Table~\ref{tab:red_seq}), our SNAPs sample contains 47 clusters for which we can define the red sequence. From this set we exclude MACSJ0947.2+7623 because its BCG colour is suspect, due to an internal telescope reflection in the F814W image that extends into the BCG. For the remaining 46 clusters, we compute the colour offset of the BCG in units of the Gaussian width of the respective red sequence. We find several noticeably blue outliers; these appear in Fig.~\ref{fig:BCG_offset_sig}, which normalizes the offset by the red sequence width. If we consider absolute colour offsets, only the most extreme offset (from MACSJ1447.4+0827, with $\Delta$mag $= -0.46$) approaches the size of those noted by \citet{Bildfell2008}; the two next largest colour offsets (from SMACSJ0549.3$-$6205 and MACSJ0547.0$-$3904) are $-0.15$ and $-0.14$, respectively. We attribute this difference to a disparity in the colours used to define the red sequence: unlike our F606W and F814W passbands (at our clusters' median redshift of $z=0.36$), Bildfell and collaborators' $g'$ and $r'$ filters almost perfectly straddle the 4000\AA\ break (at their typical redshift of $z \sim 0.25$); hence, their $g'-r'$ colours will be more sensitive to ongoing star formation. Considering the observed colour offsets  in terms of $1\sigma$ red-sequence width, we find that 4 out of 46 BCGs (9 per cent) fall more than $2.5\sigma$ below the red sequence. \citet{Green2016}, using a somewhat different methodology, observe a similar fraction of BCGs to be offset in $g-r$.

While the BCG colour offset exhibits no obvious evolution with redshift (see Fig.~\ref{fig:BCG_offset_sig}), it clearly correlates with the presence of observable structure in the BCG. Only three BCGs feature offsets in excess of $-3\sigma$ (MACSJ0547.0$-$3904, SMACSJ0549.3$-$6205, and MACSJ1447.4+0827), and each of them displays significant internal structure in the form of striking blue filaments and apparent knots of star formation (see Figs.~\ref{fig:BCG_offset_sig} and \ref{fig:BCG_cutouts}). The BCG of MACSJ0547.0-3904  also exhibits a point-like core in our SNAPshot images, which coincides with an X-ray point source in our \textit{Chandra} image of the cluster and thus strongly suggests nuclear activity. In addition to these extreme examples, nine other BCGs (marked in Fig.~\ref{fig:BCG_offset_sig} and shown in Fig.~\ref{fig:BCG_cutouts}) are not significantly offset from the red sequence but nonetheless show internal structure, dust lanes, and/or dark spots, including the aforementioned BCGs of MACSJ0257.7$-$2209 and MACSJ1006.9+3200. Finally, MACSJ0308.9+2645 exhibits multiple galaxies near or within the halo of its BCG, suggesting either an extreme case of projection of unrelated cluster members, or imminent galactic cannibalism.

\begin{figure}
\includegraphics[width=0.5\textwidth]{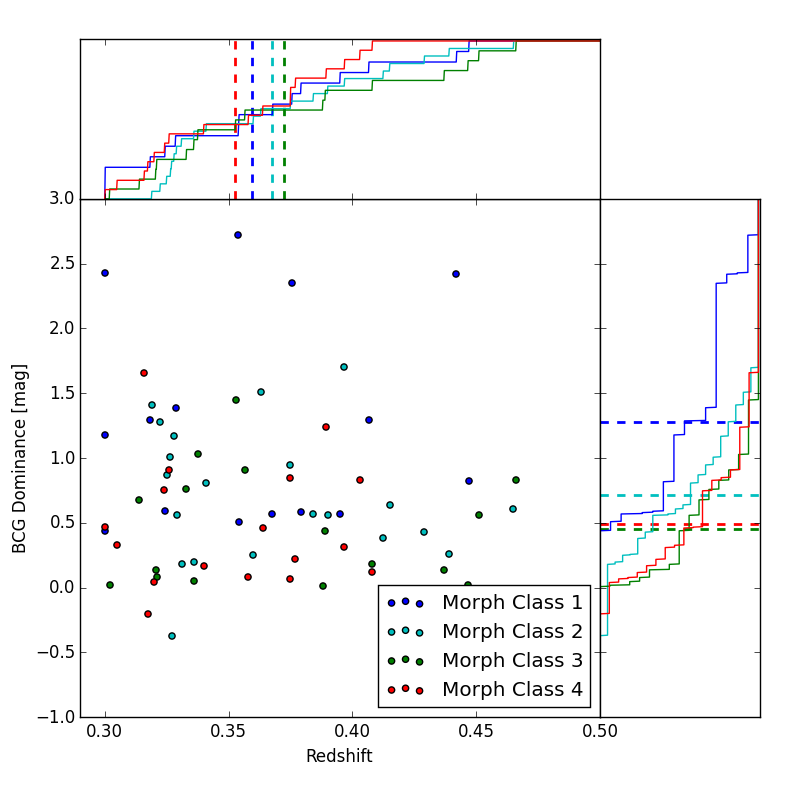}
\caption{BCG dominance as a function of redshift for the four cluster morphology classes described in Section~\ref{sec:data}. For each of the four morphology classes, the cumulative distribution functions of redshift and BCG dominance, as well as the corresponding mean values (dotted lines), appear in the top and right panels, respectively. }
\label{fig:BCGdom_cls}
\end{figure}

Finally, we explore trends in the BCG dominance (defined as the magnitude difference between the BCG and the second-brightest cluster galaxy). No correlation with redshift is observed, either for the full sample or for any subset when split according to the relaxation state of the host cluster, as measured by morphology class (Fig.~\ref{fig:BCGdom_cls}). Neither the cumulative redshift distribution functions nor the mean redshifts for the four morphology classes (top panel of Fig.~\ref{fig:BCGdom_cls}) show any sign of progression from disturbed to relaxed (or vice versa); applying two-sided Kolmogorov-Smirnov tests to the redshift distributions of the various morphology classes, one finds the largest difference between classes 2 and 4 ($D=0.27$), which is, however, insignificant ($p = 0.44$). Clear differences, however, are observed for the degree of BCG dominance between different morphology classes (e.g., $D=0.58$ for classes 1 and 4, corresponding to $p=0.005$); this outcome is unsurprising given that BCG dominance is one criterion for determining cluster morphology.
In addition, the roughly equal number of clusters in each morphology class demonstrates that the frequency of and timescale for cluster mergers is at most of the same order of magnitude as the cosmic time probed by this survey.

\section{Evolution of the Red-Sequence Slope}
\label{sec:rs_slope}
\begin{figure}
    \leavevmode\epsfxsize=9cm\epsfbox{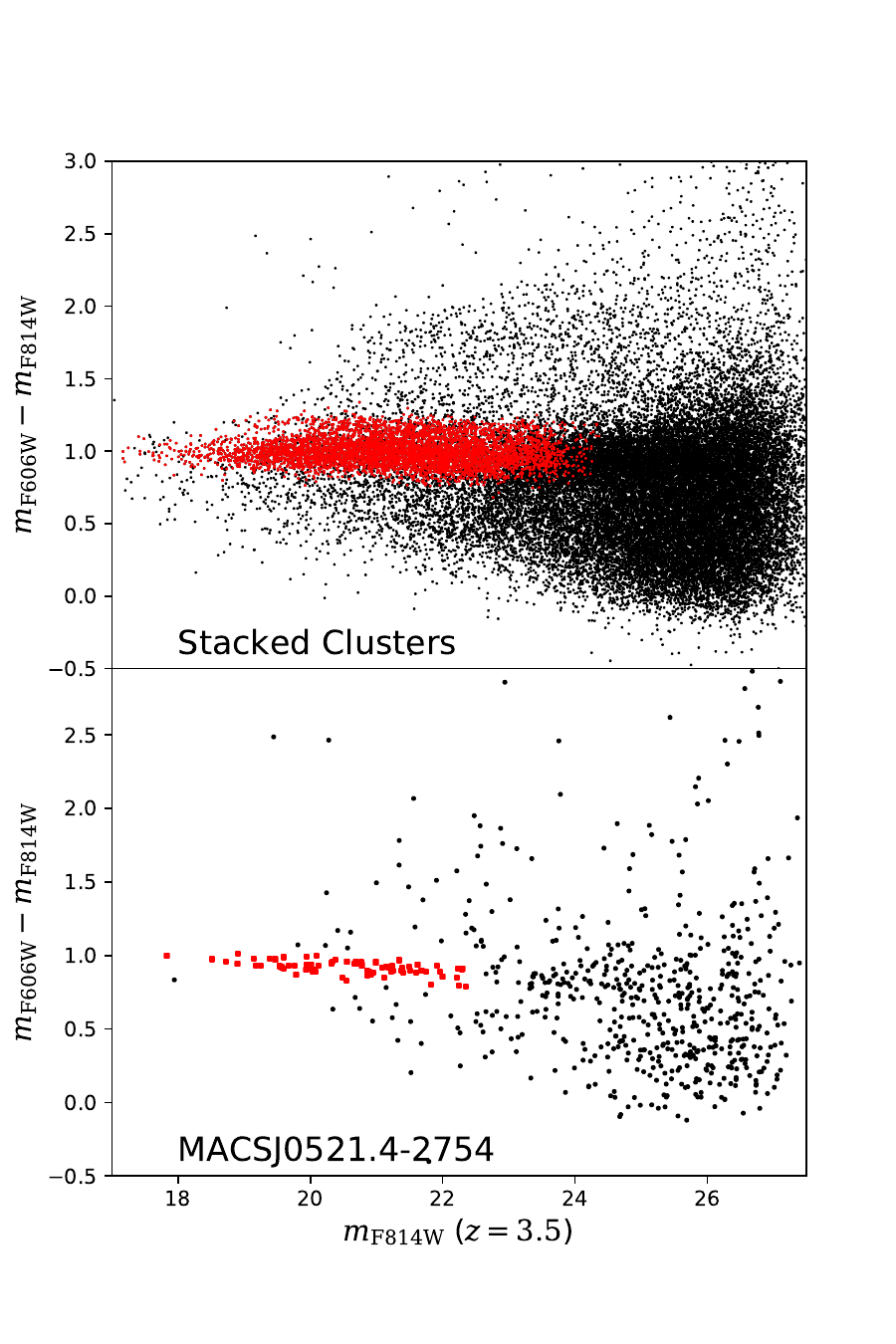}
    \caption{Colour-magnitude diagram based on F606W and F814W magnitudes illustrating the red sequence -- both for the stack of all clusters (top panel), and for one arbitrarily chosen cluster (MACSJ0521.4$-$2754, bottom panel). For the top panel, the F814W magnitudes have been $K$-corrected to a redshift of 0.35. Black points show galaxies in the ACS field of view; red squares show presumed cluster members, defined by an offset of less than $3\sigma$ from the red sequence for this cluster. Note (in the top panel) that the width of the stacked red sequence is inflated by variations in galactic extinction and reddening, with the most extreme cases appearing almost fully above the main red-sequence band.} \label{fig:rss_ex}
\end{figure}

Elliptical galaxies in clusters tend to form a tight `red sequence' in colour-magnitude space (Fig.~\ref{fig:rss_ex}). Tentative red sequences have been observed in clusters (or proto-clusters) at redshifts as high as $z \sim 2$ (e.g., \citealt{Tanaka2010, Spitler2012, Andreon2014}). 

The red sequence has been interpreted as a mass-metallicity relation, in the sense that bright, massive galaxies lose fewer of their metals to the intergalactic medium \citep{KodamaArimoto1997}; in this scenario, variations in stellar age would be the source of scatter along the sequence \citep{BowerLuceyEllis1992, Jaffe2011}. Other authors \citep{DeLucia2007, Smith2008, Stott2009} propose an age contribution to the red-sequence slope. In this scenario, faint galaxies migrate to the red sequence as they fall into the cluster and are quenched by interaction with intracluster gas; these recently quenched galaxies are at first bluer than the older, brighter cluster members, but by the same token they redden more rapidly. Hence, the red sequence slope would in this scenario flatten over time, i.e., would be more negative at higher redshift.

The observed slope does indeed evolve with redshift \citep{LopezCruz1997, Gladders1998, LopezCruz2004}; however, some -- and perhaps all -- of this evolution is the result of $K$ correction. In order to isolate any physical evolution, whether caused by a changing mass-metallicity relation or by an age contribution, we must thus transform the slopes to the cluster rest frame.

\begin{table*}
\begin{tabular}{lccccc}
\hline
 & \multicolumn{3}{c}{\dotfill Red Sequence\dotfill} & \multicolumn{2}{c}{\dotfill BCG\rule{0cm}{15pt}\dotfill} \\
 Name & Slope ($b$) & Zero-point ($a$) & half-width ($\sigma$) & $m_{606}$ & $m_{814}$\\ \hline
 MACSJ0011.7$-$1523 & $-0.027 \pm 0.004$ & $1.003 \pm 0.080$ & $0.042 \pm 0.004$ & 19.49 & 18.40\rule[-3pt]{0cm}{0ex}\\
MACSJ0032.1$+$1808 & $-0.026 \pm 0.003$ & $1.114 \pm 0.064$ & $0.039 \pm 0.004$ & 19.70 & 18.50\rule[-3pt]{0cm}{0ex}\\
MACSJ0035.4$-$2015 & $-0.027 \pm 0.004$ & $0.992 \pm 0.088$ & $0.036 \pm 0.004$ & 18.92 & 17.82\rule[-3pt]{0cm}{0ex}\\
MACSJ0110.1$+$3211 & $-0.023 \pm 0.004$ & $0.983 \pm 0.081$ & $0.025 \pm 0.003$ & 19.09 & 18.04\rule[-3pt]{0cm}{0ex}\\
MACSJ0140.0$-$0555 & $-0.025 \pm 0.005$ & $1.144 \pm 0.109$ & $0.024 \pm 0.003$ & 19.68 & 18.44\rule[-3pt]{0cm}{0ex}\\
MACSJ0152.5$-$2852 & $-0.029 \pm 0.003$ & $1.046 \pm 0.066$ & $0.035 \pm 0.003$ & 20.09 & 18.93\rule[-3pt]{0cm}{0ex}\\
MACSJ0257.6$-$2209 & $-0.023 \pm 0.004$ & $0.934 \pm 0.081$ & $0.033 \pm 0.004$ & 17.80 & 16.81\rule[-3pt]{0cm}{0ex}\\
MACSJ0308.9$+$2645 & $-0.028 \pm 0.004$ & $1.186 \pm 0.078$ & $0.038 \pm 0.004$ & 19.09 & 17.89\rule[-3pt]{0cm}{0ex}\\
MACSJ0451.9$+$0006 & $-0.027 \pm 0.004$ & $1.113 \pm 0.091$ & $0.033 \pm 0.003$ & 19.72 & 18.51\rule[-3pt]{0cm}{0ex}\\
MACSJ0521.4$-$2754 & $-0.025 \pm 0.006$ & $0.908 \pm 0.120$ & $0.036 \pm 0.005$ & 18.83 & 17.83\rule[-3pt]{0cm}{0ex}\\
MACSJ0547.0$-$3904 & $-0.025 \pm 0.007$ & $0.949 \pm 0.157$ & $0.029 \pm 0.005$ & 18.57 & 17.68\rule[-3pt]{0cm}{0ex}\\
MACSJ0712.3$+$5931 & $-0.021 \pm 0.004$ & $0.970 \pm 0.096$ & $0.026 \pm 0.004$ & 18.53 & 17.51\rule[-3pt]{0cm}{0ex}\\
MACSJ0845.4$+$0327 & $-0.018 \pm 0.004$ & $0.955 \pm 0.085$ & $0.037 \pm 0.004$ & 18.85 & 17.79\rule[-3pt]{0cm}{0ex}\\
MACSJ0916.1$-$0023 & $-0.032 \pm 0.005$ & $0.885 \pm 0.112$ & $0.053 \pm 0.004$ & 18.89 & 17.90\rule[-3pt]{0cm}{0ex}\\
MACSJ0947.2$+$7623 & $-0.036 \pm 0.003$ & $1.024 \pm 0.068$ & $0.008 \pm 0.004$ & 18.36 & 17.01\rule[-3pt]{0cm}{0ex}\\
MACSJ0949.8$+$1708 & $-0.025 \pm 0.003$ & $1.014 \pm 0.061$ & $0.022 \pm 0.003$ & 19.16 & 18.09\rule[-3pt]{0cm}{0ex}\\
MACSJ1006.9$+$3200 & $-0.025 \pm 0.005$ & $1.008 \pm 0.116$ & $0.047 \pm 0.004$ & 18.90 & 17.77\rule[-3pt]{0cm}{0ex}\\
MACSJ1115.2$+$5320 & $-0.042 \pm 0.004$ & $1.123 \pm 0.095$ & $0.044 \pm 0.004$ & 19.31 & 18.10\rule[-3pt]{0cm}{0ex}\\
MACSJ1124.5$+$4351 & $-0.023 \pm 0.004$ & $0.988 \pm 0.092$ & $0.026 \pm 0.003$ & 19.72 & 18.74\rule[-3pt]{0cm}{0ex}\\
MACSJ1133.2$+$5008 & $-0.001 \pm 0.009$ & $1.003 \pm 0.201$ & $0.043 \pm 0.007$ & 19.13 & 18.09\rule[-3pt]{0cm}{0ex}\\
MACSJ1142.4$+$5831 & $-0.016 \pm 0.004$ & $0.904 \pm 0.080$ & $0.038 \pm 0.004$ & 17.66 & 16.69\rule[-3pt]{0cm}{0ex}\\
MACSJ1226.8$+$2153C & $-0.034 \pm 0.005$ & $1.072 \pm 0.119$ & $0.032 \pm 0.005$ & 19.90 & 18.74\rule[-3pt]{0cm}{0ex}\\
MACSJ1236.9$+$6311 & $-0.024 \pm 0.004$ & $0.878 \pm 0.081$ & $0.028 \pm 0.003$ & 18.49 & 17.54\rule[-3pt]{0cm}{0ex}\\
MACSJ1258.0$+$4702 & $-0.023 \pm 0.007$ & $0.952 \pm 0.149$ & $0.029 \pm 0.004$ & 19.15 & 18.15\rule[-3pt]{0cm}{0ex}\\
MACSJ1319.9$+$7003 & $-0.028 \pm 0.004$ & $0.911 \pm 0.095$ & $0.035 \pm 0.005$ & 18.34 & 17.34\rule[-3pt]{0cm}{0ex}\\
MACSJ1328.2$+$5244 & $-0.030 \pm 0.005$ & $0.895 \pm 0.100$ & $0.035 \pm 0.004$ & 18.81 & 17.84\rule[-3pt]{0cm}{0ex}\\
MACSJ1354.6$+$7715 & $-0.020 \pm 0.003$ & $1.036 \pm 0.070$ & $0.025 \pm 0.001$ & 19.14 & 18.09\rule[-3pt]{0cm}{0ex}\\
MACSJ1447.4$+$0827 & $-0.042 \pm 0.005$ & $1.018 \pm 0.112$ & $0.037 \pm 0.005$ & 17.75 & 17.02\rule[-3pt]{0cm}{0ex}\\
MACSJ1452.9$+$5802 & $-0.021 \pm 0.003$ & $0.910 \pm 0.064$ & $0.029 \pm 0.003$ & 18.68 & 17.69\rule[-3pt]{0cm}{0ex}\\
MACSJ1621.3$+$3810 & $-0.029 \pm 0.006$ & $1.146 \pm 0.121$ & $0.036 \pm 0.004$ & 19.90 & 18.79\rule[-3pt]{0cm}{0ex}\\
MACSJ1644.9$+$0139 & $-0.030 \pm 0.004$ & $1.003 \pm 0.087$ & $0.036 \pm 0.004$ & 18.97 & 17.89\rule[-3pt]{0cm}{0ex}\\
MACSJ1652.3$+$5534 & $-0.017 \pm 0.006$ & $0.922 \pm 0.117$ & $0.041 \pm 0.006$ & 18.67 & 17.69\rule[-3pt]{0cm}{0ex}\\
MACSJ1731.6$+$2252 & $-0.022 \pm 0.004$ & $1.038 \pm 0.083$ & $0.035 \pm 0.003$ & 18.41 & 17.30\rule[-3pt]{0cm}{0ex}\\
MACSJ1738.1$+$6006 & $-0.028 \pm 0.003$ & $0.976 \pm 0.075$ & $0.034 \pm 0.003$ & 18.83 & 17.78\rule[-3pt]{0cm}{0ex}\\
MACSJ1752.0$+$4440 & $-0.015 \pm 0.005$ & $0.954 \pm 0.103$ & $0.045 \pm 0.005$ & 19.49 & 18.43\rule[-3pt]{0cm}{0ex}\\
MACSJ1806.8$+$2931 & $-0.017 \pm 0.005$ & $0.930 \pm 0.113$ & $0.034 \pm 0.004$ & 19.06 & 18.03\rule[-3pt]{0cm}{0ex}\\
MACSJ2050.7$+$0123 & $-0.023 \pm 0.006$ & $1.002 \pm 0.138$ & $0.053 \pm 0.008$ & 18.80 & 17.70\rule[-3pt]{0cm}{0ex}\\
MACSJ2051.1$+$0215 & $-0.034 \pm 0.004$ & $1.013 \pm 0.083$ & $0.032 \pm 0.004$ & 18.72 & 17.62\rule[-3pt]{0cm}{0ex}\\
MACSJ2135.2$-$0102 & $-0.019 \pm 0.004$ & $0.960 \pm 0.080$ & $0.018 \pm 0.002$ & 18.46 & 17.43\rule[-3pt]{0cm}{0ex}\\
MACSJ2149.3$+$0951 & $-0.006 \pm 0.009$ & $1.031 \pm 0.183$ & $0.050 \pm 0.011$ & 19.42 & 18.36\rule[-3pt]{0cm}{0ex}\\
MACSJ2241.8$+$1732 & $-0.018 \pm 0.003$ & $0.918 \pm 0.060$ & $0.027 \pm 0.003$ & 18.86 & 17.87\rule[-3pt]{0cm}{0ex}\\
SMACSJ0234.7$-$5831 & $-0.033 \pm 0.005$ & $1.054 \pm 0.110$ & $0.038 \pm 0.005$ & 19.87 & 18.78\rule[-3pt]{0cm}{0ex}\\
SMACSJ0549.3$-$6205 & $-0.033 \pm 0.005$ & $1.013 \pm 0.106$ & $0.039 \pm 0.004$ & 18.87 & 17.90\rule[-3pt]{0cm}{0ex}\\
SMACSJ0600.2$-$4353 & $-0.020 \pm 0.005$ & $0.952 \pm 0.101$ & $0.032 \pm 0.004$ & 18.29 & 17.27\rule[-3pt]{0cm}{0ex}\\
SMACSJ0723.3$-$7327 & $-0.030 \pm 0.005$ & $1.171 \pm 0.108$ & $0.052 \pm 0.005$ & 20.35 & 19.14\rule[-3pt]{0cm}{0ex}\\
SMACSJ2031.8$-$4036 & $-0.021 \pm 0.004$ & $0.965 \pm 0.079$ & $0.030 \pm 0.003$ & 18.81 & 17.77\rule[-3pt]{0cm}{0ex}\\
SMACSJ2131.1$-$4019 & $-0.033 \pm 0.008$ & $1.107 \pm 0.162$ & $0.045 \pm 0.005$ & 19.25 & 18.10\rule[-3pt]{0cm}{0ex}\\
\hline
\end{tabular}
\caption{Red sequence and BCG colour information. The red sequence is defined in the space ($m_\mathrm{F606W} - m_\mathrm{F814W}$) vs. $m_\mathrm{F814W}$, with functional form $C_{RS} = a + b \times (m_\mathrm{F814W} - 21.0)$. Magnitude uncertainties $\la .01$.} \label{tab:red_seq}
\end{table*}

Since our SNAPshot sample includes 47 clusters with both F814W and F606W coverage (see Table~\ref{tab:red_seq}), we use these clusters to investigate the evolution of the red-sequence slope, in these colours, from $z = 0.3$ to 0.5. We establish linear fits to the red sequence for each cluster iteratively down to a limit that is four magnitudes fainter than the mean of the second- and third-brightest cluster members. For a meaningful interpretation of the observed evolution of the slope, we must disentangle the effects of $K$ corrections from any intrinsic evolution of cluster or galaxy properties (e.g., in metallicity or stellar age). We thus transform the observed F606W$-$F814W slope to a rest-frame $U-V$ slope using the prescription from Appendix~B of \citet{Mei2009}, also employed by \citet{Cerulo2016}. This approach uses 42 synthetic stellar population models from the \citet{BC03} collection. These models include 7 formation redshifts (spaced evenly from $z=2$ to 5), three metallicities ($0.4Z_\odot$, $Z_\odot$, and $2.5Z_\odot$), and two star formation histories (instantaneous burst, and exponentially declining star formation with 1-Gyr $e$-folding). For cluster redshifts ranging from $z = 0.2$ to 0.6 we calculate the rest-frame $U-V$ colours and the observer-frame F606W$-$F814W colours for each model and then determine the best-fit linear relationship
\begin{equation}
C_{\rm{UV,rf}} = A(z) + B(z) \cdot C_{\rm{ACS,obs}}, \nonumber
\end{equation}
where $C_{\rm{ACS,obs}}$ denotes F606W$-$F814W colour. We thus obtain the colour conversion factors $A(z)$ and $B(z)$ as functions of redshift.

\begin{figure}
    \leavevmode\epsfxsize=9cm\epsfbox{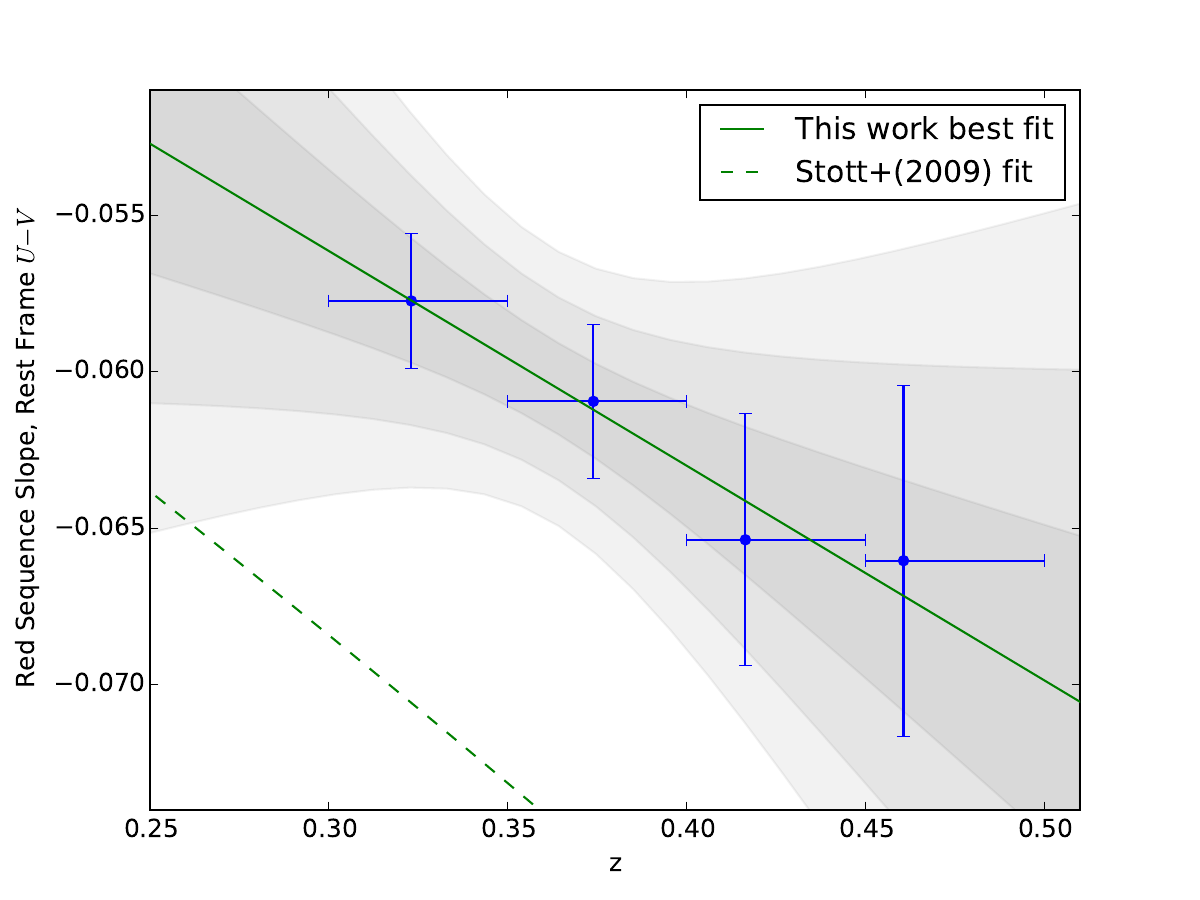}
    \caption{Red-sequence slope (rest-frame $U-V$) for 47 MACS clusters as a function of redshift. Horizontal error bars represent the bin widths ($\Delta z = 0.05$), and the horizontal placement of the data points shows the mean redshifts of the clusters in each bin. The solid green line shows the best-fit linear expression for the evolution of the slope; grey contours show the $1\sigma$-, $2\sigma$- and $3\sigma$-confidence regions for the fit. Note that our fit differs by only $2\sigma$ from a finding of no evolution. The dashed line shows the fit of \citet{Stott2009} -- see their data plotted in Fig.~\ref{fig:rss_evo_mon}.} \label{fig:rss_evo_rf} 
\end{figure}

The red-sequence slopes $m$ (and their uncertainties $\sigma_m$) in rest-frame and ACS colours are then related as follows:
\begin{eqnarray}
m_{\rm{rf}} & = &B(z) \cdot m_{\rm{ACS}}, \nonumber \\
\sigma_{\rm{m,rf}} &=& B(z) \cdot \sigma_{\rm{m,ACS}}. \nonumber
\end{eqnarray}
Converting the observed slopes into rest-frame $U-V$ slopes using this prescription, we find only weak intrinsic evolution that is in fact consistent with (i.e., differing by only two sigma from) no evolution for the $0.3 \le z \le 0.5$ range (Fig.~\ref{fig:rss_evo_rf}).

\begin{figure}
    \leavevmode\epsfxsize=9cm\epsfbox{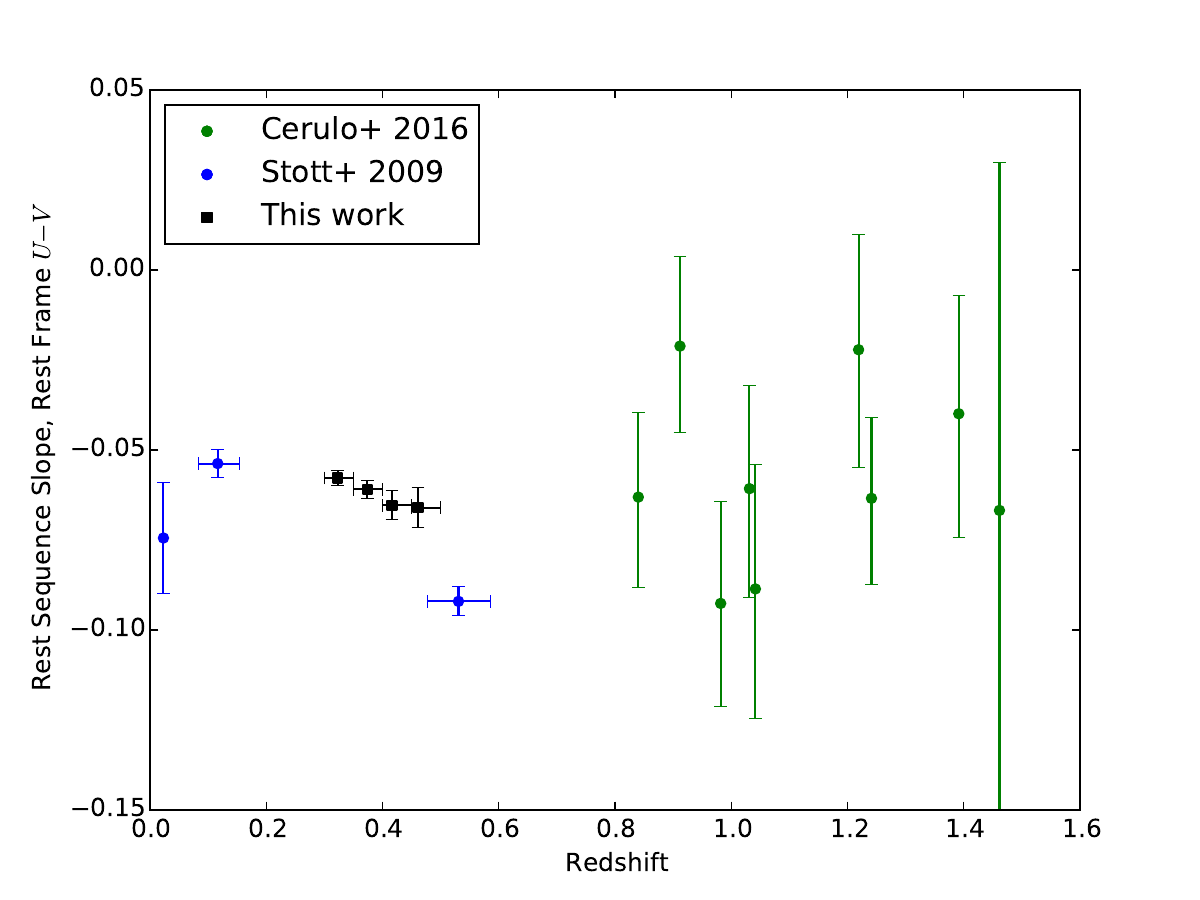}
    \caption{Our rest-frame red-sequence slope determinations compared with observations
    by \citet{Stott2009} and \citet{Cerulo2016}. The data below $z \la 0.6$ seem to show a weak decrease of slope with redshift. On the other hand, the higher-redshift data suggest no evolution since $z \sim 1.5$, subject to the admittedly large error bars. Fig.~\ref{fig:rss_evo_rf} specifies the binning of our data.} \label{fig:rss_evo_mon}
\end{figure}

\citet{Stott2009} analyse red-sequence slopes using the Coma cluster, a set of LARCS clusters (Las Campanas/AAT Rich Cluster Survey; \citealp{Pimbblet2001, Pimbblet2006}) at $z \sim 0.1$, and a set of MACS clusters at $z \sim 0.5$. (\citealt{Stott2009}'s set of MACS clusters partially overlaps ours.) They find significant evolution of the rest-frame slope with redshift, roughly consistent with (though lower than) our best-fit line shown in Fig.~\ref{fig:rss_evo_rf}. However, unlike theirs, our results are consistent with no evolution. These disparate results highlight the difficulty of comparing red-sequence slopes from different studies; challenges include not only intrinsic statistical scatter in the results but also systematic differences in fitting algorithms, photometric techniques, and parameter definitions. Acknowleding this caveat, we plot our results alongside those of \citet{Stott2009} in Fig.~\ref{fig:rss_evo_mon} and also show  observational data from the corresponding plot of \citet{Cerulo2016}'s fig.~4.
    
We note that our conclusion of weak or no evolution of the red-sequence slope is consistent with the general pattern of the data out to $z \sim 1.5$. By themselves, however, the data out to $z\sim 0.7$ suggest a weak flattening of the slope over time. Although the large uncertainties of corresponding results for individual high-redshift clusters limit their power to constrain any evolution, the data at $z>0.7$ suggest that a red sequence was already in place in clusters as distant as $z \sim 1.5$, and that the evolution of the red-sequence slope since that time was modest at best. Furthermore, we note the success of photometric redshift estimators that assume passive evolution of the red sequence following a high-redshift burst of star formation \citep{Song2012, Bleem2015, Ade2015}; the typical error of these estimators is only a few per cent. Acknowledging the aforementioned caveats, we thus conclude that a strong age contribution to the evolution of the red-sequence slope in massive clusters is not favoured by the existing data.

\section{X-ray Properties}
\label{sec:Xcorrel}

High-resolution X-ray data have been obtained with  \textit{Chandra} for 45 clusters in our sample. Table~\ref{tab:X_ray} lists the point-source-corrected X-ray luminosity $L_{\rm X}$ for the 0.01--2.4 keV band; it also lists the position of the peak of the X-ray surface brightness, measured after adaptive smoothing using the \textsc{Asmooth} algorithm \citep{Ebeling2006}. While a full analysis of the available X-ray data is far beyond the scope of this paper, we use the mentioned X-ray characteristics to explore two correlations with optical cluster properties derived from our SNAPshot data: the $L_{\rm X}$-richness relation and the X-ray/optical alignment of our cluster targets.

\begin{table}
\begin{tabular}{lcccc}
\hline \\[-6mm]
& \hspace{-7mm}$L_{\rm X}$\rule{0cm}{15pt} & \multicolumn{2}{c}{\hspace{-4mm}X-ray peak} & \\
Name	&  \hspace{-7mm}($10^{44}$ erg s$^{-1}$) & \hspace{-4mm}R.A. & Dec. & \hspace{-3mm} Rich$^c$\\ \hline \\[-4mm]
MACSJ0011.7$-$1523 & \hspace{-7mm}$\;\;9.5 \pm 0.1$ & \hspace{-3mm}00:11:42.88 & \hspace{-3mm}$-$15:23:21.8 & 50\rule[-3pt]{0cm}{0ex}\\
MACSJ0027.8$+$2616 & \hspace{-7mm}$\;\;3.8 \pm 0.2$ & \hspace{-3mm}00:27:45.37 & \hspace{-3mm}$+$26:16:25.7 & --\rule[-3pt]{0cm}{0ex}\\
MACSJ0035.4$-$2015 & \hspace{-7mm}$13.0 \pm 0.1$ & \hspace{-3mm}00:35:26.45 & \hspace{-3mm}$-$20:15:48.3 & 45\rule[-3pt]{0cm}{0ex}\\
MACSJ0140.0$-$0555 & \hspace{-7mm}$\;\;8.0 \pm 0.2$ & \hspace{-3mm}01:40:01.32 & \hspace{-3mm}$-$05:55:07.6 & 50\rule[-3pt]{0cm}{0ex}\\
MACSJ0150.3$-$1005 & \hspace{-7mm}$\;\;6.0 \pm 0.1$ & \hspace{-3mm}01:50:21.30 & \hspace{-3mm}$-$10:05:29.8 & --\rule[-3pt]{0cm}{0ex}\\
MACSJ0152.5$-$2852 & \hspace{-7mm}$\;\;8.2 \pm 0.2$ & \hspace{-3mm}01:52:34.46 & \hspace{-3mm}$-$28:53:36.3 & 53\rule[-3pt]{0cm}{0ex}\\
MACSJ0159.8$-$0849 & \hspace{-7mm}$17.3 \pm 0.2$ & \hspace{-3mm}01:59:49.41 & \hspace{-3mm}$-$08:49:58.5 & --\rule[-3pt]{0cm}{0ex}\\
MACSJ0308.9$+$2645 & \hspace{-7mm}$15.9 \pm 0.2$ & \hspace{-3mm}03:08:55.81 & \hspace{-3mm}$+$26:45:37.5 & 58\rule[-3pt]{0cm}{0ex}\\
MACSJ0404.6$+$1109 & \hspace{-7mm}$\;\;4.6 \pm 0.2$ & \hspace{-3mm}04:04:33.34 & \hspace{-3mm}$+$11:07:58.1 & --\rule[-3pt]{0cm}{0ex}\\
MACSJ0451.9$+$0006 & \hspace{-7mm}$\;\;7.3 \pm 0.2$ & \hspace{-3mm}04:51:54.40 & \hspace{-3mm}$+$00:06:20.3 & 34\rule[-3pt]{0cm}{0ex}\\
MACSJ0520.7$-$1328 & \hspace{-7mm}$\;\;9.3 \pm 0.1$ & \hspace{-3mm}05:20:42.03 & \hspace{-3mm}$-$13:28:50.0 & --\rule[-3pt]{0cm}{0ex}\\
MACSJ0547.0$-$3904 & \hspace{-7mm}$\;\;6.3 \pm 0.1$ & \hspace{-3mm}05:47:01.51 & \hspace{-3mm}$-$39:04:26.3 & 27\rule[-3pt]{0cm}{0ex}\\
MACSJ0712.3$+$5931 & \hspace{-7mm}$\;\;3.6 \pm 0.1$ & \hspace{-3mm}07:12:20.75 & \hspace{-3mm}$+$59:32:20.3 & 33\rule[-3pt]{0cm}{0ex}\\
MACSJ0913.7$+$4056 & \hspace{-7mm}$11.3 \pm 0.1$ & \hspace{-3mm}09:13:45.48 & \hspace{-3mm}$+$40:56:27.5 & --\rule[-3pt]{0cm}{0ex}\\
MACSJ0947.2$+$7623 & \hspace{-7mm}$22.3 \pm 0.3$ & \hspace{-3mm}09:47:13.04 & \hspace{-3mm}$+$76:23:14.2 & 31\rule[-3pt]{0cm}{0ex}\\
MACSJ0949.8$+$1708 & \hspace{-7mm}$11.3 \pm 0.2$ & \hspace{-3mm}09:49:51.70 & \hspace{-3mm}$+$17:07:08.2 & 36\rule[-3pt]{0cm}{0ex}\\
MACSJ1006.9$+$3200 & \hspace{-7mm}$\;\;6.9 \pm 0.2$ & \hspace{-3mm}10:06:54.57 & \hspace{-3mm}$+$32:01:39.3 & 31\rule[-3pt]{0cm}{0ex}\\
MACSJ1105.7$-$1014 & \hspace{-7mm}$\;\;6.9 \pm 0.2$ & \hspace{-3mm}11:05:46.58 & \hspace{-3mm}$-$10:14:38.9 & --\rule[-3pt]{0cm}{0ex}\\
MACSJ1115.2$+$5320 & \hspace{-7mm}$\;\;9.6 \pm 0.3$ & \hspace{-3mm}11:15:15.86 & \hspace{-3mm}$+$53:19:52.8 & 38\rule[-3pt]{0cm}{0ex}\\
MACSJ1115.8$+$0129 & \hspace{-7mm}$16.1 \pm 0.2$ & \hspace{-3mm}11:15:51.97 & \hspace{-3mm}$+$01:29:55.3 & --\rule[-3pt]{0cm}{0ex}\\
MACSJ1142.4$+$5831 & \hspace{-7mm}$\;\;7.7 \pm 0.1$ & \hspace{-3mm}11:42:24.01 & \hspace{-3mm}$+$58:31:59.7 & 37\rule[-3pt]{0cm}{0ex}\\
MACSJ1206.2$-$0847 & \hspace{-7mm}$21.1 \pm 0.2$ & \hspace{-3mm}12:06:12.16 & \hspace{-3mm}$-$08:48:01.4 & --\rule[-3pt]{0cm}{0ex}\\
MACSJ1226.8$+$2153C$^a$ & \hspace{-7mm}$\;\;1.2 \pm 0.1$ & \hspace{-3mm}12:26:41.20 & \hspace{-3mm}$+$21:52:58.4 & 36\rule[-3pt]{0cm}{0ex}\\
MACSJ1236.9$+$6311 & \hspace{-7mm}$\;\;6.9 \pm 0.1$ & \hspace{-3mm}12:36:58.89 & \hspace{-3mm}$+$63:11:12.0 & 36\rule[-3pt]{0cm}{0ex}\\
MACSJ1319.9$+$7003 & \hspace{-7mm}$\;\;5.0 \pm 0.1$ & \hspace{-3mm}13:20:08.44 & \hspace{-3mm}$+$70:04:36.5 & 36\rule[-3pt]{0cm}{0ex}\\
MACSJ1354.6$+$7715 & \hspace{-7mm}$\;\;7.0 \pm 0.1$ & \hspace{-3mm}13:54:42.71 & \hspace{-3mm}$+$77:15:17.1 & 25\rule[-3pt]{0cm}{0ex}\\
MACSJ1359.1$-$1929 & \hspace{-7mm}$\;\;5.7 \pm 0.2$ & \hspace{-3mm}13:59:10.23 & \hspace{-3mm}$-$19:29:24.9 & --\rule[-3pt]{0cm}{0ex}\\
MACSJ1427.6$-$2521 & \hspace{-7mm}$\;\;5.6 \pm 0.2$ & \hspace{-3mm}14:27:39.43 & \hspace{-3mm}$-$25:21:02.5 & --\rule[-3pt]{0cm}{0ex}\\
MACSJ1452.9$+$5802 & \hspace{-7mm}$\;\,\;\;6.0 \pm 0.1^b$ & \hspace{-3mm}14:52:57.57 & \hspace{-3mm}$+$58:02:57.1 & 47\rule[-3pt]{0cm}{0ex}\\
MACSJ1621.3$+$3810 & \hspace{-7mm}$\;\;8.6 \pm 0.2$ & \hspace{-3mm}16:21:24.84 & \hspace{-3mm}$+$38:10:08.5 & 32\rule[-3pt]{0cm}{0ex}\\
MACSJ1731.6$+$2252 & \hspace{-7mm}$\;\;8.4 \pm 0.2$ & \hspace{-3mm}17:31:39.07 & \hspace{-3mm}$+$22:51:51.9 & 43\rule[-3pt]{0cm}{0ex}\\
MACSJ2003.4$-$2322 & \hspace{-7mm}$\;\;8.7 \pm 0.1$ & \hspace{-3mm}20:03:25.40 & \hspace{-3mm}$-$23:24:55.5 & --\rule[-3pt]{0cm}{0ex}\\
MACSJ2046.0$-$3430 & \hspace{-7mm}$\;\;9.0 \pm 0.2$ & \hspace{-3mm}20:46:00.50 & \hspace{-3mm}$-$34:30:17.2 & --\rule[-3pt]{0cm}{0ex}\\
MACSJ2135.2$-$0102 & \hspace{-7mm}$\;\;6.4 \pm 0.1$ & \hspace{-3mm}21:35:11.19 & \hspace{-3mm}$-$01:02:55.8 & 29\rule[-3pt]{0cm}{0ex}\\
MACSJ2211.7$-$0349 & \hspace{-7mm}$26.3 \pm 0.3$ & \hspace{-3mm}22:11:46.00 & \hspace{-3mm}$-$03:49:47.3 & --\rule[-3pt]{0cm}{0ex}\\
MACSJ2229.7$-$2755 & \hspace{-7mm}$11.0 \pm 0.1$ & \hspace{-3mm}22:29:45.24 & \hspace{-3mm}$-$27:55:37.2 & --\rule[-3pt]{0cm}{0ex}\\
MACSJ2243.3$-$0935 & \hspace{-7mm}$\;\;9.0 \pm 0.1$ & \hspace{-3mm}22:43:21.05 & \hspace{-3mm}$-$09:35:42.6 & --\rule[-3pt]{0cm}{0ex}\\
MACSJ2245.0$+$2637 & \hspace{-7mm}$\;\;9.0 \pm 0.1$ & \hspace{-3mm}22:45:04.57 & \hspace{-3mm}$+$26:38:04.7 & --\rule[-3pt]{0cm}{0ex}\\
SMACSJ0018.9$-$4051 & \hspace{-7mm}$\;\;3.4 \pm 0.1$ & \hspace{-3mm}00:19:00.83 & \hspace{-3mm}$-$40:51:55.7 & --\rule[-3pt]{0cm}{0ex}\\
SMACSJ0040.8$-$4407 & \hspace{-7mm}$\;\;7.6 \pm 0.2$ & \hspace{-3mm}00:40:50.28 & \hspace{-3mm}$-$44:07:52.4 & --\rule[-3pt]{0cm}{0ex}\\
SMACSJ0234.7$-$5831 & \hspace{-7mm}$\;\;8.4 \pm 0.2$ & \hspace{-3mm}02:34:41.88 & \hspace{-3mm}$-$58:31:25.1 & 34\rule[-3pt]{0cm}{0ex}\\
SMACSJ0304.3$-$4401 & \hspace{-7mm}$\;\;7.1 \pm 0.2$ & \hspace{-3mm}03:04:16.75 & \hspace{-3mm}$-$44:01:32.3 & --\rule[-3pt]{0cm}{0ex}\\
SMACSJ0439.2$-$4600 & \hspace{-7mm}$\;\;4.5 \pm 0.1$ & \hspace{-3mm}04:39:14.06 & \hspace{-3mm}$-$46:00:50.4 & --\rule[-3pt]{0cm}{0ex}\\
SMACSJ0723.3$-$7327 & \hspace{-7mm}$15.6 \pm 0.2$ & \hspace{-3mm}07:23:18.23 & \hspace{-3mm}$-$73:27:16.0 & 36\rule[-3pt]{0cm}{0ex}\\
SMACSJ2031.8$-$4036 & \hspace{-7mm}$\;\;8.7 \pm 0.2$ & \hspace{-3mm}20:31:52.78 & \hspace{-3mm}$-$40:37:24.2 & 24\rule[-3pt]{0cm}{0ex}\\
 \hline\\[-6mm]
\end{tabular}
\caption{\textit{Chandra} X-ray properties of our SNAP target clusters. $L_{\rm X}$ is the point-source-corrected luminosity in the 0.1--2.4 keV band.\label{tab:X_ray}\newline
$^a$ This system is a  triple cluster (see \protect\citealp{Mann2012}), of which our SNAPshots cover only one component. The X-ray peak position and X-ray luminosity apply to this component only.\newline
$^b$ Lower bound; target placed on chip gap.\newline
$^c$ Optical richness, defined as the number of galaxies within the magnitude interval [m$_3$, m$_3$+2]. The numbers in this column are the raw counts (of red sequence members), before application of the corrections described in the text.
}
\end{table}

\subsection{$L_{\rm X}$--richness relation}
Since we need  \textit{Chandra} X-ray data and \textit{HST} SNAPshot images in both the F606W and F814W passbands to explore this relation, our sample is reduced to 22 clusters. 

In recognition of the dominance of ellipticals \citep[e.g.,][]{DeLucia07,Koester07} within the relatively small field of view of our SNAPshot data, we use the red sequence to determine cluster membership and define all galaxies within $3\sigma$ of the red sequence to be cluster members. This definition essentially eliminates the need to correct for interlopers from the fore- or background, in particular since projection effects are already greatly reduced by the relatively small angular extent\footnote{At our targets' redshifts, our \textit{HST} data cover only the cluster cores, i.e., a radius of 300 to 400 $h^{-1}$ kpc.} of our cluster targets on the sky \citep{YeeCruz99,Saro2013}. Deviating slightly from the approach taken in Section~\ref{sec:rs_slope}, we limit the resulting sample of cluster members, for each cluster, by imposing a maximal magnitude differential relative to the brightness of the third-brightest galaxy (m$_3$). Specifically, we follow \citet{abell58} and define optical richness as the number of galaxies within the magnitude interval [m$_3$, m$_3$+2].

\begin{figure}
    \includegraphics[width=0.5\textwidth]{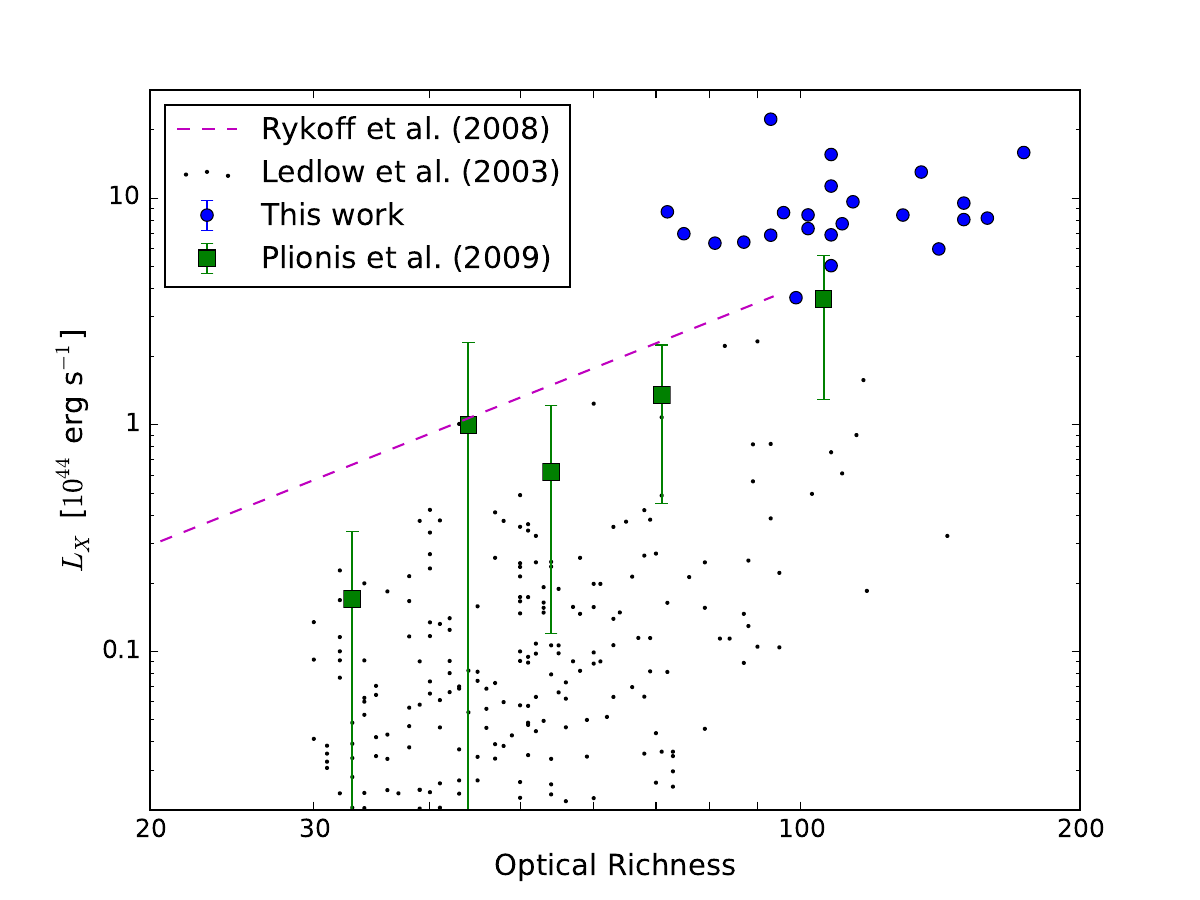}
    \caption{Correlation of X-ray luminosity (0.1--2.4 keV) with optical richness, defined as the number of red-sequence members for our work and that of \citet{Rykoff2008}, and as Abell richness in the studies by \citet{Ledlow2003} and \citet{Plionis2009}. Our data show a rather weak correlation ($r=0.49$) by themselves but provide valuable leverage when combined with literature data for less massive and more nearby clusters.
    Note that the statistical error bars on our data are too small to be visible. Adjustments applied in order to allow a meaningful comparison between these data sets are described in the text.}
\label{fig:X_num}
\end{figure}

Since MACS is, by design, limited to highly X-ray luminous clusters that tend to feature a commensurately high optical richness, the range in both $L_{\rm X}$ and richness of our sample is too small to allow a determination of the $L_{\rm X}$--richness relation from our data alone. However, our data provide a valuable complement to existing work at lower redshift for less massive clusters, such as the studies by \citet{Ledlow2003} and \citet{Plionis2009}; both of these studies employ subsamples of the Abell cluster catalogue and thus use Abell's richness counts, which employ the same magnitude interval as we do here ([m$_3$, m$_3$+2]). However, the 1.5-$h^{-1}$Mpc radius out to which galaxies contribute to Abell richness is far larger than the 300- to 400-$h^{-1}$kpc (radius) covered by our \textit{HST} data. In order to account for this discrepancy, we apply a global correction factor of 3 to our richness estimates, appropriate for massive clusters at $z\sim 0.35$ (assuming the universal satellite number density profile of \citealt{Budzynski12}).

We show the resulting $L_{\rm X}$--richness relation in Fig.~\ref{fig:X_num}.
The comparison of our results with those of \citet{Ledlow2003} and \citet{Plionis2009} illustrates another systematic effect, namely the contamination of the Abell's richness measurements by superimposed galaxy groups or clusters  \citep{Sutherland88,YeeCruz99}; such contamination artificially boosts the optical richness of Abell clusters at any given X-ray luminosity.
As another comparison with previous work, we also consider the $L_{\rm X}$--richness relation determined by \citet{Rykoff2008} for the maxBCG clusters, using red-sequence galaxy counts down to $0.4L^*$ and out to a radius of 750 $h^{-1}$kpc. After adjusting their results to satisfy Abell's richness definition (in terms of depth relative to m$_3$ and radial extent), we find that the  best-fit $L_{\rm X}$--richness relation derived by \citet{Rykoff2008} (dashed magenta line in Fig.~\ref{fig:X_num}) agrees well with our data points.

We conclude that \textit{HST} SNAPshots of MACS clusters significantly extend the dynamic range within which one can observationally establish the $L_{\rm X}$--richness relation. A comprehensive, quantitative analysis which includes the large body of work in the existing literature will, however, require careful accounting for a number of systematic effects in the definition and determination of optical richness: these effects include the fact that observational datasets reflect different optical passbands, over a range of cluster redshifts, in various magnitude ranges, out to different radii, and with distinct approaches to fore- and background contamination corrections.

\subsection{X-ray / optical offsets}

\begin{figure}
    \leavevmode\epsfxsize=9cm\epsfbox{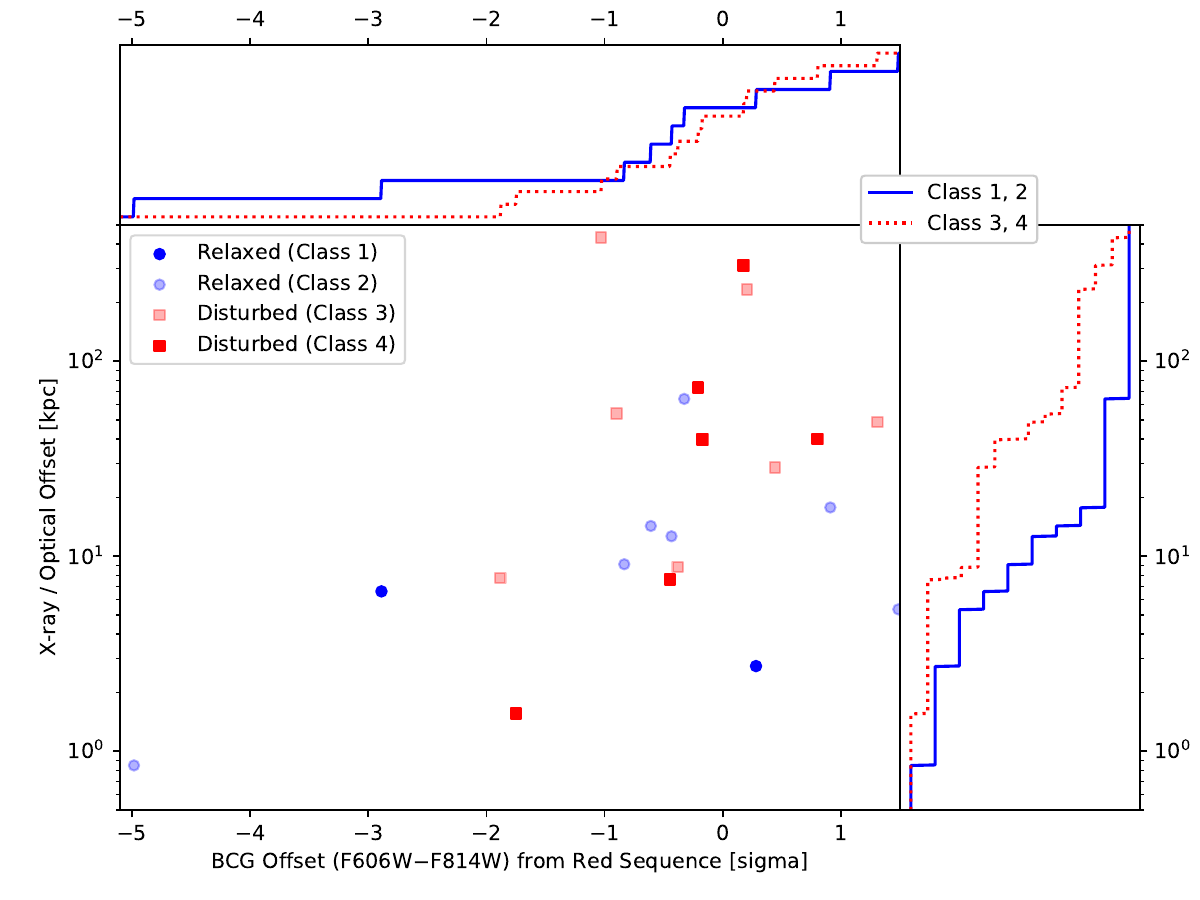}
    \caption{Offset (kpc) between the locations of the BCG and the peak of the adaptively smoothed X-ray surface brightness, as a function of the BCG's colour offset (F606W$-$F814W) from the red sequence of its host cluster. The colour and shape of the various symbols indicate morphology class. The top and right-hand panels show the cumulative distribution functions for relaxed and disturbed clusters (solid and dotted lines, respectively).}
\label{fig:X_offset_ACS}
\end{figure}

For relaxed clusters, the location of the peak of the X-ray emission (Table~\ref{tab:X_ray}) should coincide with the BCG, while for merging/disturbed clusters the different collisional properties of gas and galaxies can cause a significant X-ray/optical offset \citep[e.g.,][]{Harvey2015, Wittman2017}. However, a meaningful physical interpretation of the segregation of gas and galaxies in cluster mergers depends critically on a correct assignment of corresponding BCGs and X-ray peaks -- particularly non-trivial in complex mergers that involve more than two subclusters. Given that the offsets exhibited by some of our clusters are very large (in excess of 100 kpc), we therefore caution that these large offsets may primarily reflect ambiguity regarding the choice of a single BCG for cluster mergers. Where multiple BCG candidates exist, Table~\ref{tab:BCG_G2} lists the brightest such candidate; however, in four cases\footnote{MACSJ1226.8$+$2153C, MACSJ2003.4$-$2322, SMACSJ0234.7 $-$5831, and SMACSJ2031.8$-$4036} the peak X-ray emission is associated with a slightly less luminous candidate, resulting in a large X-ray/optical offset that is physically meaningful only inasmuch as it indicates the complexity of those clusters. In another case\footnote{MACSJ2243.3$-$0935} at least four plausible BCG candidates exist, and the X-ray peak does not coincide with any of them. In order to reveal the correct associations of BCGs and ICM halos, one would require a careful analysis of the distribution of dark and luminous matter in order to establish the merger history and trajectories of the individual subclusters.

We previously examined (Section~\ref{sec:BCG}) the connection between a BCG's colour and its internal structure; we found that the two bluest BCGs in our sample (with a red-sequence colour offset $\geq 5\sigma$) exhibit pronounced internal structure (Fig.~\ref{fig:BCG_cutouts}) and reside in relaxed clusters (MACSJ1447.4+0827 and MACSJ0547.0$-$3904; see also Figs.~\ref{fig:BCG_cutouts} and \ref{fig:BCG_offset_sig}). We do note that SNAPshot selection bias may affect these findings.

We now explore the relationship between a BCG's colour (its offset from its host cluster's red sequence) and its physical offset from the cluster's X-ray peak. By necessity, this investigation is limited to the clusters for which both X-ray and F606W$-$F814W data are available. As shown in Fig.~\ref{fig:X_offset_ACS}, only two such clusters host BCGs significantly (more than $2.5\sigma$) bluer than their red sequence;\footnote{Cf. Fig.~\ref{fig:BCG_offset_sig}, which displays a superset of the clusters under consideration here.} both of them exhibit relaxed morphologies characterized by small X-ray/optical offsets. These results agree with the analysis of \citet{Sanderson2009}, who demonstrate that relaxed clusters are associated with stronger cool cores and greater BCG activity. In general, it appears that a relaxed morphology allows intracluster-medium cooling to focus gas accretion onto a single massive galaxy and thus revive at least some star formation. Note, however, that a relaxed host cluster and a small X-ray/optical offset are necessary but not sufficient criteria for the presence of an actively evolving BCG. As Fig.~\ref{fig:X_offset_ACS} demonstrates, many MACS clusters with small or moderate X-ray/optical offsets and/or relaxed morphologies do not host noticeably blue BCGs.

\begin{figure}
    \leavevmode\epsfxsize=9cm\epsfbox{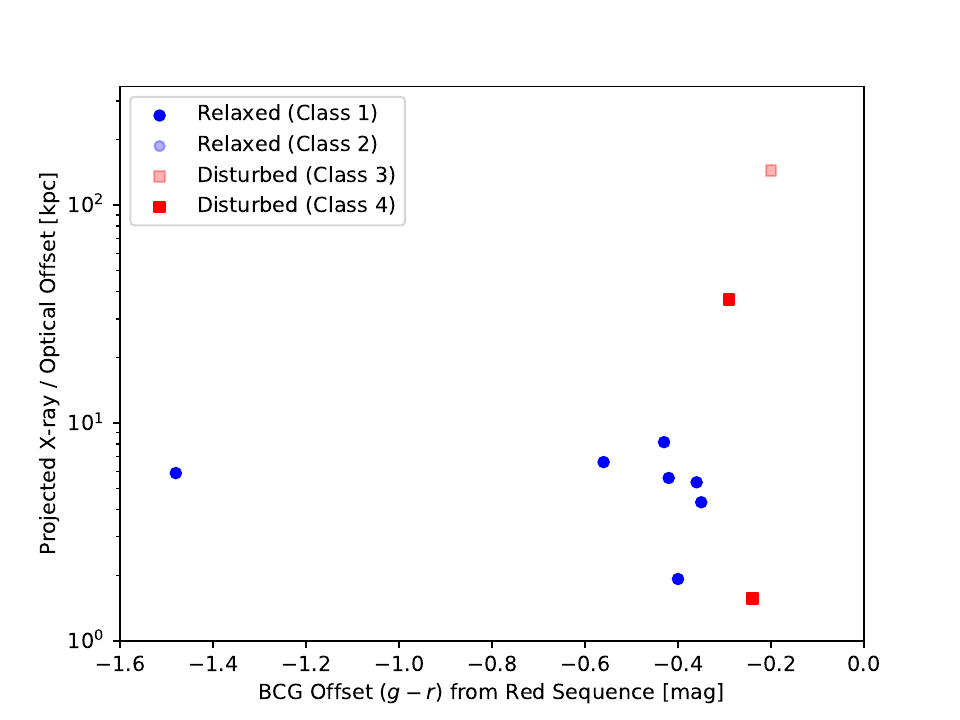}
    \caption{Offset (kpc) between BCG and X-ray peak position, as a function of the BCG's colour offset ($g-r$) from its host cluster's red sequence. The $g-r$ data come from \citet{Green2016}, and in two cases (MACSJ0404.6$+$1109 and MACSJ2243.3$-$0935) we use \citeauthor{Green2016}'s BCG identification instead of ours for the sake of consistency. }
\label{fig:X_offset_Gn}
\end{figure}

However, (as noted in Section~\ref{sec:BCG}) the F606W$-$F814W colour is not as sensitive to BCG star formation (at these redshifts) as the $g-r$ colour. \citet{Green2016} note the strong association between optical emission lines and significant colour offsets from the red sequence in these bands. Since their sample has ten clusters in common with ours\footnote{The limited overlap is due to the fact that the groundbased imaging used by \citeauthor{Green2016} is not deep enough to allow the red sequence to be robustly established for MACS clusters at the high end of our redshift range.} (Table A1 of \citealt{Green2016}), we use their data to plot, in Fig.~\ref{fig:X_offset_Gn},  the X-ray/optical offset against the BCG $g-r$ offset for these clusters\footnote{In two of these cases \citeauthor{Green2016}'s BCG identification differs from ours; for the sake of consistency, in this figure we use their BCG to mark the cluster's optical centre.}. Although this subsample is now even smaller, we find a similar trend: the two clusters with large physical offsets ($\ga 20$ kpc) show only a small colour offset, whereas the the clusters hosting blue BCGs are relaxed and exhibit good X-ray/optical alignment. In addition, the high resolution of our \textit{HST} images reveals compelling evidence of star formation in these BCGs, as indicated in Fig.~\ref{fig:BCG_offset_sig} and shown in Figs.~\ref{fig:BCG_cutouts} and \ref{fig:BCG_cutouts2}.

\section{Legacy Value}
\label{sec:legacy}

Although we hope to have demonstrated the significant discovery potential and wide applicability of the data collected in this SNAPshot survey of massive, X-ray selected clusters (see, in particular, Figs.~\ref{fig:arcs}, \ref{fig:arcs2}, \ref{fig:BCG_cutouts} and \ref{fig:BCG_cutouts2}), the true legacy value of this project is perhaps most convincingly reflected in the extensive and diverse research it has already facilitated. We therefore briefly review some of the numerous studies conducted by the extragalactic community based on these data.

Among the individual discoveries, a particularly striking object  is the so-called Cosmic Eye \citep{Smail2007} in MACSJ2135.2$-$0102, comprising two bright arcs generated by galaxy-galaxy lensing (panel 38 Fig.~\ref{fig:arcs2}). The high magnification ($\sim 30$) of the background ($z=3.07$) Lyman-break galaxy permits detailed study of its properties, including the determination of gas fractions, stellar mass, and star-formation efficiency \citep{Coppin2007}; determination of disk rotation \citep{Stark2008}; detection of polycyclic aromatic hydrocarbon emission \citep{Siana2009}; determination of properties of its interstellar medium \citep{Quider2010}; constraints on molecular gas mass \citep{Riechers2010}; and characterization of extinction law and star formation history \citep{Sklias2014}. \citet{Dye2007}, investigating the lens itself, find two spatially distinct components in the lensing galaxy, one visible, the other dark.

Focusing on other strong-lensing events discovered by MACS SNAPshot observations, \citet{Jones2010} analyse star formation and spatially resolved dynamics out to $z=3.1$; \citet{Swinbank2011} study the kinematics of the interstellar medium in a galaxy at $z = 2.3$; and \citet{Christensen2012} perform a direct measure of oxygen abundance out to $z=3.5$. \citet{Damjanov2013} report a population of compact galaxies at cluster redshifts (0.2--0.6), and \citet{Stark2014} rely on lensing to determine the stellar mass and star formation rate in low-luminosity/low-mass galaxies out to $z \sim 3$. More recent works include \citet{Newman2015}'s discovery of a massive, recently quenched galaxy at $z = 2.6$; \citet{Schaerer2015}'s analysis of ISM and star formation properties at $z \sim 2$; \citet{Thomson2015}'s study of cool molecular gas and star formation at $z = 2.3$, resolving scales down to 100 pc; and \citet{Patricio2016}'s radiative transfer model for a young $L^*$ galaxy at $z \sim 3.5$. \citet{Zitrin2012} illustrates the sheer richness of the lensing accomplished by some of these clusters, finding 47 multiply-lensed images of 12 distinct sources behind MACSJ1206.2$-$0847. In addition, \citet{Repp2016}'s seach for high-redshift galaxies behind these clusters finds $\sim 20$ Lyman break galaxies at $z \sim 7$--9. Thus the lensing power of these massive clusters has been fruitful in multiple areas of research.

MACS clusters from the SNAPshot sample also serve as laboratories for the study of galaxy evolution. Again exploiting strong-lensing amplification,  \citet{Livermore2012, Livermore2015} study the evolution of star formation density and luminosity from $z = 1$--4. \citet{Abramson2013} investigate starbursts and quenching, while multiple authors study the effects of active galactic nuclei on their host galaxies and the surrounding medium \citep{Cavagnolo2011, HL2012, Ehlert2015}. In addition, \citet{Zitrin2017} reports the discovery of a rare cluster shell galaxy system in the process of formation in MACSJ1319.9+7003. Also at the cluster redshift, \citet{Ebeling2014} and \citet{McPartland2016} discover numerous dramatic examples of ram-pressure stripping in MACS SNAPshot data (`jellyfish galaxies') -- a most spectacular (and violent) phase of galaxy evolution.

Studies of the MACS SNAPshot clusters as a whole have also been fruitful. \citet{Ebeling2009} conclude that the cluster MACSJ1206.2$-$0847 is undergoing a line-of-sight merger event, and \citet{Ho2012} analyse the merger history of MACS J0140.0$-$0555. Both \citet{Stott2007} and \citet{DePropris2013} study the evolution of the cluster member population, and \citet{Horesh2010} compare the arc-production efficiencies of X-ray selected and optically selected clusters. The colour information and high resolution afforded by the \textit{HST} images also allow construction of dark matter mass profiles for these clusters (\citealp{Richard2015, Zitrin2016}; see also Richard et al. in prep.).

The SNAPshot images also permit study of more general cosmological questions. \citet{Gilmore2009} investigate the feasibility of constraining the dark energy equation of state by stacking images from strongly lensed clusters, while \citet{Harvey2015} constrain the self-interaction cross-section of dark matter by considering cluster mergers.

Last, but certainly not least, our \textit{HST} SNAPshot surveys of MACS clusters serve as pathfinder missions for more focused efforts. Just as MACS first discovered three out of the six Hubble Frontier Fields\footnote{http://www.stsci.edu/hst/campaigns/frontier-fields/} targets (another two are MACS re-discoveries), this particular SNAPshot survey has provided a significant number of targets for the \textit{HST} legacy programme CLASH (Cluster Lensing And Supernova survey with Hubble -- \citealp{Postman2012}) and has also contributed heavily to the more recent RELICS\footnote{https://relics.stsci.edu/} (REionization LensIng Cluster Survey) project.

Finally, we point out that, by virtue of the information listed in Table~\ref{tab:clusters}, this paper also constitutes the largest release of MACS cluster redshifts to the community to date, complementing the partial  releases published by \citet{Ebeling2007,Ebeling2010} and \citet{Mann2012}. 

\section{Conclusion}
\label{sec:concl}

The value of SNAPshot surveys rests largely on the richness and diversity of data obtained with a relatively small investment of \textit{HST} observing time. This paper provides examples illustrating the wide range of astrophysical topics investigable with this data set, either alone or in combination with other observations. As a broad overview of selected topics (rather than an exhaustive catalog of all of them, or an in-depth study of any), our treatment is cursory by necessity. Section~\ref{sec:legacy} has reviewed other, more in-depth treatments of various areas of research using these data. 

Nevertheless, even our cursory analysis demonstrates the power of SNAPshot observations specifically for investigations relying on strong gravitational lensing, where the plethora of spectacular arcs and multiple-image systems discovered by our project (Figs.~\ref{fig:arcs}, \ref{fig:arcs2}) advances three distinct fields at once by (a) allowing the mapping of all gravitational matter, dark or luminous, in massive clusters; (b) identifying the best targets for in-depth study of highly magnified background galaxies; and (c) helping to constrain the properties of the first populations of galaxies at $z>6$. Similarly, the high-resolution images of BCGs in massive clusters at $z>0.3$ obtained by our SNAPshots (Figs.~\ref{fig:BCG_cutouts}, \ref{fig:BCG_cutouts2}) ideally complement existing and ongoing studies aimed at identifying the interplay of accretion, star formation, AGN feedback, and mergers in the formation and growth of these extreme stellar systems. Although far from exhaustive, the work presented here has also revealed three distinct lines of evidence (rest-frame slope of the red sequence, BCG colours, distribution and relative frequency of cluster morphology classes) supporting the notion that the general features of massive cluster morphology were in place well before $z = 0.5$, i.e., well over 5 Gyr ago, with no strong evolution in the $z=0.3-0.5$ range. Ongoing \textit{HST} SNAPshot observations of eMACS clusters at $z>0.5$ \citep{EMACS} -- which probe the population of extremely X-ray luminous clusters out to $z \sim 1$ -- stand to shed light on the yet earlier history of these exceptional objects. Finally, by probing the previously poorly sampled regime of very massive clusters at intermediate redshift, the MACS SNAPshot survey also adds crucial leverage to scaling relations such as that between X-ray luminosity and optical richness (Fig.~\ref{fig:X_num}).

Most importantly, however, our \textit{HST} SNAPshot surveys of MACS clusters have established their value as pathfinder missions for the entire extragalactic community. This programme has created a legacy dataset of high-resolution images of some of the most extreme galaxy clusters known. It thus has enabled countless discoveries, and it will continue to facilitate further in-depth study of galaxy evolution and structure formation, from the cluster redshifts to lensing-assisted probes of the elusive era of re-ionization.

\section*{Acknowledgements}
We thank Alastair Edge for inspiring conversations and helpful comments on all matters BCG; we also appreciate insightful suggestions from both Ian Smail and an anonymous referee. The authors gratefully acknowledge financial support from STScI grants GO-10491, GO-10875, GO-12166, and GO-12884. 

The scientific results reported in this article are based on observations made with the NASA/ESA Hubble Space Telescope, obtained at the Space Telescope Science Institute, which is operated by the Association of Universities for Research in Astronomy, Inc., under NASA contract NAS 5-26555. The results are also based in part on observations made by the Chandra X-ray Observatory.

In addition, some of the data presented herein were obtained at the W.M. Keck Observatory, which is operated as a scientific partnership among the California Institute of Technology, the University of California, and the National Aeronautics and Space Administration. The Observatory was made possible by the generous financial support of the W.M. Keck Foundation.

Finally, the authors wish to recognize and acknowledge the very significant cultural role and reverence that the summit of Mauna Kea has always had within the indigenous Hawaiian community.  We are most fortunate to have the opportunity to conduct observations from this mountain.

\bibliographystyle{mn2e}
\bibliography{MACS_Summary_edt}

\begin{thebibliography}{}
\makeatletter
\relax
\def\mn@urlcharsother{\let\do\@makeother \do\$\do\&\do\#\do\^\do\_\do\%\do\~}
\def\mn@doi{\begingroup\mn@urlcharsother \@ifnextchar [ {\mn@doi@}
  {\mn@doi@[]}}
\def\mn@doi@[#1]#2{\def\@tempa{#1}\ifx\@tempa\@empty \href
  {http://dx.doi.org/#2} {doi:#2}\else \href {http://dx.doi.org/#2} {#1}\fi
  \endgroup}
\def\mn@eprint#1#2{\mn@eprint@#1:#2::\@nil}
\def\mn@eprint@arXiv#1{\href {http://arxiv.org/abs/#1} {{\tt arXiv:#1}}}
\def\mn@eprint@dblp#1{\href {http://dblp.uni-trier.de/rec/bibtex/#1.xml}
  {dblp:#1}}
\def\mn@eprint@#1:#2:#3:#4\@nil{\def\@tempa {#1}\def\@tempb {#2}\def\@tempc
  {#3}\ifx \@tempc \@empty \let \@tempc \@tempb \let \@tempb \@tempa \fi \ifx
  \@tempb \@empty \def\@tempb {arXiv}\fi \@ifundefined
  {mn@eprint@\@tempb}{\@tempb:\@tempc}{\expandafter \expandafter \csname
  mn@eprint@\@tempb\endcsname \expandafter{\@tempc}}}

\bibitem[\protect\citeauthoryear{{Abell}}{{Abell}}{1958}]{abell58}
{Abell} G.~O.,  1958, \mn@doi [\apjs] {10.1086/190036}, \href
  {http://adsabs.harvard.edu/abs/1958ApJS....3..211A} {3, 211}

\bibitem[\protect\citeauthoryear{{Abramson}, {Dressler}, {Gladders}, {Oemler},
  {Poggianti}, {Monson}, {Persson}  et~al.}{{Abramson}
  et~al.}{2013}]{Abramson2013}
{Abramson} L.~E.,  {Dressler} A.,  {Gladders} M.~D.,  {Oemler} Jr. A.,
  {Poggianti} B.~M.,  {Monson} A.,  {Persson} E.,   et~al., 2013, \mn@doi
  [\apj] {10.1088/0004-637X/777/2/124}, \href
  {http://adsabs.harvard.edu/abs/2013ApJ...777..124A} {777, 124}

\bibitem[\protect\citeauthoryear{{Andreon}, {Newman}, {Trinchieri}, {Raichoor},
  {Ellis}  \& {Treu}}{{Andreon} et~al.}{2014}]{Andreon2014}
{Andreon} S.,  {Newman} A.~B.,  {Trinchieri} G.,  {Raichoor} A.,  {Ellis}
  R.~S.,   {Treu} T.,  2014, \mn@doi [\aap] {10.1051/0004-6361/201323077},
  \href {http://adsabs.harvard.edu/abs/2014A%26A...565A.120A} {565, A120}

\bibitem[\protect\citeauthoryear{{Avila}}{{Avila}}{2017}]{ACS}
{Avila} R.~J.,  2017, {Advanced Camera for Surveys Instrument Handbook for
  Cycle 25 v. 16.0}

\bibitem[\protect\citeauthoryear{{Bartelmann}, {Huss}, {Colberg}, {Jenkins}  \&
  {Pearce}}{{Bartelmann} et~al.}{1998}]{Bartelmann1998}
{Bartelmann} M.,  {Huss} A.,  {Colberg} J.~M.,  {Jenkins} A.,   {Pearce} F.~R.,
   1998, \aap, \href {http://adsabs.harvard.edu/abs/1998A%26A...330....1B}
  {330, 1}

\bibitem[\protect\citeauthoryear{{Begelman}, {Blandford}  \& {Rees}}{{Begelman}
  et~al.}{1980}]{Begelman1980}
{Begelman} M.~C.,  {Blandford} R.~D.,   {Rees} M.~J.,  1980, \mn@doi [\nat]
  {10.1038/287307a0}, \href {http://adsabs.harvard.edu/abs/1980Natur.287..307B}
  {287, 307}

\bibitem[\protect\citeauthoryear{{Bertin} \& {Arnouts}}{{Bertin} \&
  {Arnouts}}{1996}]{SExtractor}
{Bertin} E.,  {Arnouts} S.,  1996, \mn@doi [\aaps] {10.1051/aas:1996164}, \href
  {http://adsabs.harvard.edu/abs/1996A%26AS..117..393B} {117, 393}

\bibitem[\protect\citeauthoryear{{Bildfell}, {Hoekstra}, {Babul}  \&
  {Mahdavi}}{{Bildfell} et~al.}{2008}]{Bildfell2008}
{Bildfell} C.,  {Hoekstra} H.,  {Babul} A.,   {Mahdavi} A.,  2008, \mn@doi
  [\mnras] {10.1111/j.1365-2966.2008.13699.x}, \href
  {http://adsabs.harvard.edu/abs/2008MNRAS.389.1637B} {389, 1637}

\bibitem[\protect\citeauthoryear{{Bleem}, {Stalder}, {de Haan}, {Aird},
  {Allen}, {Applegate}, {Ashby}  et~al.}{{Bleem} et~al.}{2015}]{Bleem2015}
{Bleem} L.~E.,  {Stalder} B.,  {de Haan} T.,  {Aird} K.~A.,  {Allen} S.~W.,
  {Applegate} D.~E.,  {Ashby} M.~L.~N.,   et~al., 2015, \mn@doi [\apjs]
  {10.1088/0067-0049/216/2/27}, \href
  {http://adsabs.harvard.edu/abs/2015ApJS..216...27B} {216, 27}

\bibitem[\protect\citeauthoryear{{B{\"o}hringer} et~al.,}{{B{\"o}hringer}
  et~al.}{2004}]{Bohringer2004}
{B{\"o}hringer} H.,  et~al., 2004, \mn@doi [\aap] {10.1051/0004-6361:20034484},
  \href {http://adsabs.harvard.edu/abs/2004A%26A...425..367B} {425, 367}

\bibitem[\protect\citeauthoryear{{Boller}, {Freyberg}, {Tr{\"u}mper}, {Haberl},
  {Voges}  \& {Nandra}}{{Boller} et~al.}{2016}]{Boller2016}
{Boller} T.,  {Freyberg} M.~J.,  {Tr{\"u}mper} J.,  {Haberl} F.,  {Voges} W.,
  {Nandra} K.,  2016, \mn@doi [\aap] {10.1051/0004-6361/201525648}, \href
  {http://adsabs.harvard.edu/abs/2016A%26A...588A.103B} {588, A103}

\bibitem[\protect\citeauthoryear{{Bower}, {Lucey}  \& {Ellis}}{{Bower}
  et~al.}{1992}]{BowerLuceyEllis1992}
{Bower} R.~G.,  {Lucey} J.~R.,   {Ellis} R.~S.,  1992, \mn@doi [\mnras]
  {10.1093/mnras/254.4.601}, \href
  {http://adsabs.harvard.edu/abs/1992MNRAS.254..601B} {254, 601}

\bibitem[\protect\citeauthoryear{{Brada{\v c}}, {Allen}, {Treu}, {Ebeling},
  {Massey}, {Morris}, {von der Linden}  \& {Applegate}}{{Brada{\v c}}
  et~al.}{2008}]{Bradac2008}
{Brada{\v c}} M.,  {Allen} S.~W.,  {Treu} T.,  {Ebeling} H.,  {Massey} R.,
  {Morris} R.~G.,  {von der Linden} A.,   {Applegate} D.,  2008, \mn@doi [\apj]
  {10.1086/591246}, \href {http://adsabs.harvard.edu/abs/2008ApJ...687..959B}
  {687, 959}

\bibitem[\protect\citeauthoryear{{Bruzual} \& {Charlot}}{{Bruzual} \&
  {Charlot}}{2003}]{BC03}
{Bruzual} G.,  {Charlot} S.,  2003, \mn@doi [\mnras]
  {10.1046/j.1365-8711.2003.06897.x}, \href
  {http://adsabs.harvard.edu/abs/2003MNRAS.344.1000B} {344, 1000}

\bibitem[\protect\citeauthoryear{{Budzynski}, {Koposov}, {McCarthy}, {McGee}
  \& {Belokurov}}{{Budzynski} et~al.}{2012}]{Budzynski12}
{Budzynski} J.~M.,  {Koposov} S.~E.,  {McCarthy} I.~G.,  {McGee} S.~L.,
  {Belokurov} V.,  2012, \mn@doi [\mnras] {10.1111/j.1365-2966.2012.20663.x},
  \href {http://adsabs.harvard.edu/abs/2012MNRAS.423..104B} {423, 104}

\bibitem[\protect\citeauthoryear{{Cavagnolo}, {McNamara}, {Wise}, {Nulsen},
  {Br{\"u}ggen}, {Gitti}  \& {Rafferty}}{{Cavagnolo}
  et~al.}{2011}]{Cavagnolo2011}
{Cavagnolo} K.~W.,  {McNamara} B.~R.,  {Wise} M.~W.,  {Nulsen} P.~E.~J.,
  {Br{\"u}ggen} M.,  {Gitti} M.,   {Rafferty} D.~A.,  2011, \mn@doi [\apj]
  {10.1088/0004-637X/732/2/71}, \href
  {http://adsabs.harvard.edu/abs/2011ApJ...732...71C} {732, 71}

\bibitem[\protect\citeauthoryear{{Cerulo} et~al.,}{{Cerulo}
  et~al.}{2016}]{Cerulo2016}
{Cerulo} P.,  et~al., 2016, \mn@doi [\mnras] {10.1093/mnras/stw080}, \href
  {http://adsabs.harvard.edu/abs/2016MNRAS.457.2209C} {457, 2209}

\bibitem[\protect\citeauthoryear{{Christensen}, {Laursen}, {Richard}, {Hjorth},
  {Milvang-Jensen}, {Dessauges-Zavadsky}, {Limousin}  et~al.}{{Christensen}
  et~al.}{2012}]{Christensen2012}
{Christensen} L.,  {Laursen} P.,  {Richard} J.,  {Hjorth} J.,  {Milvang-Jensen}
  B.,  {Dessauges-Zavadsky} M.,  {Limousin} M.,   et~al., 2012, \mn@doi
  [\mnras] {10.1111/j.1365-2966.2012.22007.x}, \href
  {http://adsabs.harvard.edu/abs/2012MNRAS.427.1973C} {427, 1973}

\bibitem[\protect\citeauthoryear{{Coppin}, {Swinbank}, {Neri}, {Cox}, {Smail},
  {Ellis}, {Geach}  et~al.}{{Coppin} et~al.}{2007}]{Coppin2007}
{Coppin} K.~E.~K.,  {Swinbank} A.~M.,  {Neri} R.,  {Cox} P.,  {Smail} I.,
  {Ellis} R.~S.,  {Geach} J.~E.,   et~al., 2007, \mn@doi [\apj]
  {10.1086/519789}, \href {http://adsabs.harvard.edu/abs/2007ApJ...665..936C}
  {665, 936}

\bibitem[\protect\citeauthoryear{{Dalal}, {Holder}  \& {Hennawi}}{{Dalal}
  et~al.}{2004}]{Dalal2004}
{Dalal} N.,  {Holder} G.,   {Hennawi} J.~F.,  2004, \mn@doi [\apj]
  {10.1086/420960}, \href {http://adsabs.harvard.edu/abs/2004ApJ...609...50D}
  {609, 50}

\bibitem[\protect\citeauthoryear{{Damjanov}, {Chilingarian}, {Hwang}  \&
  {Geller}}{{Damjanov} et~al.}{2013}]{Damjanov2013}
{Damjanov} I.,  {Chilingarian} I.,  {Hwang} H.~S.,   {Geller} M.~J.,  2013,
  \mn@doi [\apjl] {10.1088/2041-8205/775/2/L48}, \href
  {http://adsabs.harvard.edu/abs/2013ApJ...775L..48D} {775, L48}

\bibitem[\protect\citeauthoryear{{De Lucia} \& {Blaizot}}{{De Lucia} \&
  {Blaizot}}{2007}]{DeLucia07}
{De Lucia} G.,  {Blaizot} J.,  2007, \mn@doi [\mnras]
  {10.1111/j.1365-2966.2006.11287.x}, \href
  {http://adsabs.harvard.edu/abs/2007MNRAS.375....2D} {375, 2}

\bibitem[\protect\citeauthoryear{{De Lucia} et~al.,}{{De Lucia}
  et~al.}{2007}]{DeLucia2007}
{De Lucia} G.,  et~al., 2007, \mn@doi [\mnras]
  {10.1111/j.1365-2966.2006.11199.x}, \href
  {http://adsabs.harvard.edu/abs/2007MNRAS.374..809D} {374, 809}

\bibitem[\protect\citeauthoryear{{De Propris}, {Phillipps}  \& {Bremer}}{{De
  Propris} et~al.}{2013}]{DePropris2013}
{De Propris} R.,  {Phillipps} S.,   {Bremer} M.~N.,  2013, \mn@doi [\mnras]
  {10.1093/mnras/stt1262}, \href
  {http://adsabs.harvard.edu/abs/2013MNRAS.434.3469D} {434, 3469}

\bibitem[\protect\citeauthoryear{{Dressel}}{{Dressel}}{2017}]{WFC3}
{Dressel} L.,  2017, {Wide Field Camera 3 Instrument Handbook, Version 9.0}

\bibitem[\protect\citeauthoryear{{Dye}, {Smail}, {Swinbank}, {Ebeling}  \&
  {Edge}}{{Dye} et~al.}{2007}]{Dye2007}
{Dye} S.,  {Smail} I.,  {Swinbank} A.~M.,  {Ebeling} H.,   {Edge} A.~C.,  2007,
  \mn@doi [\mnras] {10.1111/j.1365-2966.2007.11960.x}, \href
  {http://adsabs.harvard.edu/abs/2007MNRAS.379..308D} {379, 308}

\bibitem[\protect\citeauthoryear{{Ebeling}, {Edge}, {Bohringer}, {Allen},
  {Crawford}, {Fabian}, {Voges}  \& {Huchra}}{{Ebeling}
  et~al.}{1998}]{Ebeling1998}
{Ebeling} H.,  {Edge} A.~C.,  {Bohringer} H.,  {Allen} S.~W.,  {Crawford}
  C.~S.,  {Fabian} A.~C.,  {Voges} W.,   {Huchra} J.~P.,  1998, \mn@doi
  [\mnras] {10.1046/j.1365-8711.1998.01949.x}, \href
  {http://adsabs.harvard.edu/abs/1998MNRAS.301..881E} {301, 881}

\bibitem[\protect\citeauthoryear{{Ebeling}, {Edge}, {Allen}, {Crawford},
  {Fabian}  \& {Huchra}}{{Ebeling} et~al.}{2000}]{Ebeling2000}
{Ebeling} H.,  {Edge} A.~C.,  {Allen} S.~W.,  {Crawford} C.~S.,  {Fabian}
  A.~C.,   {Huchra} J.~P.,  2000, \mn@doi [\mnras]
  {10.1046/j.1365-8711.2000.03549.x}, \href
  {http://adsabs.harvard.edu/abs/2000MNRAS.318..333E} {318, 333}

\bibitem[\protect\citeauthoryear{{Ebeling}, {Edge}  \& {Henry}}{{Ebeling}
  et~al.}{2001}]{Ebeling2001}
{Ebeling} H.,  {Edge} A.~C.,   {Henry} J.~P.,  2001, \mn@doi [\apj]
  {10.1086/320958}, \href {http://adsabs.harvard.edu/abs/2001ApJ...553..668E}
  {553, 668}

\bibitem[\protect\citeauthoryear{{Ebeling}, {White}  \& {Rangarajan}}{{Ebeling}
  et~al.}{2006}]{Ebeling2006}
{Ebeling} H.,  {White} D.~A.,   {Rangarajan} F.~V.~N.,  2006, \mn@doi [\mnras]
  {10.1111/j.1365-2966.2006.10135.x}, \href
  {http://adsabs.harvard.edu/abs/2006MNRAS.368...65E} {368, 65}

\bibitem[\protect\citeauthoryear{{Ebeling}, {Barrett}, {Donovan}, {Ma}, {Edge}
  \& {van Speybroeck}}{{Ebeling} et~al.}{2007}]{Ebeling2007}
{Ebeling} H.,  {Barrett} E.,  {Donovan} D.,  {Ma} C.-J.,  {Edge} A.~C.,   {van
  Speybroeck} L.,  2007, \mn@doi [\apjl] {10.1086/518603}, \href
  {http://adsabs.harvard.edu/abs/2007ApJ...661L..33E} {661, L33}

\bibitem[\protect\citeauthoryear{{Ebeling}, {Ma}, {Kneib}, {Jullo}, {Courtney},
  {Barrett}, {Edge}  et~al.}{{Ebeling} et~al.}{2009}]{Ebeling2009}
{Ebeling} H.,  {Ma} C.~J.,  {Kneib} J.-P.,  {Jullo} E.,  {Courtney} N.~J.~D.,
  {Barrett} E.,  {Edge} A.~C.,   et~al., 2009, \mn@doi [\mnras]
  {10.1111/j.1365-2966.2009.14502.x}, \href
  {http://adsabs.harvard.edu/abs/2009MNRAS.395.1213E} {395, 1213}

\bibitem[\protect\citeauthoryear{{Ebeling}, {Edge}, {Mantz}, {Barrett},
  {Henry}, {Ma}  \& {van Speybroeck}}{{Ebeling} et~al.}{2010}]{Ebeling2010}
{Ebeling} H.,  {Edge} A.~C.,  {Mantz} A.,  {Barrett} E.,  {Henry} J.~P.,  {Ma}
  C.~J.,   {van Speybroeck} L.,  2010, \mn@doi [\mnras]
  {10.1111/j.1365-2966.2010.16920.x}, \href
  {http://adsabs.harvard.edu/abs/2010MNRAS.407...83E} {407, 83}

\bibitem[\protect\citeauthoryear{{Ebeling}, {Edge}, {Burgett}, {Chambers},
  {Hodapp}, {Huber}, {Kaiser}  et~al.}{{Ebeling} et~al.}{2013}]{EMACS}
{Ebeling} H.,  {Edge} A.~C.,  {Burgett} W.~S.,  {Chambers} K.~C.,  {Hodapp}
  K.~W.,  {Huber} M.~E.,  {Kaiser} N.,   et~al., 2013, \mn@doi [\mnras]
  {10.1093/mnras/stt387}, \href
  {http://adsabs.harvard.edu/abs/2013MNRAS.432...62E} {432, 62}

\bibitem[\protect\citeauthoryear{{Ebeling}, {Stephenson}  \& {Edge}}{{Ebeling}
  et~al.}{2014}]{Ebeling2014}
{Ebeling} H.,  {Stephenson} L.~N.,   {Edge} A.~C.,  2014, \mn@doi [\apjl]
  {10.1088/2041-8205/781/2/L40}, \href
  {http://adsabs.harvard.edu/abs/2014ApJ...781L..40E} {781, L40}

\bibitem[\protect\citeauthoryear{{Edge}}{{Edge}}{2001}]{Edge2001}
{Edge} A.~C.,  2001, \mn@doi [\mnras] {10.1046/j.1365-8711.2001.04802.x}, \href
  {http://adsabs.harvard.edu/abs/2001MNRAS.328..762E} {328, 762}

\bibitem[\protect\citeauthoryear{{Edwards}, {Hudson}, {Balogh}  \&
  {Smith}}{{Edwards} et~al.}{2007}]{Edwards2007}
{Edwards} L.~O.~V.,  {Hudson} M.~J.,  {Balogh} M.~L.,   {Smith} R.~J.,  2007,
  \mn@doi [\mnras] {10.1111/j.1365-2966.2007.11910.x}, \href
  {http://adsabs.harvard.edu/abs/2007MNRAS.379..100E} {379, 100}

\bibitem[\protect\citeauthoryear{{Ehlert}, {Allen}, {Brandt}, {Canning}, {Luo},
  {Mantz}, {Morris}  et~al.}{{Ehlert} et~al.}{2015}]{Ehlert2015}
{Ehlert} S.,  {Allen} S.~W.,  {Brandt} W.~N.,  {Canning} R.~E.~A.,  {Luo} B.,
  {Mantz} A.,  {Morris} R.~G.,   et~al., 2015, \mn@doi [\mnras]
  {10.1093/mnras/stu2091}, \href
  {http://adsabs.harvard.edu/abs/2015MNRAS.446.2709E} {446, 2709}

\bibitem[\protect\citeauthoryear{{Gilmore} \& {Natarajan}}{{Gilmore} \&
  {Natarajan}}{2009}]{Gilmore2009}
{Gilmore} J.,  {Natarajan} P.,  2009, \mn@doi [\mnras]
  {10.1111/j.1365-2966.2009.14612.x}, \href
  {http://adsabs.harvard.edu/abs/2009MNRAS.396..354G} {396, 354}

\bibitem[\protect\citeauthoryear{{Gladders}, {L{\'o}pez-Cruz}, {Yee}  \&
  {Kodama}}{{Gladders} et~al.}{1998}]{Gladders1998}
{Gladders} M.~D.,  {L{\'o}pez-Cruz} O.,  {Yee} H.~K.~C.,   {Kodama} T.,  1998,
  \mn@doi [\apj] {10.1086/305858}, \href
  {http://adsabs.harvard.edu/abs/1998ApJ...501..571G} {501, 571}

\bibitem[\protect\citeauthoryear{{Gladders}, {Hoekstra}, {Yee}, {Hall}  \&
  {Barrientos}}{{Gladders} et~al.}{2003}]{Gladders2003}
{Gladders} M.~D.,  {Hoekstra} H.,  {Yee} H.~K.~C.,  {Hall} P.~B.,
  {Barrientos} L.~F.,  2003, \mn@doi [\apj] {10.1086/376518}, \href
  {http://adsabs.harvard.edu/abs/2003ApJ...593...48G} {593, 48}

\bibitem[\protect\citeauthoryear{{Green}, {Edge}, {Stott}, {Ebeling},
  {Burgett}, {Chambers}, {Draper}  et~al.}{{Green} et~al.}{2016}]{Green2016}
{Green} T.~S.,  {Edge} A.~C.,  {Stott} J.~P.,  {Ebeling} H.,  {Burgett} W.~S.,
  {Chambers} K.~C.,  {Draper} P.~W.,   et~al., 2016, \mn@doi [\mnras]
  {10.1093/mnras/stw1338}, \href
  {http://adsabs.harvard.edu/abs/2016MNRAS.461..560G} {461, 560}

\bibitem[\protect\citeauthoryear{{Harvey}, {Massey}, {Kitching}, {Taylor}  \&
  {Tittley}}{{Harvey} et~al.}{2015}]{Harvey2015}
{Harvey} D.,  {Massey} R.,  {Kitching} T.,  {Taylor} A.,   {Tittley} E.,  2015,
  \mn@doi [Science] {10.1126/science.1261381}, \href
  {http://adsabs.harvard.edu/abs/2015Sci...347.1462H} {347, 1462}

\bibitem[\protect\citeauthoryear{{Hlavacek-Larrondo}, {Fabian}, {Edge},
  {Ebeling}, {Sanders}, {Hogan}  \& {Taylor}}{{Hlavacek-Larrondo}
  et~al.}{2012}]{HL2012}
{Hlavacek-Larrondo} J.,  {Fabian} A.~C.,  {Edge} A.~C.,  {Ebeling} H.,
  {Sanders} J.~S.,  {Hogan} M.~T.,   {Taylor} G.~B.,  2012, \mn@doi [\mnras]
  {10.1111/j.1365-2966.2011.20405.x}, \href
  {http://adsabs.harvard.edu/abs/2012MNRAS.421.1360H} {421, 1360}

\bibitem[\protect\citeauthoryear{{Ho}, {Ebeling}  \& {Richard}}{{Ho}
  et~al.}{2012}]{Ho2012}
{Ho} I.-T.,  {Ebeling} H.,   {Richard} J.,  2012, \mn@doi [\mnras]
  {10.1111/j.1365-2966.2012.21806.x}, \href
  {http://adsabs.harvard.edu/abs/2012MNRAS.426.1992H} {426, 1992}

\bibitem[\protect\citeauthoryear{{Horesh}, {Ofek}, {Maoz}, {Bartelmann},
  {Meneghetti}  \& {Rix}}{{Horesh} et~al.}{2005}]{Horesh2005}
{Horesh} A.,  {Ofek} E.~O.,  {Maoz} D.,  {Bartelmann} M.,  {Meneghetti} M.,
  {Rix} H.-W.,  2005, \mn@doi [\apj] {10.1086/466519}, \href
  {http://adsabs.harvard.edu/abs/2005ApJ...633..768H} {633, 768}

\bibitem[\protect\citeauthoryear{{Horesh}, {Maoz}, {Ebeling}, {Seidel}  \&
  {Bartelmann}}{{Horesh} et~al.}{2010}]{Horesh2010}
{Horesh} A.,  {Maoz} D.,  {Ebeling} H.,  {Seidel} G.,   {Bartelmann} M.,  2010,
  \mn@doi [\mnras] {10.1111/j.1365-2966.2010.16763.x}, \href
  {http://adsabs.harvard.edu/abs/2010MNRAS.406.1318H} {406, 1318}

\bibitem[\protect\citeauthoryear{{Horesh}, {Maoz}, {Hilbert}  \&
  {Bartelmann}}{{Horesh} et~al.}{2011}]{Horesh2011}
{Horesh} A.,  {Maoz} D.,  {Hilbert} S.,   {Bartelmann} M.,  2011, \mn@doi
  [\mnras] {10.1111/j.1365-2966.2011.19293.x}, \href
  {http://adsabs.harvard.edu/abs/2011MNRAS.418...54H} {418, 54}

\bibitem[\protect\citeauthoryear{{Hudson} \& {Ebeling}}{{Hudson} \&
  {Ebeling}}{1997}]{HudsonEbeling1997}
{Hudson} M.~J.,  {Ebeling} H.,  1997, \mn@doi [\apj] {10.1086/303904}, \href
  {http://adsabs.harvard.edu/abs/1997ApJ...479..621H} {479, 621}

\bibitem[\protect\citeauthoryear{{Jaff{\'e}}, {Arag{\'o}n-Salamanca}, {De
  Lucia}, {Jablonka}, {Rudnick}, {Saglia}  \& {Zaritsky}}{{Jaff{\'e}}
  et~al.}{2011}]{Jaffe2011}
{Jaff{\'e}} Y.~L.,  {Arag{\'o}n-Salamanca} A.,  {De Lucia} G.,  {Jablonka} P.,
  {Rudnick} G.,  {Saglia} R.,   {Zaritsky} D.,  2011, \mn@doi [\mnras]
  {10.1111/j.1365-2966.2010.17445.x}, \href
  {http://adsabs.harvard.edu/abs/2011MNRAS.410..280J} {410, 280}

\bibitem[\protect\citeauthoryear{{Johnstone}, {Fabian}  \&
  {Nulsen}}{{Johnstone} et~al.}{1987}]{Johnstone1987}
{Johnstone} R.~M.,  {Fabian} A.~C.,   {Nulsen} P.~E.~J.,  1987, \mn@doi
  [\mnras] {10.1093/mnras/224.1.75}, \href
  {http://adsabs.harvard.edu/abs/1987MNRAS.224...75J} {224, 75}

\bibitem[\protect\citeauthoryear{{Jones}, {Swinbank}, {Ellis}, {Richard}  \&
  {Stark}}{{Jones} et~al.}{2010}]{Jones2010}
{Jones} T.~A.,  {Swinbank} A.~M.,  {Ellis} R.~S.,  {Richard} J.,   {Stark}
  D.~P.,  2010, \mn@doi [\mnras] {10.1111/j.1365-2966.2010.16378.x}, \href
  {http://adsabs.harvard.edu/abs/2010MNRAS.404.1247J} {404, 1247}

\bibitem[\protect\citeauthoryear{{Kashlinsky}, {Atrio-Barandela}, {Ebeling},
  {Edge}  \& {Kocevski}}{{Kashlinsky} et~al.}{2010}]{Kashlinsky2010}
{Kashlinsky} A.,  {Atrio-Barandela} F.,  {Ebeling} H.,  {Edge} A.,   {Kocevski}
  D.,  2010, \mn@doi [\apjl] {10.1088/2041-8205/712/1/L81}, \href
  {http://adsabs.harvard.edu/abs/2010ApJ...712L..81K} {712, L81}

\bibitem[\protect\citeauthoryear{{Kneib} \& {Natarajan}}{{Kneib} \&
  {Natarajan}}{2011}]{KneibNatarajan2011}
{Kneib} J.-P.,  {Natarajan} P.,  2011, \mn@doi [\aapr]
  {10.1007/s00159-011-0047-3}, \href
  {http://adsabs.harvard.edu/abs/2011A%26ARv..19...47K} {19, 47}

\bibitem[\protect\citeauthoryear{{Kocevski}, {Ebeling}, {Mullis}  \&
  {Tully}}{{Kocevski} et~al.}{2007}]{Kocevski2007}
{Kocevski} D.~D.,  {Ebeling} H.,  {Mullis} C.~R.,   {Tully} R.~B.,  2007,
  \mn@doi [\apj] {10.1086/513303}, \href
  {http://adsabs.harvard.edu/abs/2007ApJ...662..224K} {662, 224}

\bibitem[\protect\citeauthoryear{{Kodama} \& {Arimoto}}{{Kodama} \&
  {Arimoto}}{1997}]{KodamaArimoto1997}
{Kodama} T.,  {Arimoto} N.,  1997, \aap, \href
  {http://adsabs.harvard.edu/abs/1997A%26A...320...41K} {320, 41}

\bibitem[\protect\citeauthoryear{{Koester} et~al.,}{{Koester}
  et~al.}{2007}]{Koester07}
{Koester} B.~P.,  et~al., 2007, \mn@doi [\apj] {10.1086/509599}, \href
  {http://adsabs.harvard.edu/abs/2007ApJ...660..239K} {660, 239}

\bibitem[\protect\citeauthoryear{{Laine}, {van der Marel}, {Lauer}, {Postman},
  {O'Dea}  \& {Owen}}{{Laine} et~al.}{2003}]{Laine2003}
{Laine} S.,  {van der Marel} R.~P.,  {Lauer} T.~R.,  {Postman} M.,  {O'Dea}
  C.~P.,   {Owen} F.~N.,  2003, \mn@doi [\aj] {10.1086/345823}, \href
  {http://adsabs.harvard.edu/abs/2003AJ....125..478L} {125, 478}

\bibitem[\protect\citeauthoryear{{Lauer} \& {Postman}}{{Lauer} \&
  {Postman}}{1994}]{LauerPostman1994}
{Lauer} T.~R.,  {Postman} M.,  1994, \mn@doi [\apj] {10.1086/173997}, \href
  {http://adsabs.harvard.edu/abs/1994ApJ...425..418L} {425, 418}

\bibitem[\protect\citeauthoryear{{Ledlow}, {Voges}, {Owen}  \&
  {Burns}}{{Ledlow} et~al.}{2003}]{Ledlow2003}
{Ledlow} M.~J.,  {Voges} W.,  {Owen} F.~N.,   {Burns} J.~O.,  2003, \mn@doi
  [\aj] {10.1086/379670}, \href
  {http://adsabs.harvard.edu/abs/2003AJ....126.2740L} {126, 2740}

\bibitem[\protect\citeauthoryear{{Livermore}, {Jones}, {Richard}, {Bower},
  {Ellis}, {Swinbank}, {Rigby}  et~al.}{{Livermore}
  et~al.}{2012}]{Livermore2012}
{Livermore} R.~C.,  {Jones} T.,  {Richard} J.,  {Bower} R.~G.,  {Ellis} R.~S.,
  {Swinbank} A.~M.,  {Rigby} J.~R.,   et~al., 2012, \mn@doi [\mnras]
  {10.1111/j.1365-2966.2012.21900.x}, \href
  {http://adsabs.harvard.edu/abs/2012MNRAS.427..688L} {427, 688}

\bibitem[\protect\citeauthoryear{{Livermore}, {Jones}, {Richard}, {Bower},
  {Swinbank}, {Yuan}, {Edge}  et~al.}{{Livermore} et~al.}{2015}]{Livermore2015}
{Livermore} R.~C.,  {Jones} T.~A.,  {Richard} J.,  {Bower} R.~G.,  {Swinbank}
  A.~M.,  {Yuan} T.-T.,  {Edge} A.~C.,   et~al., 2015, \mn@doi [\mnras]
  {10.1093/mnras/stv686}, \href
  {http://adsabs.harvard.edu/abs/2015MNRAS.450.1812L} {450, 1812}

\bibitem[\protect\citeauthoryear{{L{\'o}pez-Cruz}}{{L{\'o}pez-Cruz}}{1997}]{LopezCruz1997}
{L{\'o}pez-Cruz} O.,  1997, PhD thesis, , University of Toronto (LC97), (1997)

\bibitem[\protect\citeauthoryear{{L{\'o}pez-Cruz}, {Barkhouse}  \&
  {Yee}}{{L{\'o}pez-Cruz} et~al.}{2004}]{LopezCruz2004}
{L{\'o}pez-Cruz} O.,  {Barkhouse} W.~A.,   {Yee} H.~K.~C.,  2004, \mn@doi
  [\apj] {10.1086/423664}, \href
  {http://adsabs.harvard.edu/abs/2004ApJ...614..679L} {614, 679}

\bibitem[\protect\citeauthoryear{{Mann} \& {Ebeling}}{{Mann} \&
  {Ebeling}}{2012}]{Mann2012}
{Mann} A.~W.,  {Ebeling} H.,  2012, \mn@doi [\mnras]
  {10.1111/j.1365-2966.2011.20170.x}, \href
  {http://adsabs.harvard.edu/abs/2012MNRAS.420.2120M} {420, 2120}

\bibitem[\protect\citeauthoryear{{Mantz}, {Allen}, {Ebeling}  \&
  {Rapetti}}{{Mantz} et~al.}{2008}]{Mantz2008}
{Mantz} A.,  {Allen} S.~W.,  {Ebeling} H.,   {Rapetti} D.,  2008, \mn@doi
  [\mnras] {10.1111/j.1365-2966.2008.13311.x}, \href
  {http://adsabs.harvard.edu/abs/2008MNRAS.387.1179M} {387, 1179}

\bibitem[\protect\citeauthoryear{{Mantz}, {Allen}, {Rapetti}  \&
  {Ebeling}}{{Mantz} et~al.}{2010a}]{Mantz2010a}
{Mantz} A.,  {Allen} S.~W.,  {Rapetti} D.,   {Ebeling} H.,  2010a, \mn@doi
  [\mnras] {10.1111/j.1365-2966.2010.16992.x}, \href
  {http://adsabs.harvard.edu/abs/2010MNRAS.406.1759M} {406, 1759}

\bibitem[\protect\citeauthoryear{{Mantz}, {Allen}, {Ebeling}, {Rapetti}  \&
  {Drlica-Wagner}}{{Mantz} et~al.}{2010b}]{Mantz2010b}
{Mantz} A.,  {Allen} S.~W.,  {Ebeling} H.,  {Rapetti} D.,   {Drlica-Wagner} A.,
   2010b, \mn@doi [\mnras] {10.1111/j.1365-2966.2010.16993.x}, \href
  {http://adsabs.harvard.edu/abs/2010MNRAS.406.1773M} {406, 1773}

\bibitem[\protect\citeauthoryear{{Mantz}, {Allen}, {Morris}, {Rapetti},
  {Applegate}, {Kelly}, {von der Linden}  \& {Schmidt}}{{Mantz}
  et~al.}{2014}]{Mantz2014}
{Mantz} A.~B.,  {Allen} S.~W.,  {Morris} R.~G.,  {Rapetti} D.~A.,  {Applegate}
  D.~E.,  {Kelly} P.~L.,  {von der Linden} A.,   {Schmidt} R.~W.,  2014,
  \mn@doi [\mnras] {10.1093/mnras/stu368}, \href
  {http://adsabs.harvard.edu/abs/2014MNRAS.440.2077M} {440, 2077}

\bibitem[\protect\citeauthoryear{{Markevitch}, {Gonzalez}, {Clowe},
  {Vikhlinin}, {Forman}, {Jones}, {Murray}  \& {Tucker}}{{Markevitch}
  et~al.}{2004}]{Markevitch2004}
{Markevitch} M.,  {Gonzalez} A.~H.,  {Clowe} D.,  {Vikhlinin} A.,  {Forman} W.,
   {Jones} C.,  {Murray} S.,   {Tucker} W.,  2004, \mn@doi [\apj]
  {10.1086/383178}, \href {http://adsabs.harvard.edu/abs/2004ApJ...606..819M}
  {606, 819}

\bibitem[\protect\citeauthoryear{{Marriage}, {Acquaviva}, {Ade}, {Aguirre},
  {Amiri}, {Appel}, {Barrientos}  et~al.}{{Marriage}
  et~al.}{2011}]{Marriage2011}
{Marriage} T.~A.,  {Acquaviva} V.,  {Ade} P.~A.~R.,  {Aguirre} P.,  {Amiri} M.,
   {Appel} J.~W.,  {Barrientos} L.~F.,   et~al., 2011, \mn@doi [\apj]
  {10.1088/0004-637X/737/2/61}, \href
  {http://adsabs.harvard.edu/abs/2011ApJ...737...61M} {737, 61}

\bibitem[\protect\citeauthoryear{{McPartland}, {Ebeling}, {Roediger}  \&
  {Blumenthal}}{{McPartland} et~al.}{2016}]{McPartland2016}
{McPartland} C.,  {Ebeling} H.,  {Roediger} E.,   {Blumenthal} K.,  2016,
  \mn@doi [\mnras] {10.1093/mnras/stv2508}, \href
  {http://adsabs.harvard.edu/abs/2016MNRAS.455.2994M} {455, 2994}

\bibitem[\protect\citeauthoryear{{Mei}, {Holden}, {Blakeslee}, {Ford}, {Franx},
  {Homeier}, {Illingworth}  et~al.}{{Mei} et~al.}{2009}]{Mei2009}
{Mei} S.,  {Holden} B.~P.,  {Blakeslee} J.~P.,  {Ford} H.~C.,  {Franx} M.,
  {Homeier} N.~L.,  {Illingworth} G.~D.,   et~al., 2009, \mn@doi [\apj]
  {10.1088/0004-637X/690/1/42}, \href
  {http://adsabs.harvard.edu/abs/2009ApJ...690...42M} {690, 42}

\bibitem[\protect\citeauthoryear{{Meneghetti}, {Bartelmann}  \&
  {Moscardini}}{{Meneghetti} et~al.}{2003}]{Meneghetti2003}
{Meneghetti} M.,  {Bartelmann} M.,   {Moscardini} L.,  2003, \mn@doi [\mnras]
  {10.1046/j.1365-8711.2003.06276.x}, \href
  {http://adsabs.harvard.edu/abs/2003MNRAS.340..105M} {340, 105}

\bibitem[\protect\citeauthoryear{{Meneghetti}, {Fedeli}, {Pace},
  {Gottl{\"o}ber}  \& {Yepes}}{{Meneghetti} et~al.}{2010}]{Meneghetti2010}
{Meneghetti} M.,  {Fedeli} C.,  {Pace} F.,  {Gottl{\"o}ber} S.,   {Yepes} G.,
  2010, \mn@doi [\aap] {10.1051/0004-6361/201014098}, \href
  {http://adsabs.harvard.edu/abs/2010A%26A...519A..90M} {519, A90}

\bibitem[\protect\citeauthoryear{{Meneghetti}, {Fedeli}, {Zitrin},
  {Bartelmann}, {Broadhurst}, {Gottl{\"o}ber}, {Moscardini}  \&
  {Yepes}}{{Meneghetti} et~al.}{2011}]{Meneghetti2011}
{Meneghetti} M.,  {Fedeli} C.,  {Zitrin} A.,  {Bartelmann} M.,  {Broadhurst}
  T.,  {Gottl{\"o}ber} S.,  {Moscardini} L.,   {Yepes} G.,  2011, \mn@doi
  [\aap] {10.1051/0004-6361/201016040}, \href
  {http://adsabs.harvard.edu/abs/2011A%26A...530A..17M} {530, A17}

\bibitem[\protect\citeauthoryear{{Meneghetti}, {Bartelmann}, {Dahle}  \&
  {Limousin}}{{Meneghetti} et~al.}{2013}]{Meneghetti2013}
{Meneghetti} M.,  {Bartelmann} M.,  {Dahle} H.,   {Limousin} M.,  2013, \mn@doi
  [\ssr] {10.1007/s11214-013-9981-x}, \href
  {http://adsabs.harvard.edu/abs/2013SSRv..177...31M} {177, 31}

\bibitem[\protect\citeauthoryear{{Merten}, {Coe}, {Dupke}, {Massey}, {Zitrin},
  {Cypriano}, {Okabe}  et~al.}{{Merten} et~al.}{2011}]{Merten2011}
{Merten} J.,  {Coe} D.,  {Dupke} R.,  {Massey} R.,  {Zitrin} A.,  {Cypriano}
  E.~S.,  {Okabe} N.,   et~al., 2011, \mn@doi [\mnras]
  {10.1111/j.1365-2966.2011.19266.x}, \href
  {http://adsabs.harvard.edu/abs/2011MNRAS.417..333M} {417, 333}

\bibitem[\protect\citeauthoryear{{Newman}, {Belli}  \& {Ellis}}{{Newman}
  et~al.}{2015}]{Newman2015}
{Newman} A.~B.,  {Belli} S.,   {Ellis} R.~S.,  2015, \mn@doi [\apjl]
  {10.1088/2041-8205/813/1/L7}, \href
  {http://adsabs.harvard.edu/abs/2015ApJ...813L...7N} {813, L7}

\bibitem[\protect\citeauthoryear{{Oguri}, {Lee}  \& {Suto}}{{Oguri}
  et~al.}{2003}]{Oguri2003}
{Oguri} M.,  {Lee} J.,   {Suto} Y.,  2003, \mn@doi [\apj] {10.1086/379223},
  \href {http://adsabs.harvard.edu/abs/2003ApJ...599....7O} {599, 7}

\bibitem[\protect\citeauthoryear{{Oke} \& {Gunn}}{{Oke} \&
  {Gunn}}{1983}]{OkeGunn1983}
{Oke} J.~B.,  {Gunn} J.~E.,  1983, \mn@doi [\apj] {10.1086/160817}, \href
  {http://adsabs.harvard.edu/abs/1983ApJ...266..713O} {266, 713}

\bibitem[\protect\citeauthoryear{{Patr{\'{\i}}cio}, {Richard}, {Verhamme},
  {Wisotzki}, {Brinchmann}, {Turner}, {Christensen}  et~al.}{{Patr{\'{\i}}cio}
  et~al.}{2016}]{Patricio2016}
{Patr{\'{\i}}cio} V.,  {Richard} J.,  {Verhamme} A.,  {Wisotzki} L.,
  {Brinchmann} J.,  {Turner} M.~L.,  {Christensen} L.,   et~al., 2016, \mn@doi
  [\mnras] {10.1093/mnras/stv2859}, \href
  {http://adsabs.harvard.edu/abs/2016MNRAS.456.4191P} {456, 4191}

\bibitem[\protect\citeauthoryear{{Pimbblet}, {Smail}, {Edge}, {Couch}, {O'Hely}
   \& {Zabludoff}}{{Pimbblet} et~al.}{2001}]{Pimbblet2001}
{Pimbblet} K.~A.,  {Smail} I.,  {Edge} A.~C.,  {Couch} W.~J.,  {O'Hely} E.,
  {Zabludoff} A.~I.,  2001, \mn@doi [\mnras]
  {10.1046/j.1365-8711.2001.04759.x}, \href
  {http://adsabs.harvard.edu/abs/2001MNRAS.327..588P} {327, 588}

\bibitem[\protect\citeauthoryear{{Pimbblet}, {Smail}, {Edge}, {O'Hely}, {Couch}
   \& {Zabludoff}}{{Pimbblet} et~al.}{2006}]{Pimbblet2006}
{Pimbblet} K.~A.,  {Smail} I.,  {Edge} A.~C.,  {O'Hely} E.,  {Couch} W.~J.,
  {Zabludoff} A.~I.,  2006, \mn@doi [\mnras]
  {10.1111/j.1365-2966.2005.09892.x}, \href
  {http://adsabs.harvard.edu/abs/2006MNRAS.366..645P} {366, 645}

\bibitem[\protect\citeauthoryear{{Planck Collaboration}, {Ade}, {Aghanim},
  {Armitage-Caplan}, {Arnaud}, {Ashdown}, {Atrio-Barandela}  et~al.}{{Planck
  Collaboration} et~al.}{2014}]{PC2014}
{Planck Collaboration} {Ade} P.~A.~R.,  {Aghanim} N.,  {Armitage-Caplan} C.,
  {Arnaud} M.,  {Ashdown} M.,  {Atrio-Barandela} F.,   et~al., 2014, \mn@doi
  [\aap] {10.1051/0004-6361/201321521}, \href
  {http://adsabs.harvard.edu/abs/2014A%26A...571A..20P} {571, A20}

\bibitem[\protect\citeauthoryear{{Planck Collaboration}, {Ade}, {Aghanim},
  {Arnaud}, {Ashdown}, {Aumont}, {Baccigalupi}  et~al.}{{Planck Collaboration}
  et~al.}{2015}]{Ade2015}
{Planck Collaboration} {Ade} P.~A.~R.,  {Aghanim} N.,  {Arnaud} M.,  {Ashdown}
  M.,  {Aumont} J.,  {Baccigalupi} C.,   et~al., 2015, \mn@doi [\aap]
  {10.1051/0004-6361/201424674}, \href
  {http://adsabs.harvard.edu/abs/2015A%26A...582A..29P} {582, A29}

\bibitem[\protect\citeauthoryear{{Planck Collaboration}, {Ade}, {Aghanim},
  {Arg{\"u}eso}, {Arnaud}, {Ashdown}, {Aumont}  et~al.}{{Planck Collaboration}
  et~al.}{2016}]{Planck2016}
{Planck Collaboration} {Ade} P.~A.~R.,  {Aghanim} N.,  {Arg{\"u}eso} F.,
  {Arnaud} M.,  {Ashdown} M.,  {Aumont} J.,   et~al., 2016, \mn@doi [\aap]
  {10.1051/0004-6361/201526914}, \href
  {http://adsabs.harvard.edu/abs/2016A%26A...594A..26P} {594, A26}

\bibitem[\protect\citeauthoryear{{Plionis}, {Tovmassian}  \&
  {Andernach}}{{Plionis} et~al.}{2009}]{Plionis2009}
{Plionis} M.,  {Tovmassian} H.~M.,   {Andernach} H.,  2009, \mn@doi [\mnras]
  {10.1111/j.1365-2966.2009.14507.x}, \href
  {http://adsabs.harvard.edu/abs/2009MNRAS.395....2P} {395, 2}

\bibitem[\protect\citeauthoryear{{Postman}, {Coe}, {Ben{\'{\i}}tez}, {Bradley},
  {Broadhurst}, {Donahue}, {Ford}  et~al.}{{Postman}
  et~al.}{2012}]{Postman2012}
{Postman} M.,  {Coe} D.,  {Ben{\'{\i}}tez} N.,  {Bradley} L.,  {Broadhurst} T.,
   {Donahue} M.,  {Ford} H.,   et~al., 2012, \mn@doi [\apjs]
  {10.1088/0067-0049/199/2/25}, \href
  {http://adsabs.harvard.edu/abs/2012ApJS..199...25P} {199, 25}

\bibitem[\protect\citeauthoryear{{Quider}, {Shapley}, {Pettini}, {Steidel}  \&
  {Stark}}{{Quider} et~al.}{2010}]{Quider2010}
{Quider} A.~M.,  {Shapley} A.~E.,  {Pettini} M.,  {Steidel} C.~C.,   {Stark}
  D.~P.,  2010, \mn@doi [\mnras] {10.1111/j.1365-2966.2009.16005.x}, \href
  {http://adsabs.harvard.edu/abs/2010MNRAS.402.1467Q} {402, 1467}

\bibitem[\protect\citeauthoryear{{Quillen}, {Zufelt}, {Park}, {O'Dea}, {Baum},
  {Privon}, {Noel-Storr}  et~al.}{{Quillen} et~al.}{2008}]{Quillen2008}
{Quillen} A.~C.,  {Zufelt} N.,  {Park} J.,  {O'Dea} C.~P.,  {Baum} S.~A.,
  {Privon} G.,  {Noel-Storr} J.,   et~al., 2008, \mn@doi [\apjs]
  {10.1086/525560}, \href {http://adsabs.harvard.edu/abs/2008ApJS..176...39Q}
  {176, 39}

\bibitem[\protect\citeauthoryear{{Reichert}, {B{\"o}hringer}, {Fassbender}  \&
  {M{\"u}hlegger}}{{Reichert} et~al.}{2011}]{Reichert2011}
{Reichert} A.,  {B{\"o}hringer} H.,  {Fassbender} R.,   {M{\"u}hlegger} M.,
  2011, \mn@doi [\aap] {10.1051/0004-6361/201116861}, \href
  {http://adsabs.harvard.edu/abs/2011A%26A...535A...4R} {535, A4}

\bibitem[\protect\citeauthoryear{{Repp}, {Ebeling}  \& {Richard}}{{Repp}
  et~al.}{2016}]{Repp2016}
{Repp} A.,  {Ebeling} H.,   {Richard} J.,  2016, \mn@doi [\mnras]
  {10.1093/mnras/stw002}, \href
  {http://adsabs.harvard.edu/abs/2016MNRAS.457.1399R} {457, 1399}

\bibitem[\protect\citeauthoryear{{Richard}, {Patricio}, {Martinez}, {Bacon},
  {Cl{\'e}ment}, {Weilbacher}, {Soto}  et~al.}{{Richard}
  et~al.}{2015}]{Richard2015}
{Richard} J.,  {Patricio} V.,  {Martinez} J.,  {Bacon} R.,  {Cl{\'e}ment} B.,
  {Weilbacher} P.,  {Soto} K.,   et~al., 2015, \mn@doi [\mnras]
  {10.1093/mnrasl/slu150}, \href
  {http://adsabs.harvard.edu/abs/2015MNRAS.446L..16R} {446, L16}

\bibitem[\protect\citeauthoryear{{Riechers}, {Carilli}, {Walter}  \&
  {Momjian}}{{Riechers} et~al.}{2010}]{Riechers2010}
{Riechers} D.~A.,  {Carilli} C.~L.,  {Walter} F.,   {Momjian} E.,  2010,
  \mn@doi [\apjl] {10.1088/2041-8205/724/2/L153}, \href
  {http://adsabs.harvard.edu/abs/2010ApJ...724L.153R} {724, L153}

\bibitem[\protect\citeauthoryear{{Rykoff}, {McKay}, {Becker}, {Evrard},
  {Johnston}, {Koester}, {Rozo}  et~al.}{{Rykoff} et~al.}{2008}]{Rykoff2008}
{Rykoff} E.~S.,  {McKay} T.~A.,  {Becker} M.~R.,  {Evrard} A.,  {Johnston}
  D.~E.,  {Koester} B.~P.,  {Rozo} E.,   et~al., 2008, \mn@doi [\apj]
  {10.1086/527537}, \href {http://adsabs.harvard.edu/abs/2008ApJ...675.1106R}
  {675, 1106}

\bibitem[\protect\citeauthoryear{{Sanderson}, {Edge}  \& {Smith}}{{Sanderson}
  et~al.}{2009}]{Sanderson2009}
{Sanderson} A.~J.~R.,  {Edge} A.~C.,   {Smith} G.~P.,  2009, \mn@doi [\mnras]
  {10.1111/j.1365-2966.2009.15214.x}, \href
  {http://adsabs.harvard.edu/abs/2009MNRAS.398.1698S} {398, 1698}

\bibitem[\protect\citeauthoryear{{Saro}, {Mohr}, {Bazin}  \& {Dolag}}{{Saro}
  et~al.}{2013}]{Saro2013}
{Saro} A.,  {Mohr} J.~J.,  {Bazin} G.,   {Dolag} K.,  2013, \mn@doi [\apj]
  {10.1088/0004-637X/772/1/47}, \href
  {http://adsabs.harvard.edu/abs/2013ApJ...772...47S} {772, 47}

\bibitem[\protect\citeauthoryear{{Schaerer}, {Boone}, {Jones},
  {Dessauges-Zavadsky}, {Sklias}, {Zamojski}, {Cava}  et~al.}{{Schaerer}
  et~al.}{2015}]{Schaerer2015}
{Schaerer} D.,  {Boone} F.,  {Jones} T.,  {Dessauges-Zavadsky} M.,  {Sklias}
  P.,  {Zamojski} M.,  {Cava} A.,   et~al., 2015, \mn@doi [\aap]
  {10.1051/0004-6361/201425542}, \href
  {http://adsabs.harvard.edu/abs/2015A%26A...576L...2S} {576, L2}

\bibitem[\protect\citeauthoryear{{Schindler}, {Castillo-Morales}, {De
  Filippis}, {Schwope}  \& {Wambsganss}}{{Schindler}
  et~al.}{2001}]{Schindler2001}
{Schindler} S.,  {Castillo-Morales} A.,  {De Filippis} E.,  {Schwope} A.,
  {Wambsganss} J.,  2001, \mn@doi [\aap] {10.1051/0004-6361:20010990}, \href
  {http://adsabs.harvard.edu/abs/2001A%26A...376L..27S} {376, L27}

\bibitem[\protect\citeauthoryear{{Siana} et~al.,}{{Siana}
  et~al.}{2009}]{Siana2009}
{Siana} B.,  et~al., 2009, \mn@doi [\apj] {10.1088/0004-637X/698/2/1273}, \href
  {http://adsabs.harvard.edu/abs/2009ApJ...698.1273S} {698, 1273}

\bibitem[\protect\citeauthoryear{{Sklias}, {Zamojski}, {Schaerer},
  {Dessauges-Zavadsky}, {Egami}, {Rex}, {Rawle}  et~al.}{{Sklias}
  et~al.}{2014}]{Sklias2014}
{Sklias} P.,  {Zamojski} M.,  {Schaerer} D.,  {Dessauges-Zavadsky} M.,  {Egami}
  E.,  {Rex} M.,  {Rawle} T.,   et~al., 2014, \mn@doi [\aap]
  {10.1051/0004-6361/201322424}, \href
  {http://adsabs.harvard.edu/abs/2014A%26A...561A.149S} {561, A149}

\bibitem[\protect\citeauthoryear{{Smail}, {Swinbank}, {Richard}, {Ebeling},
  {Kneib}, {Edge}, {Stark}  et~al.}{{Smail} et~al.}{2007}]{Smail2007}
{Smail} I.,  {Swinbank} A.~M.,  {Richard} J.,  {Ebeling} H.,  {Kneib} J.-P.,
  {Edge} A.~C.,  {Stark} D.,   et~al., 2007, \mn@doi [\apjl] {10.1086/510902},
  \href {http://adsabs.harvard.edu/abs/2007ApJ...654L..33S} {654, L33}

\bibitem[\protect\citeauthoryear{{Smith} et~al.,}{{Smith}
  et~al.}{2008}]{Smith2008}
{Smith} R.~J.,  et~al., 2008, \mn@doi [\mnras]
  {10.1111/j.1745-3933.2008.00469.x}, \href
  {http://adsabs.harvard.edu/abs/2008MNRAS.386L..96S} {386, L96}

\bibitem[\protect\citeauthoryear{{Song}, {Mohr}, {Barkhouse}, {Warren}, {Dolag}
   \& {Rude}}{{Song} et~al.}{2012}]{Song2012}
{Song} J.,  {Mohr} J.~J.,  {Barkhouse} W.~A.,  {Warren} M.~S.,  {Dolag} K.,
  {Rude} C.,  2012, \mn@doi [\apj] {10.1088/0004-637X/747/1/58}, \href
  {http://adsabs.harvard.edu/abs/2012ApJ...747...58S} {747, 58}

\bibitem[\protect\citeauthoryear{{Spitler}, {Labb{\'e}}, {Glazebrook},
  {Persson}, {Monson}, {Papovich}, {Tran}  et~al.}{{Spitler}
  et~al.}{2012}]{Spitler2012}
{Spitler} L.~R.,  {Labb{\'e}} I.,  {Glazebrook} K.,  {Persson} S.~E.,  {Monson}
  A.,  {Papovich} C.,  {Tran} K.-V.~H.,   et~al., 2012, \mn@doi [\apjl]
  {10.1088/2041-8205/748/2/L21}, \href
  {http://adsabs.harvard.edu/abs/2012ApJ...748L..21S} {748, L21}

\bibitem[\protect\citeauthoryear{{Stanek}, {Evrard}, {B{\"o}hringer},
  {Schuecker}  \& {Nord}}{{Stanek} et~al.}{2006}]{Stanek2006}
{Stanek} R.,  {Evrard} A.~E.,  {B{\"o}hringer} H.,  {Schuecker} P.,   {Nord}
  B.,  2006, \mn@doi [\apj] {10.1086/506248}, \href
  {http://adsabs.harvard.edu/abs/2006ApJ...648..956S} {648, 956}

\bibitem[\protect\citeauthoryear{{Stark}, {Swinbank}, {Ellis}, {Dye}, {Smail}
  \& {Richard}}{{Stark} et~al.}{2008}]{Stark2008}
{Stark} D.~P.,  {Swinbank} A.~M.,  {Ellis} R.~S.,  {Dye} S.,  {Smail} I.~R.,
  {Richard} J.,  2008, \mn@doi [\nat] {10.1038/nature07294}, \href
  {http://adsabs.harvard.edu/abs/2008Natur.455..775S} {455, 775}

\bibitem[\protect\citeauthoryear{{Stark}, {Richard}, {Siana}, {Charlot},
  {Freeman}, {Gutkin}, {Wofford}  et~al.}{{Stark} et~al.}{2014}]{Stark2014}
{Stark} D.~P.,  {Richard} J.,  {Siana} B.,  {Charlot} S.,  {Freeman} W.~R.,
  {Gutkin} J.,  {Wofford} A.,   et~al., 2014, \mn@doi [\mnras]
  {10.1093/mnras/stu1618}, \href
  {http://adsabs.harvard.edu/abs/2014MNRAS.445.3200S} {445, 3200}

\bibitem[\protect\citeauthoryear{{Stott}, {Smail}, {Edge}, {Ebeling}, {Smith},
  {Kneib}  \& {Pimbblet}}{{Stott} et~al.}{2007}]{Stott2007}
{Stott} J.~P.,  {Smail} I.,  {Edge} A.~C.,  {Ebeling} H.,  {Smith} G.~P.,
  {Kneib} J.-P.,   {Pimbblet} K.~A.,  2007, \mn@doi [\apj] {10.1086/514329},
  \href {http://adsabs.harvard.edu/abs/2007ApJ...661...95S} {661, 95}

\bibitem[\protect\citeauthoryear{{Stott}, {Pimbblet}, {Edge}, {Smith}  \&
  {Wardlow}}{{Stott} et~al.}{2009}]{Stott2009}
{Stott} J.~P.,  {Pimbblet} K.~A.,  {Edge} A.~C.,  {Smith} G.~P.,   {Wardlow}
  J.~L.,  2009, \mn@doi [\mnras] {10.1111/j.1365-2966.2009.14477.x}, \href
  {http://adsabs.harvard.edu/abs/2009MNRAS.394.2098S} {394, 2098}

\bibitem[\protect\citeauthoryear{{Stott}, {Hickox}, {Edge}, {Collins},
  {Hilton}, {Harrison}, {Romer}  et~al.}{{Stott} et~al.}{2012}]{Stott2012}
{Stott} J.~P.,  {Hickox} R.~C.,  {Edge} A.~C.,  {Collins} C.~A.,  {Hilton} M.,
  {Harrison} C.~D.,  {Romer} A.~K.,   et~al., 2012, \mn@doi [\mnras]
  {10.1111/j.1365-2966.2012.20764.x}, \href
  {http://adsabs.harvard.edu/abs/2012MNRAS.422.2213S} {422, 2213}

\bibitem[\protect\citeauthoryear{{Sunyaev} \& {Zeldovich}}{{Sunyaev} \&
  {Zeldovich}}{1972}]{SZ1972}
{Sunyaev} R.~A.,  {Zeldovich} Y.~B.,  1972, Comments on Astrophysics and Space
  Physics, \href {http://adsabs.harvard.edu/abs/1972CoASP...4..173S} {4, 173}

\bibitem[\protect\citeauthoryear{{Sutherland}}{{Sutherland}}{1988}]{Sutherland88}
{Sutherland} W.,  1988, \mn@doi [\mnras] {10.1093/mnras/234.1.159}, \href
  {http://adsabs.harvard.edu/abs/1988MNRAS.234..159S} {234, 159}

\bibitem[\protect\citeauthoryear{{Swinbank}, {Papadopoulos}, {Cox}, {Krips},
  {Ivison}, {Smail}, {Thomson}  et~al.}{{Swinbank} et~al.}{2011}]{Swinbank2011}
{Swinbank} A.~M.,  {Papadopoulos} P.~P.,  {Cox} P.,  {Krips} M.,  {Ivison}
  R.~J.,  {Smail} I.,  {Thomson} A.~P.,   et~al., 2011, \mn@doi [\apj]
  {10.1088/0004-637X/742/1/11}, \href
  {http://adsabs.harvard.edu/abs/2011ApJ...742...11S} {742, 11}

\bibitem[\protect\citeauthoryear{{Tanaka}, {De Breuck}, {Venemans}  \&
  {Kurk}}{{Tanaka} et~al.}{2010}]{Tanaka2010}
{Tanaka} M.,  {De Breuck} C.,  {Venemans} B.,   {Kurk} J.,  2010, \mn@doi
  [\aap] {10.1051/0004-6361/200913939}, \href
  {http://adsabs.harvard.edu/abs/2010A%26A...518A..18T} {518, A18}

\bibitem[\protect\citeauthoryear{{Thomas}, {Saglia}, {Bender}, {Erwin}  \&
  {Fabricius}}{{Thomas} et~al.}{2014}]{Thomas2014}
{Thomas} J.,  {Saglia} R.~P.,  {Bender} R.,  {Erwin} P.,   {Fabricius} M.,
  2014, \mn@doi [\apj] {10.1088/0004-637X/782/1/39}, \href
  {http://adsabs.harvard.edu/abs/2014ApJ...782...39T} {782, 39}

\bibitem[\protect\citeauthoryear{{Thomson}, {Ivison}, {Owen}, {Danielson},
  {Swinbank}  \& {Smail}}{{Thomson} et~al.}{2015}]{Thomson2015}
{Thomson} A.~P.,  {Ivison} R.~J.,  {Owen} F.~N.,  {Danielson} A.~L.~R.,
  {Swinbank} A.~M.,   {Smail} I.,  2015, \mn@doi [\mnras]
  {10.1093/mnras/stv118}, \href
  {http://adsabs.harvard.edu/abs/2015MNRAS.448.1874T} {448, 1874}

\bibitem[\protect\citeauthoryear{{Torri}, {Meneghetti}, {Bartelmann},
  {Moscardini}, {Rasia}  \& {Tormen}}{{Torri} et~al.}{2004}]{Torri2004}
{Torri} E.,  {Meneghetti} M.,  {Bartelmann} M.,  {Moscardini} L.,  {Rasia} E.,
   {Tormen} G.,  2004, \mn@doi [\mnras] {10.1111/j.1365-2966.2004.07508.x},
  \href {http://adsabs.harvard.edu/abs/2004MNRAS.349..476T} {349, 476}

\bibitem[\protect\citeauthoryear{{Voges}, {Aschenbach}, {Boller},
  {Br{\"a}uninger}, {Briel}, {Burkert}, {Dennerl}  et~al.}{{Voges}
  et~al.}{1999}]{Voges1999}
{Voges} W.,  {Aschenbach} B.,  {Boller} T.,  {Br{\"a}uninger} H.,  {Briel} U.,
  {Burkert} W.,  {Dennerl} K.,   et~al., 1999, \aap, \href
  {http://adsabs.harvard.edu/abs/1999A%26A...349..389V} {349, 389}

\bibitem[\protect\citeauthoryear{{Wen} \& {Han}}{{Wen} \&
  {Han}}{2013}]{Wen2013}
{Wen} Z.~L.,  {Han} J.~L.,  2013, \mn@doi [\mnras] {10.1093/mnras/stt1581},
  \href {http://adsabs.harvard.edu/abs/2013MNRAS.436..275W} {436, 275}

\bibitem[\protect\citeauthoryear{{Werner}, {Oonk}, {Sun}, {Nulsen}, {Allen},
  {Canning}, {Simionescu}  et~al.}{{Werner} et~al.}{2014}]{Werner2014}
{Werner} N.,  {Oonk} J.~B.~R.,  {Sun} M.,  {Nulsen} P.~E.~J.,  {Allen} S.~W.,
  {Canning} R.~E.~A.,  {Simionescu} A.,   et~al., 2014, \mn@doi [\mnras]
  {10.1093/mnras/stu006}, \href
  {http://adsabs.harvard.edu/abs/2014MNRAS.439.2291W} {439, 2291}

\bibitem[\protect\citeauthoryear{{Williamson}, {Benson}, {High}, {Vanderlinde},
  {Ade}, {Aird}, {Andersson}  et~al.}{{Williamson}
  et~al.}{2011}]{Williamson2011}
{Williamson} R.,  {Benson} B.~A.,  {High} F.~W.,  {Vanderlinde} K.,  {Ade}
  P.~A.~R.,  {Aird} K.~A.,  {Andersson} K.,   et~al., 2011, \mn@doi [\apj]
  {10.1088/0004-637X/738/2/139}, \href
  {http://adsabs.harvard.edu/abs/2011ApJ...738..139W} {738, 139}

\bibitem[\protect\citeauthoryear{{Wittman}, {Golovich}  \& {Dawson}}{{Wittman}
  et~al.}{2017}]{Wittman2017}
{Wittman} D.,  {Golovich} N.,   {Dawson} W.~A.,  2017, preprint, \href
  {http://adsabs.harvard.edu/abs/2017arXiv170105877W} {} (\mn@eprint {arXiv}
  {1701.05877})

\bibitem[\protect\citeauthoryear{{Xu}, {Postman}, {Meneghetti}, {Seitz},
  {Zitrin}, {Merten}, {Maoz}  et~al.}{{Xu} et~al.}{2016}]{Xu2016}
{Xu} B.,  {Postman} M.,  {Meneghetti} M.,  {Seitz} S.,  {Zitrin} A.,  {Merten}
  J.,  {Maoz} D.,   et~al., 2016, \mn@doi [\apj] {10.3847/0004-637X/817/2/85},
  \href {http://adsabs.harvard.edu/abs/2016ApJ...817...85X} {817, 85}

\bibitem[\protect\citeauthoryear{{Yee} \& {L{\'o}pez-Cruz}}{{Yee} \&
  {L{\'o}pez-Cruz}}{1999}]{YeeCruz99}
{Yee} H.~K.~C.,  {L{\'o}pez-Cruz} O.,  1999, \mn@doi [\aj] {10.1086/300837},
  \href {http://adsabs.harvard.edu/abs/1999AJ....117.1985Y} {117, 1985}

\bibitem[\protect\citeauthoryear{{Zaritsky} \& {Gonzalez}}{{Zaritsky} \&
  {Gonzalez}}{2003}]{Zaritsky2003}
{Zaritsky} D.,  {Gonzalez} A.~H.,  2003, \mn@doi [\apj] {10.1086/345601}, \href
  {http://adsabs.harvard.edu/abs/2003ApJ...584..691Z} {584, 691}

\bibitem[\protect\citeauthoryear{{Zhang}, {Reiprich}, {Schneider}, {Clerc},
  {Merloni}, {Schwope}, {Borm}  et~al.}{{Zhang} et~al.}{2017}]{Zhang2017}
{Zhang} Y.-Y.,  {Reiprich} T.~H.,  {Schneider} P.,  {Clerc} N.,  {Merloni} A.,
  {Schwope} A.,  {Borm} K.,   et~al., 2017, \mn@doi [\aap]
  {10.1051/0004-6361/201628971}, \href
  {http://adsabs.harvard.edu/abs/2017A%26A...599A.138Z} {599, A138}

\bibitem[\protect\citeauthoryear{{Zitrin}}{{Zitrin}}{2017}]{Zitrin2017}
{Zitrin} A.,  2017, \mn@doi [\apj] {10.3847/1538-4357/834/1/45}, \href
  {http://adsabs.harvard.edu/abs/2017ApJ...834...45Z} {834, 45}

\bibitem[\protect\citeauthoryear{{Zitrin} \& {Broadhurst}}{{Zitrin} \&
  {Broadhurst}}{2016}]{Zitrin2016}
{Zitrin} A.,  {Broadhurst} T.,  2016, \mn@doi [\apj]
  {10.3847/0004-637X/833/1/25}, \href
  {http://adsabs.harvard.edu/abs/2016ApJ...833...25Z} {833, 25}

\bibitem[\protect\citeauthoryear{{Zitrin}, {Rosati}, {Nonino}, {Grillo},
  {Postman}, {Coe}, {Seitz}  et~al.}{{Zitrin} et~al.}{2012}]{Zitrin2012}
{Zitrin} A.,  {Rosati} P.,  {Nonino} M.,  {Grillo} C.,  {Postman} M.,  {Coe}
  D.,  {Seitz} S.,   et~al., 2012, \mn@doi [\apj] {10.1088/0004-637X/749/2/97},
  \href {http://adsabs.harvard.edu/abs/2012ApJ...749...97Z} {749, 97}

\bibitem[\protect\citeauthoryear{{von der Linden}, {Allen}, {Applegate},
  {Kelly}, {Allen}, {Ebeling}, {Burchat}  et~al.}{{von der Linden}
  et~al.}{2014}]{Linden2014}
{von der Linden} A.,  {Allen} M.~T.,  {Applegate} D.~E.,  {Kelly} P.~L.,
  {Allen} S.~W.,  {Ebeling} H.,  {Burchat} P.~R.,   et~al., 2014, \mn@doi
  [\mnras] {10.1093/mnras/stt1945}, \href
  {http://adsabs.harvard.edu/abs/2014MNRAS.439....2V} {439, 2}

\makeatother
\end{thebibliography}

\clearpage
\begin{table*}
\begin{tabular}{lcccccccccc}
\hline\\[-6mm]
& & & & & & & & $z$ & \multicolumn{2}{c}{Morphology\rule{0cm}{15pt}}\\ 
Name & Right Ascension & Declination & F606 & F814 & F110W & F140W & $z$ & reference & X-ray & Optical\\ \hline \\[-4mm]
MACSJ0011.7$-$1523 & 00:11:42.8 & $-$15:23:22 & Y & Y & -- & -- & 0.379 & (2) & 1 & 2 \rule[-3pt]{0cm}{18pt}\\
MACSJ0027.8$+$2616 & 00:27:45.8 & $+$26:16:26 & Y & -- & Y & Y & 0.360 & (1) & 2 & 2 \rule[-3pt]{0cm}{0ex}\\
MACSJ0032.1$+$1808 & 00:32:10.6 & $+$18:07:39 & Y & Y & -- & -- & 0.377 & (1) & -- & 4 \rule[-3pt]{0cm}{0ex}\\
MACSJ0033.8$-$0751 & 00:33:52.2 & $-$07:51:12 & -- & Y & -- & -- & 0.305 & (1) & -- & 4 \rule[-3pt]{0cm}{0ex}\\
MACSJ0034.4$+$0225 & 00:34:26.1 & $+$02:25:33 & -- & Y & -- & -- & 0.388 & (1) & -- & 3 \rule[-3pt]{0cm}{0ex}\\
MACSJ0034.9$+$0234 & 00:34:57.8 & $+$02:33:32 & -- & Y & -- & -- & 0.390 & (1) & -- & 2 \rule[-3pt]{0cm}{0ex}\\
MACSJ0035.4$-$2015 & 00:35:26.1 & $-$20:15:45 & Y & Y & -- & -- & 0.353 & (2) & 3 & 2 \rule[-3pt]{0cm}{0ex}\\
MACSJ0051.6$+$2720 & 00:51:38.6 & $+$27:19:60 & -- & -- & Y & Y & 0.364 & (1) & -- & 2 \rule[-3pt]{0cm}{0ex}\\
MACSJ0110.1$+$3211 & 01:10:07.2 & $+$32:10:48 & Y & Y & -- & -- & 0.341 & (1) & -- & 2 \rule[-3pt]{0cm}{0ex}\\
MACSJ0140.0$-$0555 & 01:40:00.0 & $-$05:54:57 & Y & Y & Y & Y & 0.451 & (3) & 3 & 3 \rule[-3pt]{0cm}{0ex}\\
MACSJ0140.0$-$3410 & 01:40:05.5 & $-$34:10:38 & Y & -- & -- & -- & 0.395 & (1) & -- & 1 \rule[-3pt]{0cm}{0ex}\\
MACSJ0150.3$-$1005 & 01:50:21.2 & $-$10:05:31 & -- & Y & Y & Y & 0.363 & (1) & 1 & 1 \rule[-3pt]{0cm}{0ex}\\
MACSJ0152.5$-$2852 & 01:52:34.5 & $-$28:53:37 & Y & Y & Y & Y & 0.412 & (2) & 2 & 3 \rule[-3pt]{0cm}{0ex}\\
MACSJ0159.8$-$0849 & 01:59:49.3 & $-$08:49:59 & Y & -- & -- & -- & 0.407 & (2) & 1 & 1 \rule[-3pt]{0cm}{0ex}\\
MACSJ0242.5$-$2132 & 02:42:35.9 & $-$21:32:26 & Y & -- & -- & -- & 0.314 & (2) & 1 & 1 \rule[-3pt]{0cm}{0ex}\\
MACSJ0257.6$-$2209 & 02:57:41.1 & $-$22:09:18 & Y & Y & Y & Y & 0.322 & (2) & -- & 2 \rule[-3pt]{0cm}{0ex}\\
MACSJ0308.9$+$2645 & 03:08:57.6 & $+$26:45:33 & Y & Y & -- & -- & 0.356 & (2) & 3 & 2 \rule[-3pt]{0cm}{0ex}\\
MACSJ0404.6$+$1109 & 04:04:33.1 & $+$11:08:07 & -- & Y & -- & -- & 0.358 & (2) & 4 & 3 \rule[-3pt]{0cm}{0ex}\\
MACSJ0449.3$-$2848 & 04:49:20.7 & $-$28:49:09 & Y & -- & -- & -- & 0.327 & (1) & -- & 2 \rule[-3pt]{0cm}{0ex}\\
MACSJ0451.9$+$0006 & 04:51:54.6 & $+$00:06:18 & Y & Y & Y & Y & 0.429 & (3) & 2 & 3 \rule[-3pt]{0cm}{0ex}\\
MACSJ0520.7$-$1328 & 05:20:42.0 & $-$13:28:47 & Y & -- & Y & Y & 0.336 & (2) & 2 & 3 \rule[-3pt]{0cm}{0ex}\\
MACSJ0521.4$-$2754 & 05:21:26.2 & $-$27:54:42 & Y & Y & -- & -- & 0.314 & (1) & -- & 3 \rule[-3pt]{0cm}{0ex}\\
MACSJ0547.0$-$3904 & 05:47:01.5 & $-$39:04:26 & Y & Y & -- & -- & 0.319 & (2) & 2 & 3 \rule[-3pt]{0cm}{0ex}\\
MACSJ0600.1$-$2008 & 06:00:08.6 & $-$20:07:36 & -- & Y & -- & -- & 0.427 & (1) & -- & 4 \rule[-3pt]{0cm}{0ex}\\
MACSJ0611.8$-$3036 & 06:11:49.6 & $-$30:38:09 & -- & -- & Y & Y & 0.320 & (1) & -- & 4 \rule[-3pt]{0cm}{0ex}\\
MACSJ0712.3$+$5931 & 07:12:20.5 & $+$59:32:20 & Y & Y & Y & Y & 0.328 & (1) & 2 & 1 \rule[-3pt]{0cm}{0ex}\\
MACSJ0845.4$+$0327 & 08:45:27.8 & $+$03:27:39 & Y & Y & -- & -- & 0.329 & (1) & -- & 2 \rule[-3pt]{0cm}{0ex}\\
MACSJ0913.7$+$4056 & 09:13:45.5 & $+$40:56:28 & Y & -- & -- & -- & 0.442 & (3) & 1 & 1 \rule[-3pt]{0cm}{0ex}\\
MACSJ0916.1$-$0023 & 09:16:11.6 & $-$00:23:36 & Y & Y & Y & Y & 0.320 & (1) & -- & 4 \rule[-3pt]{0cm}{0ex}\\
MACSJ0947.2$+$7623 & 09:47:12.8 & $+$76:23:14 & Y & Y & Y & Y & 0.354 & (2) & 1 & 1 \rule[-3pt]{0cm}{0ex}\\
MACSJ0949.8$+$1708 & 09:49:51.8 & $+$17:07:10 & Y & Y & -- & -- & 0.384 & (2) & 2 & 3 \rule[-3pt]{0cm}{0ex}\\
MACSJ1006.9$+$3200 & 10:06:55.4 & $+$32:00:54 & Y & Y & -- & -- & 0.403 & (3) & 4 & 4 \rule[-3pt]{0cm}{0ex}\\
MACSJ1105.7$-$1014 & 11:05:46.8 & $-$10:14:46 & Y & -- & -- & -- & 0.415 & (3) & 2 & 2 \rule[-3pt]{0cm}{0ex}\\
MACSJ1115.2$+$5320 & 11:15:16.2 & $+$53:19:36 & Y & Y & Y & Y & 0.466 & (3) & 3 & 3 \rule[-3pt]{0cm}{0ex}\\
MACSJ1115.8$+$0129 & 11:15:51.9 & $+$01:29:55 & Y & -- & -- & -- & 0.354 & (2) & 1 & 1 \rule[-3pt]{0cm}{0ex}\\
MACSJ1124.5$+$4351 & 11:24:29.8 & $+$43:51:26 & Y & Y & Y & Y & 0.368 & (1) & -- & 1 \rule[-3pt]{0cm}{0ex}\\
MACSJ1133.2$+$5008 & 11:33:13.0 & $+$50:08:25 & Y & Y & Y & Y & 0.389 & (1) & -- & 3 \rule[-3pt]{0cm}{0ex}\\
MACSJ1141.6$-$1905 & 11:41:41.0 & $-$19:05:21 & -- & -- & Y & Y & 0.305 & (1) & -- & 3 \rule[-3pt]{0cm}{0ex}\\
MACSJ1142.4$+$5831 & 11:42:24.3 & $+$58:31:47 & Y & Y & Y & Y & 0.326 & (1) & 4 & 4 \rule[-3pt]{0cm}{0ex}\\
MACSJ1206.2$-$0847 & 12:06:12.1 & $-$08:48:03 & Y & -- & -- & -- & 0.439 & (2) & 2 & 2 \rule[-3pt]{0cm}{0ex}\\
MACSJ1226.8$+$2153C$^a$ & 12:26:42.5 & $+$21:52:55 & Y & Y & Y & Y & 0.437 & (3) & -- & 3 \rule[-3pt]{0cm}{0ex}\\
MACSJ1236.9$+$6311 & 12:37:00.6 & $+$63:11:12 & Y & Y & Y & Y & 0.302 & (1) & 3 & 2 \rule[-3pt]{0cm}{0ex}\\
MACSJ1258.0$+$4702 & 12:58:03.3 & $+$47:02:58 & Y & Y & -- & -- & 0.331 & (1) & -- & 2 \rule[-3pt]{0cm}{0ex}\\
MACSJ1319.9$+$7003 & 13:20:08.4 & $+$70:04:39 & Y & Y & Y & Y & 0.327 & (2) & 2 & 1 \rule[-3pt]{0cm}{0ex}\\
MACSJ1328.2$+$5244 & 13:28:13.6 & $+$52:43:47 & Y & Y & -- & -- & 0.321 & (1) & -- & 3 \rule[-3pt]{0cm}{0ex}\\
MACSJ1354.6$+$7715 & 13:54:25.3 & $+$77:15:35 & Y & Y & Y & Y & 0.397 & (1) & 4 & 4 \rule[-3pt]{0cm}{0ex}\\
MACSJ1359.1$-$1929 & 13:59:10.2 & $-$19:29:25 & -- & Y & Y & Y & 0.447 & (3) & 1 & 1 \rule[-3pt]{0cm}{0ex}\\
MACSJ1427.6$-$2521 & 14:27:39.5 & $-$25:21:02 & -- & Y & -- & -- & 0.318 & (2) & 1 & 1 \rule[-3pt]{0cm}{0ex}\\
MACSJ1447.4$+$0827 & 14:47:26.0 & $+$08:28:25 & Y & Y & -- & -- & 0.376 & (3) & -- & 1 \rule[-3pt]{0cm}{0ex}\\
MACSJ1452.9$+$5802 & 14:52:55.5 & $+$58:02:39 & Y & Y & -- & -- & 0.324 & (1) & -- & 4 \rule[-3pt]{0cm}{0ex}\\
MACSJ1526.7$+$1647 & 15:26:42.5 & $+$16:47:32 & Y & -- & Y & Y & 0.338 & (1) & -- & 3 \rule[-3pt]{0cm}{0ex}\\
MACSJ1551.9$-$0207 & 15:51:58.5 & $-$02:07:50 & -- & Y & Y & Y & 0.300 & (1) & -- & 1 \rule[-3pt]{0cm}{0ex}\\
MACSJ1621.3$+$3810 & 16:21:24.8 & $+$38:10:09 & Y & Y & Y & Y & 0.465 & (3) & 1 & 2 \rule[-3pt]{0cm}{0ex}\\
MACSJ1625.7$-$0830 & 16:25:45.9 & $-$08:30:55 & -- & Y & Y & Y & 0.464 & (1) & -- & 2 \rule[-3pt]{0cm}{0ex}\\
MACSJ1644.9$+$0139 & 16:45:00.8 & $+$01:40:01 & Y & Y & -- & -- & 0.336 & (1) & -- & 3 \rule[-3pt]{0cm}{0ex}\\
MACSJ1652.3$+$5534 & 16:52:18.7 & $+$55:34:58 & Y & Y & Y & Y & 0.324 & (1) & -- & 1 \rule[-3pt]{0cm}{0ex}\\
MACSJ1731.6$+$2252 & 17:31:39.0 & $+$22:52:11 & Y & Y & Y & Y & 0.389 & (2) & 4 & 4 \rule[-3pt]{0cm}{0ex}\\
MACSJ1738.1$+$6006 & 17:38:06.9 & $+$60:06:18 & Y & Y & Y & Y & 0.329 & (1) & -- & 1 \rule[-3pt]{0cm}{0ex}\\\hline
\end{tabular}
\caption{MACS clusters with \textit{Hubble} SNAPshots. For explanation of morphology codes see Section~\ref{sec:data}. Redshift references: (1) this work; (2) \citet{Ebeling2010}; (3) \citet{Mann2012}. \newline
$^a$ See (3) for overall morphology of this three-component cluster configuration.}\label{tab:clusters}
\end{table*}

\setcounter{table}{5}
\begin{table*}
\begin{tabular}{lcccccccccc}
\hline\\[-6mm]
& & & & & & & & $z$ & \multicolumn{2}{c}{Morphology\rule{0cm}{15pt}}\\ 
Name & Right Ascension & Declination & F606 & F814 & F110W & F140W & $z$ & reference & X-ray & Optical\\ \hline \\[-4mm]
MACSJ1752.0$+$4440 & 17:51:58.8 & $+$44:39:36 & Y & Y & Y & Y & 0.364 & (1) & -- & 4 \rule[-3pt]{0cm}{18pt}\\
MACSJ1806.8$+$2931 & 18:06:52.4 & $+$29:30:13 & Y & Y & -- & -- & 0.300 & (1) & -- & 1 \rule[-3pt]{0cm}{0ex}\\
MACSJ2003.4$-$2322 & 20:03:29.7 & $-$23:24:25 & -- & Y & Y & Y & 0.316 & (1) & 4 & 4 \rule[-3pt]{0cm}{0ex}\\
MACSJ2046.0$-$3430 & 20:46:00.5 & $-$34:30:18 & -- & -- & Y & Y & 0.423 & (3) & 1 & 2 \rule[-3pt]{0cm}{0ex}\\
MACSJ2050.7$+$0123 & 20:50:42.4 & $+$01:23:39 & Y & Y & Y & Y & 0.333 & (1) & -- & 3 \rule[-3pt]{0cm}{0ex}\\
MACSJ2051.1$+$0215 & 20:51:10.9 & $+$02:16:05 & Y & Y & Y & Y & 0.321 & (1) & -- & 3 \rule[-3pt]{0cm}{0ex}\\
MACSJ2134.6$-$2706 & 21:34:36.0 & $-$27:05:56 & -- & Y & -- & -- & 0.363 & (1) & -- & 2 \rule[-3pt]{0cm}{0ex}\\
MACSJ2135.2$-$0102 & 21:35:12.1 & $-$01:02:59 & Y & Y & Y & Y & 0.325 & (1) & 2 & 3 \rule[-3pt]{0cm}{0ex}\\
MACSJ2149.3$+$0951 & 21:49:19.7 & $+$09:51:37 & Y & Y & -- & -- & 0.375 & (1) & -- & 2 \rule[-3pt]{0cm}{0ex}\\
MACSJ2211.7$-$0349 & 22:11:45.9 & $-$03:49:45 & Y & -- & -- & -- & 0.397 & (2) & 2 & 2 \rule[-3pt]{0cm}{0ex}\\
MACSJ2229.7$-$2755 & 22:29:45.2 & $-$27:55:36 & -- & -- & Y & Y & 0.324 & (2) & 1 & 1 \rule[-3pt]{0cm}{0ex}\\
MACSJ2241.8$+$1732 & 22:41:56.6 & $+$17:32:43 & Y & Y & -- & -- & 0.317 & (1) & -- & 4 \rule[-3pt]{0cm}{0ex}\\
MACSJ2243.3$-$0935 & 22:43:20.4 & $-$09:35:22 & Y & -- & -- & -- & 0.447 & (2) & 3 & 4 \rule[-3pt]{0cm}{0ex}\\
MACSJ2245.0$+$2637 & 22:45:04.7 & $+$26:38:05 & -- & -- & Y & Y & 0.301 & (2) & 1 & 2 \rule[-3pt]{0cm}{0ex}\\
MACSJ2245.4$+$2808 & 22:45:24.1 & $+$28:08:01 & -- & Y & -- & -- & 0.340 & (1) & -- & 4 \rule[-3pt]{0cm}{0ex}\\
SMACSJ0018.9$-$4051 & 00:19:01.5 & $-$40:51:50 & Y & -- & -- & -- & 0.477 & (1) & 2 & -- \rule[-3pt]{0cm}{0ex}\\
SMACSJ0040.8$-$4407 & 00:40:50.1 & $-$44:07:49 & -- & -- & Y & Y & 0.363 & (1) & 2 & -- \rule[-3pt]{0cm}{0ex}\\
SMACSJ0234.7$-$5831 & 02:34:46.1 & $-$58:31:07 & Y & Y & Y & Y & 0.408 & (1) & 1 & 3 \rule[-3pt]{0cm}{0ex}\\
SMACSJ0304.3$-$4401 & 03:04:16.9 & $-$44:01:31 & Y & -- & -- & -- & 0.460 & (1) & 4 & -- \rule[-3pt]{0cm}{0ex}\\
SMACSJ0332.8$-$8452 & 03:33:06.6 & $-$84:53:42 & -- & -- & Y & Y & 0.370 & (1) & -- & -- \rule[-3pt]{0cm}{0ex}\\
SMACSJ0439.2$-$4600 & 04:39:14.0 & $-$46:00:49 & Y & -- & -- & -- & 0.320 & (1) & 1 & -- \rule[-3pt]{0cm}{0ex}\\
SMACSJ0549.3$-$6205 & 05:49:17.0 & $-$62:05:11 & Y & Y & Y & Y & 0.375 & (1) & -- & 4 \rule[-3pt]{0cm}{0ex}\\
SMACSJ0600.2$-$4353 & 06:00:17.5 & $-$43:53:19 & Y & Y & Y & Y & 0.300 & (1) & -- & 4 \rule[-3pt]{0cm}{0ex}\\
SMACSJ0723.3$-$7327 & 07:23:22.7 & $-$73:27:14 & Y & Y & -- & -- & 0.404 & (1) & 3 & 4 \rule[-3pt]{0cm}{0ex}\\
SMACSJ1519.1$-$8130 & 15:18:48.7 & $-$81:30:14 & Y & -- & -- & -- & 0.480 & (1) & -- & -- \rule[-3pt]{0cm}{0ex}\\
SMACSJ2031.8$-$4036 & 20:31:50.7 & $-$40:37:15 & Y & Y & Y & Y & 0.342 & (1) & 3 & 4 \rule[-3pt]{0cm}{0ex}\\
SMACSJ2131.1$-$4019 & 21:31:05.0 & $-$40:19:21 & Y & Y & Y & Y & 0.421 & (1) & -- & 1 \rule[-3pt]{0cm}{0ex}\\
SMACSJ2332.4$-$5358 & 23:32:27.5 & $-$53:58:29 & -- & -- & Y & Y & 0.403 & (1) & -- & -- \\ \hline 
\end{tabular}
\contcaption{MACS clusters with \textit{Hubble} SNAPshots. For explanation of morphology codes see Section~\ref{sec:data}. Redshift references: (1) this work; (2) \citet{Ebeling2010}; (3) \citet{Mann2012}.}\label{tab:clusters}
\end{table*}
\clearpage

\begin{table*}
\begin{tabular}{lccccccc}
\hline\\[-6mm]
& \multicolumn{3}{c}{\dotfill BCG\dotfill} & \multicolumn{3}{c}{\dotfill G2\dotfill\rule{0cm}{15pt}} \\
Cluster & mag$^a$ & RA & Dec & mag$^a$ & RA & Dec & band \\ \hline\\[-4mm]
MACSJ0011.7$-$1523 & $18.40$ & 00:11:42.84 & $-$15:23:21.7 & $18.98$ & 00:11:45.69 & $-$15:24:50.6 & F814W \rule[-3pt]{0cm}{18pt}\\
MACSJ0027.8$+$2616 & $19.20$ & 00:27:45.79 & $+$26:16:26.5 & $19.45$ & 00:27:43.69 & $+$26:16:21.4 & F606W \rule[-3pt]{0cm}{0ex}\\
MACSJ0032.1$+$1808 & $18.50$ & 00:32:09.41 & $+$18:06:55.7 & $18.72$ & 00:32:08.23 & $+$18:06:25.0 & F814W \rule[-3pt]{0cm}{0ex}\\
MACSJ0033.8$-$0751 & $17.57$ & 00:33:51.30 & $-$07:50:15.5 & $17.89$ & 00:33:53.14 & $-$07:52:10.5 & F814W \rule[-3pt]{0cm}{0ex}\\
MACSJ0034.4$+$0225 & $18.52$ & 00:34:28.16 & $+$02:25:22.3 & $18.53$ & 00:34:25.98 & $+$02:25:24.9 & F814W \rule[-3pt]{0cm}{0ex}\\
MACSJ0034.9$+$0234 & $17.71$ & 00:34:57.82 & $+$02:33:31.5 & $18.28$ & 00:34:56.79 & $+$02:33:18.7 & F814W \rule[-3pt]{0cm}{0ex}\\
MACSJ0035.4$-$2015 & $17.82$ & 00:35:26.12 & $-$20:15:44.9 & $19.28$ & 00:35:22.99 & $-$20:14:35.7 & F814W \rule[-3pt]{0cm}{0ex}\\
MACSJ0051.6$+$2720 & $16.32$ & 00:51:38.59 & $+$27:19:59.9 & $18.04$ & 00:51:41.62 & $+$27:20:01.5 & F110W \rule[-3pt]{0cm}{0ex}\\
MACSJ0110.1$+$3211 & $18.04$ & 01:10:07.19 & $+$32:10:48.5 & $18.85$ & 01:10:06.97 & $+$32:10:28.4 & F814W \rule[-3pt]{0cm}{0ex}\\
MACSJ0140.0$-$0555 & $18.44$ & 01:40:00.83 & $-$05:55:03.2 & $19.01$ & 01:40:03.20 & $-$05:55:21.8 & F814W \rule[-3pt]{0cm}{0ex}\\
MACSJ0140.0$-$3410 & $19.53$ & 01:40:05.48 & $-$34:10:38.3 & $20.10$ & 01:40:01.24 & $-$34:09:48.2 & F606W \rule[-3pt]{0cm}{0ex}\\
MACSJ0150.3$-$1005 & $17.29$ & 01:50:21.25 & $-$10:05:30.7 & $19.15$ & 01:50:18.40 & $-$10:05:12.0 & F814W \rule[-3pt]{0cm}{0ex}\\
MACSJ0152.5$-$2852 & $18.93$ & 01:52:34.49 & $-$28:53:37.2 & $19.01$ & 01:52:33.73 & $-$28:55:18.4 & F814W \rule[-3pt]{0cm}{0ex}\\
MACSJ0159.8$-$0849 & $18.89$ & 01:59:49.31 & $-$08:49:58.9 & $20.18$ & 01:59:58.46 & $-$08:50:07.2 & F606W \rule[-3pt]{0cm}{0ex}\\
MACSJ0242.5$-$2132 & $17.80$ & 02:42:35.94 & $-$21:32:25.9 & $20.31$ & 02:42:31.70 & $-$21:31:06.7 & F606W \rule[-3pt]{0cm}{0ex}\\
MACSJ0257.6$-$2209 & $16.81$ & 02:57:41.08 & $-$22:09:17.7 & $18.09$ & 02:57:37.01 & $-$22:10:15.4 & F814W \rule[-3pt]{0cm}{0ex}\\
MACSJ0308.9$+$2645 & $17.89$ & 03:08:55.93 & $+$26:45:37.3 & $18.80$ & 03:08:49.60 & $+$26:45:56.1 & F814W \rule[-3pt]{0cm}{0ex}\\
MACSJ0404.6$+$1109 & $18.11$ & 04:04:32.71 & $+$11:08:04.7 & $18.19$ & 04:04:33.67 & $+$11:07:53.3 & F814W \rule[-3pt]{0cm}{0ex}\\
MACSJ0449.3$-$2848 & $19.74$ & 04:49:20.72 & $-$28:49:08.8 & $19.36$ & 04:49:15.66 & $-$28:49:08.7 & F606W \rule[-3pt]{0cm}{0ex}\\
MACSJ0451.9$+$0006 & $18.51$ & 04:51:54.61 & $+$00:06:18.2 & $18.94$ & 04:51:53.99 & $+$00:06:18.2 & F814W \rule[-3pt]{0cm}{0ex}\\
MACSJ0520.7$-$1328 & $19.13$ & 05:20:42.05 & $-$13:28:46.8 & $19.33$ & 05:20:48.96 & $-$13:29:38.1 & F606W \rule[-3pt]{0cm}{0ex}\\
MACSJ0521.4$-$2754 & $17.83$ & 05:21:25.45 & $-$27:54:53.2 & $18.51$ & 05:21:25.54 & $-$27:55:14.8 & F814W \rule[-3pt]{0cm}{0ex}\\
MACSJ0547.0$-$3904 & $17.68$ & 05:47:01.52 & $-$39:04:26.4 & $19.09$ & 05:47:03.75 & $-$39:04:35.0 & F814W \rule[-3pt]{0cm}{0ex}\\
MACSJ0600.1$-$2008 & $18.51$ & 06:00:08.19 & $-$20:08:09.2 & $19.24$ & 06:00:10.25 & $-$20:07:02.6 & F814W \rule[-3pt]{0cm}{0ex}\\
MACSJ0611.8$-$3036 & $16.75$ & 06:11:50.21 & $-$30:38:55.1 & $16.47$ & 06:11:44.95 & $-$30:37:00.8 & F110W \rule[-3pt]{0cm}{0ex}\\
MACSJ0712.3$+$5931 & $17.51$ & 07:12:20.50 & $+$59:32:20.4 & $18.67$ & 07:12:20.61 & $+$59:31:32.0 & F814W \rule[-3pt]{0cm}{0ex}\\
MACSJ0845.4$+$0327 & $17.79$ & 08:45:27.77 & $+$03:27:38.8 & $18.35$ & 08:45:29.25 & $+$03:27:28.4 & F814W \rule[-3pt]{0cm}{0ex}\\
MACSJ0913.7$+$4056 & $18.31$ & 09:13:45.50 & $+$40:56:28.5 & $20.73$ & 09:13:35.53 & $+$40:56:21.3 & F606W \rule[-3pt]{0cm}{0ex}\\
MACSJ0916.1$-$0023 & $17.90$ & 09:16:09.24 & $-$00:24:16.5 & $17.94$ & 09:16:17.56 & $+$00:24:05.9 & F814W \rule[-3pt]{0cm}{0ex}\\
MACSJ0947.2$+$7623 & $17.01$ & 09:47:12.78 & $+$76:23:13.6 & $19.73$ & 09:47:14.43 & $+$76:23:27.4 & F814W \rule[-3pt]{0cm}{0ex}\\
MACSJ0949.8$+$1708 & $18.09$ & 09:49:51.80 & $+$17:07:10.5 & $18.65$ & 09:49:55.40 & $+$17:06:38.4 & F814W \rule[-3pt]{0cm}{0ex}\\
MACSJ1006.9$+$3200 & $17.77$ & 10:06:54.68 & $+$32:01:32.0 & $18.60$ & 10:06:55.27 & $+$32:00:01.1 & F814W \rule[-3pt]{0cm}{0ex}\\
MACSJ1105.7$-$1014 & $19.51$ & 11:05:46.80 & $-$10:14:46.0 & $20.15$ & 11:05:46.22 & $-$10:14:24.7 & F606W \rule[-3pt]{0cm}{0ex}\\
MACSJ1115.2$+$5320 & $18.10$ & 11:15:14.85 & $+$53:19:54.3 & $18.93$ & 11:15:18.75 & $+$53:19:48.3 & F814W \rule[-3pt]{0cm}{0ex}\\
MACSJ1115.8$+$0129 & $19.26$ & 11:15:51.89 & $+$01:29:54.7 & $19.77$ & 11:15:46.33 & $+$01:29:39.2 & F606W \rule[-3pt]{0cm}{0ex}\\
MACSJ1124.5$+$4351 & $18.74$ & 11:24:29.78 & $+$43:51:25.5 & $19.30$ & 11:24:38.23 & $+$43:51:35.2 & F814W \rule[-3pt]{0cm}{0ex}\\
MACSJ1133.2$+$5008 & $18.09$ & 11:33:13.17 & $+$50:08:39.9 & $18.53$ & 11:33:09.89 & $+$50:08:18.9 & F814W \rule[-3pt]{0cm}{0ex}\\
MACSJ1141.6$-$1905 & $17.07$ & 11:41:40.83 & $-$19:05:15.5 & $17.03$ & 11:41:40.60 & $-$19:05:28.2 & F110W \rule[-3pt]{0cm}{0ex}\\
MACSJ1142.4$+$5831 & $16.69$ & 11:42:24.80 & $+$58:32:05.5 & $17.60$ & 11:42:26.28 & $+$58:32:43.4 & F814W \rule[-3pt]{0cm}{0ex}\\
MACSJ1206.2$-$0847 & $19.78$ & 12:06:12.13 & $-$08:48:03.3 & $20.05$ & 12:06:05.37 & $-$08:49:04.7 & F606W \rule[-3pt]{0cm}{0ex}\\
MACSJ1226.8$+$2153C & $18.74$ & 12:26:38.80 & $+$21:53:22.7 & $18.88$ & 12:26:40.79 & $+$21:52:58.1 & F814W \rule[-3pt]{0cm}{0ex}\\
MACSJ1236.9$+$6311 & $17.54$ & 12:36:58.72 & $+$63:11:13.6 & $17.56$ & 12:36:59.31 & $+$63:11:11.5 & F814W \rule[-3pt]{0cm}{0ex}\\
MACSJ1258.0$+$4702 & $18.15$ & 12:58:03.29 & $+$47:02:57.6 & $18.34$ & 12:58:02.09 & $+$47:02:54.4 & F814W \rule[-3pt]{0cm}{0ex}\\
MACSJ1319.9$+$7003 & $17.34$ & 13:20:08.40 & $+$70:04:39.2 & $18.35$ & 13:20:00.78 & $+$70:03:15.1 & F814W \rule[-3pt]{0cm}{0ex}\\
MACSJ1328.2$+$5244 & $17.84$ & 13:28:12.08 & $+$52:43:18.8 & $17.98$ & 13:28:15.69 & $+$52:44:24.8 & F814W \rule[-3pt]{0cm}{0ex}\\
MACSJ1354.6$+$7715 & $18.09$ & 13:54:42.72 & $+$77:15:17.4 & $18.40$ & 13:54:34.68 & $+$77:15:48.5 & F814W \rule[-3pt]{0cm}{0ex}\\
MACSJ1359.1$-$1929 & $18.68$ & 13:59:10.25 & $-$19:29:24.8 & $19.50$ & 13:59:09.01 & $-$19:27:44.5 & F814W \rule[-3pt]{0cm}{0ex}\\
MACSJ1427.6$-$2521 & $17.41$ & 14:27:39.47 & $-$25:21:02.2 & $18.70$ & 14:27:32.34 & $-$25:21:37.6 & F814W \rule[-3pt]{0cm}{0ex}\\
MACSJ1447.4$+$0827 & $17.02$ & 14:47:26.03 & $+$08:28:24.7 & $19.37$ & 14:47:25.76 & $+$08:28:04.7 & F814W \rule[-3pt]{0cm}{0ex}\\
MACSJ1452.9$+$5802 & $17.69$ & 14:52:57.49 & $+$58:02:55.1 & $18.44$ & 14:53:01.41 & $+$58:03:24.4 & F814W \rule[-3pt]{0cm}{0ex}\\
MACSJ1526.7$+$1647 & $19.18$ & 15:26:42.67 & $+$16:47:38.7 & $20.21$ & 15:26:41.03 & $+$16:47:11.5 & F606W \rule[-3pt]{0cm}{0ex}\\
MACSJ1551.9$-$0207 & $17.56$ & 15:51:58.47 & $-$02:07:50.3 & $20.00$ & 15:51:58.83 & $-$02:08:14.8 & F814W \rule[-3pt]{0cm}{0ex}\\
MACSJ1621.3$+$3810 & $18.79$ & 16:21:24.75 & $+$38:10:08.8 & $19.40$ & 16:21:22.55 & $+$38:09:24.8 & F814W \rule[-3pt]{0cm}{0ex}\\
MACSJ1625.7$-$0830 & $19.39$ & 16:25:45.92 & $-$08:30:54.9 & $19.46$ & 16:25:48.57 & $-$08:32:02.6 & F814W \rule[-3pt]{0cm}{0ex}\\
MACSJ1644.9$+$0139 & $17.89$ & 16:45:00.42 & $+$01:39:57.0 & $17.94$ & 16:45:01.22 & $+$01:40:15.7 & F814W \rule[-3pt]{0cm}{0ex}\\
MACSJ1652.3$+$5534 & $17.69$ & 16:52:18.71 & $+$55:34:58.2 & $18.28$ & 16:52:17.54 & $+$55:34:56.7 & F814W \rule[-3pt]{0cm}{0ex}\\
MACSJ1731.6$+$2252 & $17.30$ & 17:31:39.95 & $+$22:51:58.5 & $18.54$ & 17:31:41.73 & $+$22:53:34.4 & F814W \rule[-3pt]{0cm}{0ex}\\
MACSJ1738.1$+$6006 & $17.78$ & 17:38:06.89 & $+$60:06:17.9 & $19.17$ & 17:38:10.14 & $+$60:05:52.5 & F814W \rule[-3pt]{0cm}{0ex}\\ \hline \\[-2mm]
\end{tabular}
\caption{BCG and second brightest cluster members (selection as described in Section~\ref{sec:BCG}).\label{tab:BCG_G2}\newline
$^a$ Magnitude uncertainties $\la .01$. Magnitudes measured in F814W if available; else, in F606W, if available; else, in F110W. The band used for each cluster appears in the final column of the table.
}
\end{table*}

\begin{table*}
\begin{tabular}{lccccccc}
\hline\\[-6mm]
& \multicolumn{3}{c}{\dotfill BCG\dotfill} & \multicolumn{3}{c}{\dotfill G2\dotfill\rule{0cm}{15pt}} \\
Cluster & mag$^a$ & RA & Dec & mag$^a$ & RA & Dec & band \\ \hline\\[-4mm]
MACSJ1752.0$+$4440 & $18.43$ & 17:51:53.39 & $+$44:39:13.6 & $18.89$ & 17:52:03.83 & $+$44:39:44.0 & F814W \rule[-3pt]{0cm}{18pt}\\
MACSJ1806.8$+$2931 & $18.03$ & 18:06:52.41 & $+$29:30:12.7 & $19.21$ & 18:06:47.52 & $+$29:30:05.6 & F814W \rule[-3pt]{0cm}{0ex}\\
MACSJ2003.4$-$2322 & $17.34$ & 20:03:36.77 & $-$23:25:10.6 & $17.94$ & 20:03:25.23 & $-$23:24:57.0 & F814W \rule[-3pt]{0cm}{0ex}\\
MACSJ2046.0$-$3430 & $17.11$ & 20:46:00.54 & $-$34:30:17.9 & $16.60$ & 20:45:59.37 & $-$34:29:57.8 & F110W \rule[-3pt]{0cm}{0ex}\\
MACSJ2050.7$+$0123 & $17.70$ & 20:50:43.15 & $+$01:23:28.7 & $18.47$ & 20:50:43.95 & $+$01:23:31.7 & F814W \rule[-3pt]{0cm}{0ex}\\
MACSJ2051.1$+$0215 & $17.62$ & 20:51:09.58 & $+$02:16:13.7 & $17.70$ & 20:51:12.27 & $+$02:15:58.2 & F814W \rule[-3pt]{0cm}{0ex}\\
MACSJ2134.6$-$2706 & $17.50$ & 21:34:36.00 & $-$27:05:55.8 & $19.00$ & 21:34:42.99 & $-$27:06:18.7 & F814W \rule[-3pt]{0cm}{0ex}\\
MACSJ2135.2$-$0102 & $17.43$ & 21:35:12.08 & $-$01:02:58.7 & $18.31$ & 21:35:09.69 & $-$01:01:35.6 & F814W \rule[-3pt]{0cm}{0ex}\\
MACSJ2149.3$+$0951 & $18.36$ & 21:49:19.66 & $+$09:51:37.1 & $19.30$ & 21:49:18.39 & $+$09:51:58.7 & F814W \rule[-3pt]{0cm}{0ex}\\
MACSJ2211.7$-$0349 & $18.73$ & 22:11:45.91 & $-$03:49:44.7 & $20.43$ & 22:11:45.82 & $-$03:50:47.8 & F606W \rule[-3pt]{0cm}{0ex}\\
MACSJ2229.7$-$2755 & $16.24$ & 22:29:45.22 & $-$27:55:36.3 & $17.45$ & 22:29:43.86 & $-$27:55:25.0 & F110W \rule[-3pt]{0cm}{0ex}\\
MACSJ2241.8$+$1732 & $17.87$ & 22:41:58.85 & $+$17:31:40.3 & $17.67$ & 22:41:56.25 & $+$17:32:11.6 & F814W \rule[-3pt]{0cm}{0ex}\\
MACSJ2243.3$-$0935 & $20.49$ & 22:43:19.83 & $-$09:34:51.5 & $20.51$ & 22:43:21.26 & $-$09:35:10.4 & F606W \rule[-3pt]{0cm}{0ex}\\
MACSJ2245.0$+$2637 & $15.25$ & 22:45:04.68 & $+$26:38:04.8 & $17.58$ & 22:45:10.31 & $+$26:37:53.5 & F110W \rule[-3pt]{0cm}{0ex}\\
MACSJ2245.4$+$2808 & $17.82$ & 22:45:27.76 & $+$28:09:00.2 & $17.98$ & 22:45:21.17 & $+$28:07:05.8 & F814W \rule[-3pt]{0cm}{0ex}\\
SMACSJ0018.9$-$4051$^b$ & $20.12$ & 00:19:01.54 & $-$40:51:50.5 & -- & -- & -- & F606W \rule[-3pt]{0cm}{0ex}\\
SMACSJ0040.8$-$4407$^b$ & $16.35$ & 00:40:49.94 & $-$44:07:51.0 & -- & -- & -- & F110W \rule[-3pt]{0cm}{0ex}\\
SMACSJ0234.7$-$5831 & $18.78$ & 02:34:50.76 & $-$58:30:46.9 & $18.96$ & 2:34:39.73 & $-$58:30:16.8 & F814W \rule[-3pt]{0cm}{0ex}\\
SMACSJ0304.3$-$4401$^b$ & $20.09$ & 03:04:16.85 & $-$44:01:31.7 & -- & -- & -- & F606W \rule[-3pt]{0cm}{0ex}\\
SMACSJ0332.8$-$8452$^b$ & $16.56$ & 03:33:14.18 & $-$84:53:25.2 & -- & -- & -- & F110W \rule[-3pt]{0cm}{0ex}\\
SMACSJ0439.2$-$4600$^b$ & $18.73$ & 04:39:13.96 & $-$46:00:49.1 & -- & -- & -- & F606W \rule[-3pt]{0cm}{0ex}\\
SMACSJ0549.3$-$6205 & $17.90$ & 05:49:19.98 & $-$62:05:13.6 & $18.02$ & 5:49:13.35 & $-$62:06:12.5 & F814W \rule[-3pt]{0cm}{0ex}\\
SMACSJ0600.2$-$4353 & $17.27$ & 06:00:12.96 & $-$43:53:29.4 & $18.11$ & 6:00:19.12 & $-$43:54:12.3 & F814W \rule[-3pt]{0cm}{0ex}\\
SMACSJ0723.3$-$7327 & $19.14$ & 07:23:18.42 & $-$73:27:17.2 & $19.21$ & 7:23:20.48 & $-$73:25:50.1 & F814W \rule[-3pt]{0cm}{0ex}\\
SMACSJ1519.1$-$8130$^b$ & $20.72$ & 15:18:54.97 & $-$81:30:23.7 & -- & -- & -- & F606W \rule[-3pt]{0cm}{0ex}\\
SMACSJ2031.8$-$4036 & $17.77$ & 20:31:47.82 & $-$40:36:54.3 & $18.25$ & 20:31:53.28 & $-$40:37:30.4 & F814W \rule[-3pt]{0cm}{0ex}\\
SMACSJ2131.1$-$4019 & $18.10$ & 21:31:04.95 & $-$40:19:21.2 & $19.42$ & 21:31:03.56 & $-$40:19:05.3 & F814W \rule[-3pt]{0cm}{0ex}\\
SMACSJ2332.4$-$5358$^b$ & $16.75$ & 23:32:27.52 & $-$53:58:28.1 & -- & -- & -- & F110W \rule[-8pt]{0cm}{0ex}\\ \hline
\end{tabular}
\contcaption{BCG and second brightest cluster members (selection as described in Section~\ref{sec:BCG}).\label{tab:BCG_G2}
\newline
$^a$ Magnitude uncertainties $\la .01$. Magnitudes measured in F814W if available; else, in F606W, if available; else, in F110W. The band used for each cluster appears in the final column of the table.\newline
$^b$ Colour information not available (from either \textit{HST} or ground-based imaging); thus distinguishing BCG and G2 from foreground galaxies is problematic. For these clusters we present our BCG identification tentatively, and we decline to identify G2.}
\label{lastpage}
\end{table*}

\end{document}